\documentclass[onecolumn]{aastex61}

\shorttitle{The size and shape of (10199) Chariklo.}

\begin{document}

\title{
Size and shape of Chariklo from multi-epoch stellar occultations
\footnote{Based on observations obtained at the Southern Astrophysical Research (SOAR) telescope, which is a joint project of the Minist\'{e}rio da Ci\^{e}ncia, Tecnologia, e Inova\c{c}\~{a}o (MCTI) da Rep\'{u}blica Federativa do Brasil, the U.S. National Optical Astronomy Observatory (NOAO), the University of North Carolina at Chapel Hill (UNC), and Michigan State University (MSU).}
}

\author[0000-0002-6477-1360]{R. Leiva}
\affiliation{LESIA/Observatoire de Paris, CNRS UMR 8109, Universit\'e Pierre et Marie Curie, Universit\'e Paris-Diderot, 5 place Jules Janssen, F-92195 Meudon C\'edex, France.}
\affiliation{Instituto de Astrof\'isica, Facultad de F\'isica, Pontificia Universidad Cat\'olica de Chile, Av. Vicu\~na Mackenna 4860, Santiago, Chile}
\correspondingauthor{R. Leiva}
\email{rnleiva@uc.cl}
\author[0000-0003-1995-0842]{B. Sicardy}
\affiliation{LESIA/Observatoire de Paris, CNRS UMR 8109, Universit\'e Pierre et Marie Curie, Universit\'e Paris-Diderot, 5 place Jules Janssen, F-92195 Meudon C\'edex, France.}
\author[0000-0002-1642-4065]{J. I. B. Camargo}
\affiliation{Observat\'orio Nacional/MCTIC, Rua General Jos\'e Cristino 77, RJ 20921-400, Rio de Janeiro, Brazil}
\affiliation{Laborat\'orio Interinstitucional de e-Astronomia - LIneA, Rua General Jos\'e Cristino 77, RJ 20921-400, Rio de Janeiro, Brazil}
\author{J.-L. Ortiz}
\affiliation{Instituto de Astrof\'isica de Andaluc\'ia, CSIC, Glorieta de la Astronom\'ia s/n, E-18008, Granada, Spain}
\author[0000-0002-2193-8204]{J. Desmars}
\affiliation{LESIA/Observatoire de Paris, CNRS UMR 8109, Universit\'e Pierre et Marie Curie, Universit\'e Paris-Diderot, 5 place Jules Janssen, F-92195 Meudon C\'edex, France.}
\author[0000-0001-7654-6809]{D. B\'erard}
\affiliation{LESIA/Observatoire de Paris, CNRS UMR 8109, Universit\'e Pierre et Marie Curie, Universit\'e Paris-Diderot, 5 place Jules Janssen, F-92195 Meudon C\'edex, France.}
\author{E. Lellouch}
\affiliation{LESIA, Observatoire de Paris, PSL Research University, CNRS, Sorbonne Universit\'es, UPMC Univ. Paris 06, Univ. Paris Diderot, Sorbonne Paris Cit\'e, 5 place Jules Janssen, 92195 Meudon, France}
\author[0000-0002-4058-0420]{E. Meza}
\affiliation{LESIA/Observatoire de Paris, CNRS UMR 8109, Universit\'e Pierre et Marie Curie, Universit\'e Paris-Diderot, 5 place Jules Janssen, F-92195 Meudon C\'edex, France.}
\author[0000-0003-0626-1749]{P. Kervella}
\affiliation{LESIA/Observatoire de Paris, CNRS UMR 8109, Universit\'e Pierre et Marie Curie, Universit\'e Paris-Diderot, 5 place Jules Janssen, F-92195 Meudon C\'edex, France.}
\affiliation{Unidad Mixta Internacional Franco-Chilena de Astronom\'ia (CNRS UMI 3386), Departamento de Astronom\'ia, Universidad de Chile, Camino El Observatorio 1515, Las Condes, Santiago, Chile}
\author{C. Snodgrass}
\affiliation{School of Physical Sciences, The Open University, Milton Keynes, MK7 6AA, UK}

\author{R. Duffard}
\affiliation{Instituto de Astrof\'isica de Andaluc\'ia, CSIC, Glorieta de la Astronom\'ia s/n, E-18008, Granada, Spain}

\author{N. Morales}
\affiliation{Instituto de Astrof\'isica de Andaluc\'ia, CSIC, Glorieta de la Astronom\'ia s/n, E-18008, Granada, Spain}

\author{A. R. Gomes-J\'unior}
\affiliation{Observat\'orio do Valongo/UFRJ, Ladeira Pedro Antonio 43, RJ 20.080-090, Rio de Janeiro, Brazil}

\author[0000-0002-4106-476X]{G. Benedetti-Rossi}
\affiliation{Observat\'orio Nacional/MCTIC, Rua General Jos\'e Cristino 77, RJ 20921-400, Rio de Janeiro, Brazil}
\affiliation{Laborat\'orio Interinstitucional de e-Astronomia - LIneA, Rua General Jos\'e Cristino 77, RJ 20921-400, Rio de Janeiro, Brazil}

\author{R. Vieira-Martins}
\affiliation{Observat\'orio do Valongo/UFRJ, Ladeira Pedro Antonio 43, RJ 20.080-090, Rio de Janeiro, Brazil}
\affiliation{Observat\'orio Nacional/MCTIC, Rua General Jos\'e Cristino 77, RJ 20921-400, Rio de Janeiro, Brazil}
\affiliation{Laborat\'orio Interinstitucional de e-Astronomia - LIneA, Rua General Jos\'e Cristino 77, RJ 20921-400, Rio de Janeiro, Brazil}

\author{F. Braga-Ribas}
\affiliation{Federal University of Technology- Paran\'a (UTFPR/DAFIS), Rua Sete de Setembro, 3165, CEP 80230-901, Curitiba, PR, Brazil}
\affiliation{Observat\'orio Nacional/MCTIC, Rua General Jos\'e Cristino 77, RJ 20921-400, Rio de Janeiro, Brazil}
\affiliation{Laborat\'orio Interinstitucional de e-Astronomia - LIneA, Rua General Jos\'e Cristino 77, RJ 20921-400, Rio de Janeiro, Brazil}

\author{M. Assafin}
\affiliation{Observat\'orio do Valongo/UFRJ, Ladeira Pedro Antonio 43, RJ 20.080-090, Rio de Janeiro, Brazil}

\author{B. E. Morgado}
\affiliation{Observat\'orio Nacional/MCTIC, Rua General Jos\'e Cristino 77, RJ 20921-400, Rio de Janeiro, Brazil}

\author{F. Colas}
\affiliation{IMCCE, Observatoire de Paris, PSL Research University, CNRS, Sorbonne Universit\'es, UPMC Univ. Paris 06, 77 Av. Denfert-Rochereau, F-75014, Paris, France}

\author{C. De Witt}
\affiliation{IOTA/ES, Barthold-Knaust-Strasse 8, D-30459 Hannover, Germany} 

\author{A. A. Sickafoose}
\affiliation{South African Astronomical Observatory, P.O. Box 9, 7935 Observatory, South Africa}
\affiliation{Department of Earth, Atmospheric, and Planetary Sciences, Massachusetts Institute of Technology Cambridge, MA 02139-4307, United States}

\author{H. Breytenbach}
\affiliation{South African Astronomical Observatory, P.O. Box 9, 7935 Observatory, South Africa}
\affiliation{University of Cape Town, Department of Astronomy, Rondebosch, Cape Town, 7700, South Africa}

\author[0000-0002-1480-5219]{J.-L. Dauvergne}
\affiliation{AFA/Ciel et Espace, 17 Emile Deutsch de la Meurthe, F-75014, Paris, France}
\author{P. Schoenau}
\affiliation{IOTA/ES, Barthold-Knaust-Strasse 8, D-30459 Hannover, Germany} 
\author{L. Maquet}
\affiliation{IMCCE, Observatoire de Paris, PSL Research University, CNRS, Sorbonne Universit\'es, UPMC Univ. Paris 06, 77 Av. Denfert-Rochereau, F-75014, Paris, France}
\author{K.-L. Bath}
\affiliation{IOTA/ES, Barthold-Knaust-Strasse 8, D-30459 Hannover, Germany} 
\affiliation{Internationale Amateursternwarte e. V. IAS, Hakos/Namibia and Bichler Str. 46, D-81479, Munich, Germany}
\author{H.-J. Bode}
\affiliation{IOTA/ES, Barthold-Knaust-Strasse 8, D-30459 Hannover, Germany} 
\affiliation{Internationale Amateursternwarte e. V. IAS, Hakos/Namibia and Bichler Str. 46, D-81479, Munich, Germany}
\author[0000-0002-6382-5443]{A. Cool}
\affiliation{Defence Science \& Technology Group, P.O. Box 1500, Edinburgh SA 5111, Australia}
\affiliation{The Heights Observatory, 12 Augustus St, Modbury Heights SA 5092, Australia}
\author{B. Lade}
\affiliation{Stockport Observatory, Astronomical Society of South Australia, Stockport, SA, Australia}
\affiliation{Defence Science \& Technology Group, P.O. Box 1500, Edinburgh SA 5111, Australia}
\affiliation{The Heights Observatory, 12 Augustus St, Modbury Heights SA 5092, Australia}
\author{S. Kerr}
\affiliation{Occultation Section of the Royal Astronomical Society of New Zealand (RASNZ), P.O. Box 3181, Wellington, New Zealand}
\affiliation{Astronomical Association of Queensland, 5 Curtis Street, Pimpama QLD 4209, Australia}
\author{D. Herald}
\affiliation{Occultation Section of the Royal Astronomical Society of New Zealand (RASNZ), P.O. Box 3181, Wellington, New Zealand}
\affiliation{International Occultation Timing Association (IOTA), P.O. Box 7152, Kent, WA 98042, USA}
\affiliation{Canberra Astronomical Society, Canberra, ACT, Australia}

\begin{abstract}

We use data from five stellar occultations observed between 2013 and 2016 to constrain Chariklo's size and shape, and the ring reflectivity. We consider four possible models for Chariklo (sphere, Maclaurin spheroid, tri-axial ellipsoid and Jacobi ellipsoid) and
 we use a Bayesian approach to estimate the corresponding parameters.
The spherical model has a radius $R=129\pm3$~km. 
The Maclaurin model has equatorial and polar radii
 $a=b=143^{+3}_{-6}$~km and $c=96^{+14}_{-4}$~km,
respectively, with density $970^{+300}_{-180}$~kg~m$^{-3}$.
The ellipsoidal model has semiaxes 
$a=148^{+6}_{-4}$~km,
$b=132^{+6}_{-5}$~km and
$c=102^{+10}_{-8}$~km. 
Finally, the Jacobi model has semiaxes 
$a$=157$\pm$4~km,
$b$=139$\pm$ 4~km and
$c$=86$\pm$1~km, 
and
density $796^{+2}_{-4}$~kg~m$^{-3}$ . 
Depending on the model, we obtain topographic features of 6-11~km,
typical of Saturn icy satellites with similar size and density.
We constrain Chariklo's geometric albedo between 3.1\% (sphere) and 4.9\% (ellipsoid),
while the ring $I/F$ reflectivity is less constrained between 0.6\% (Jacobi) and 8.9\% (sphere).
The ellipsoid model explains both the optical light curve and the long-term photometry variation of the system,
giving a plausible value for the geometric albedo of the ring particles of $10-15\%$.
The derived Chariklo's mass of 6-8$\times10^{18}$~kg places the rings close to the 3:1 resonance between the ring mean motion and 
Chariklo's rotation period.
\end{abstract}


\section{INTRODUCTION}

The Centaur object (10199) Chariklo is the only small object of the Solar System known so far 
to show the unambiguous presence of a ring system.
It was discovered during a ground-based stellar occultation in 2013 \citep{Braga-Ribas2014}, 
and confirmed by several subsequent observations \citep{Berard2017}.

Meanwhile, the basic physical characteristics of Chariklo remain fragmentary. 
Chariklo's radius estimations, taken from thermal measurements, 
vary from 108~km to 151~km,
with geometric albedo in the range 4-8\% \citep{Jewitt1998,Altenhoff2001,Sekiguchi2012,Bauer2013,Fornasier2014}.
The 2013 stellar occultation had a poor coverage of the main body, still providing a spheroidal shape with
 equatorial radius $a=144.9$~km and polar radius $c=114$~km. 

Rotational light curves obtained in the visible between 1997 and 2013 exhibit a variable peak-to-peak amplitude 
from non detectable in 1997 and 1999 to amplitudes of 0.11-0.13 in 2006 and 2013 respectively \citep{Davies1998,Peixinho2001,Fornasier2014,Galiazzo2016}.
This can be produced by an elongated body,  longitudinal albedo variegations or, more probably, a combination of both.
The best measurements provide a rotation period of 7.004$\pm$0.036 hr \citep{Fornasier2014}.
Spectroscopic measurements show the presence of water ice in the system \citep{Guilbert-Lepoutre2011,Duffard2014}.
Finally, no satellites have been detected around Chariklo, preventing any mass estimation.

The size, shape and density of Chariklo are important parameters to constrain the origins of both the main body and its rings.
Moreover, topographic features and/or an ellipsoidal shape may have drastic influence in the ring dynamics 
through resonances between ring mean motion and body rotation.

Here we use several stellar occultations to put constraints on the size and shape of Chariklo's main body.
This technique has been used on several TNOs and Centaur objects, including 
2002~TX$_{300}$ \citep{Elliot2010},
Eris \citep{Sicardy2011}, 
Makemake \citep{Ortiz2012}, 
Varuna \citep{Sicardy2010a}, 
Quaoar \citep{Braga-Ribas2013}, 
2002 KX$_{14}$ \citep{Alvarez-Candal2014}, 
2007 UK$_{126}$ \citep{Benedetti-Rossi2016,Schindler2016} and 
2003 AZ$_{84}$ \citep{Dias-Oliveira2017}.

Due to the very small angular size of Chariklo ($\sim80$ milliarcsecond, rings included),
prediction of stellar occultations are difficult and coverage of the shadow path is poor with only a few chords on the body per event.
In order to retrieve the full 3D structure of the body, we have to use some a priori hypotheses
 about the shape of the body (e.g. sphere, spheroid, ellipsoid).
Moreover, to assess the more probable shape parameters given the sparse data, 
we have adopted a Bayesian approach to derive posterior probability distributions for 
the radius of the spherical model,
and size, shape and orientation for the Maclaurin, ellipsoidal and Jacobi models.
The advantage of this method is that it can incorporate knowledge from complementary observations,
in a quantitative way avoiding qualitative assumptions \citep{Brown2013}.
 
In Section~\ref{sec:observations} we describe the prediction of occultations and the observations,
 resulting in a total of eight occultation chords observed during five stellar occultations between 2013 and 2016.
 In Section~\ref{sec:phy_model} we describe the rings, main body models and the implementation of 
 the Markov Chain Monte Carlo (MCMC) to derive parameter values. 
In Section~\ref{sec:results} we describe the main results,
 before discussions (Section \ref{sec:discussion}) and concluding remarks (Section \ref{sec:conclusions}).

\section{OBSERVATIONS}
\label{sec:observations}

\subsection{Prediction of stellar occultations}
\label{sec:prediction}

Stellar occultation predictions by  Chariklo for the period 2012.5-2014 were based 
on local catalogs with astrometric positions of the stars around Chariklo's path on the sky.
Moreover, improved ephemerides for Chariklo were obtained from those catalogs with
typical uncertainties of 20-30 milliarcsecond (mas) \citep{Camargo2014}.
A similar approach was used for predictions after 2014 using the Wide Field Imager (WFI) at 
the MPG 2.2-m telescope (La Silla, Chile)  and the IAG 0.6~m telescope at OPD/LNA (Pico dos Dias, Brazil). 
For the occultation of 2016 October 1, the star position was obtained from GAIA Data Release 1 \citep{GaiaCollaboration2016,GaiaCollaboration2016a}.
A total of thirteen positive occultations were observed up to 2016,
involving the detection of the rings, the body or both (see details in \citealt{Berard2017}).
Each event provided the position of Chariklo relative to the star 
with an accuracy of a few mas.
Those positions were used in turn to improve Chariklo's ephemeris,
 providing updated orbital elements in the so-called Numerical Integration of the Motion of an Asteroid (NIMA) procedure 
 \citep{Desmars2015}, and permitting better subsequent predictions.

\subsection{Observations and data reduction}
\label{sec:observations_and_data_reduction}

Among the thirteen stellar occultations observed between 2013 and 2016 \citep{Braga-Ribas2014,Berard2017},
we use those five that include the simultaneous detections of the main body and rings.
In those cases, the orientation and center of the system can be constrained as detailed in Section~\ref{sec:ring_model}.
The 2013 June 3, 2014 June 28 and 2016 October 1 events were double-chord
while the 2014 April 29 and 2016 August 8 events were single-chord.
Additionally, we consider the stations whose negative detections (that is, no occultation by the body) 
were close enough to the main body's limb such to constrain its extension.
In consequence, 
from the occultation of 2013 June 3 we use the negative detections from Ponta Grossa and Cerro Burek \citep{Braga-Ribas2014},
from 2014 April 29 we use a negative detection from Springbok,
and from 2014 June 28 we use the negative detection from Hakos.
Table~\ref{tab:observations} gives the circumstances of the observations
between 2014 and 2016 used in this work.
Details of observations from other sites and stellar occultations not used in this work
are given in \cite{Braga-Ribas2014} and \cite{Berard2017}.

\begin{figure}
\epsscale{.8}
\plotone{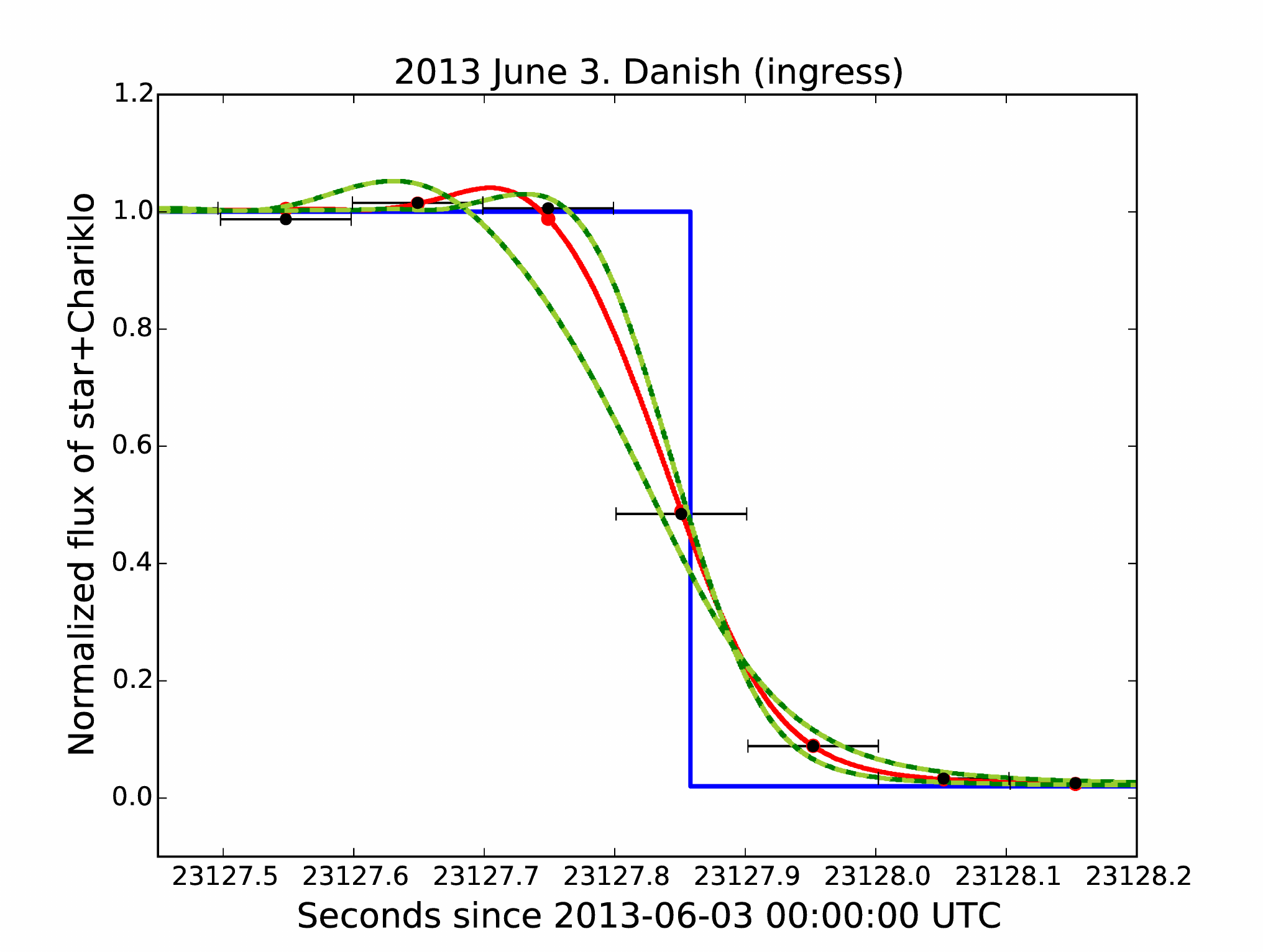}
\plotone{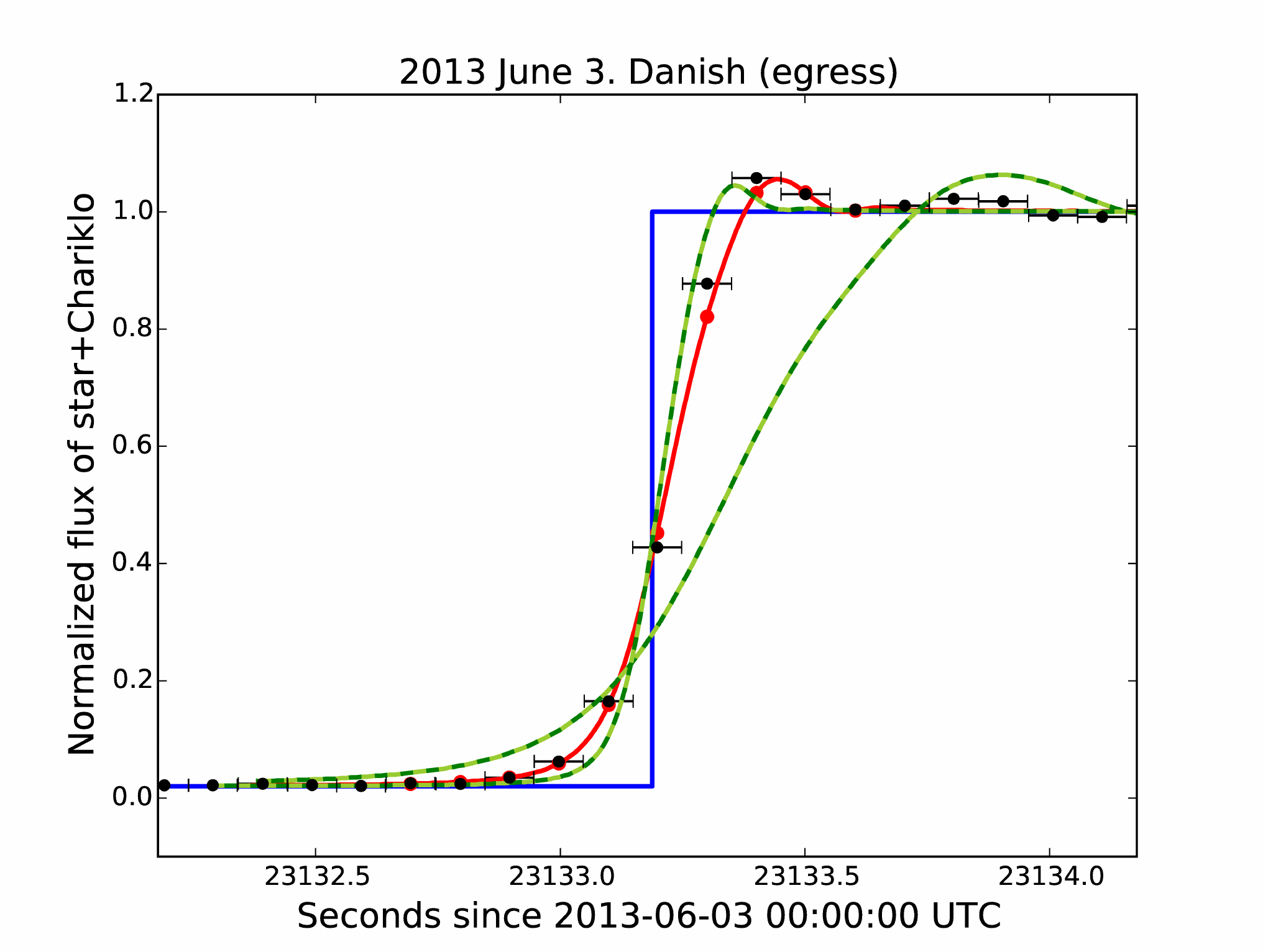}
\caption{
Occultation light curve of 2013 June 3 at Danish telescope at ingress (top) and egress (bottom).
Black dots are the data points with horizontal bars indicating the time interval of acquisition (the exposure time). 
Blue continuous line is the geometric best solution indicating the limb of the object,
 from where we obtain the occultation times given in Table~\ref{tab:timing_body}.
Red continuous line is the best solution after applying the limb diffraction model. 
The high SNR and cadence of the light curve from Danish telescope allows to determine 
the orientation between the occultation chord and the normal to the local limb, given by the angles 
$\alpha_{\rm ing}=60^{\circ}$ and  $\alpha_{\rm egr}=73^{\circ}$ 
at ingress and egress respectively (see Figure~\ref{fig:20130603_danish_local_limb_fit}).
For illustration, the green lines show the limb profiles for $\pm10^{\circ}$ with respect to
the best-fit values $\alpha_{\rm ing}$ and $\alpha_{\rm egr}$,
showing a clear departure from the data (see Section~\ref{sec:occultation_timing} for details).
}
\label{fig:light_curves1}
\end{figure}

\begin{figure*}
\gridline{
\fig{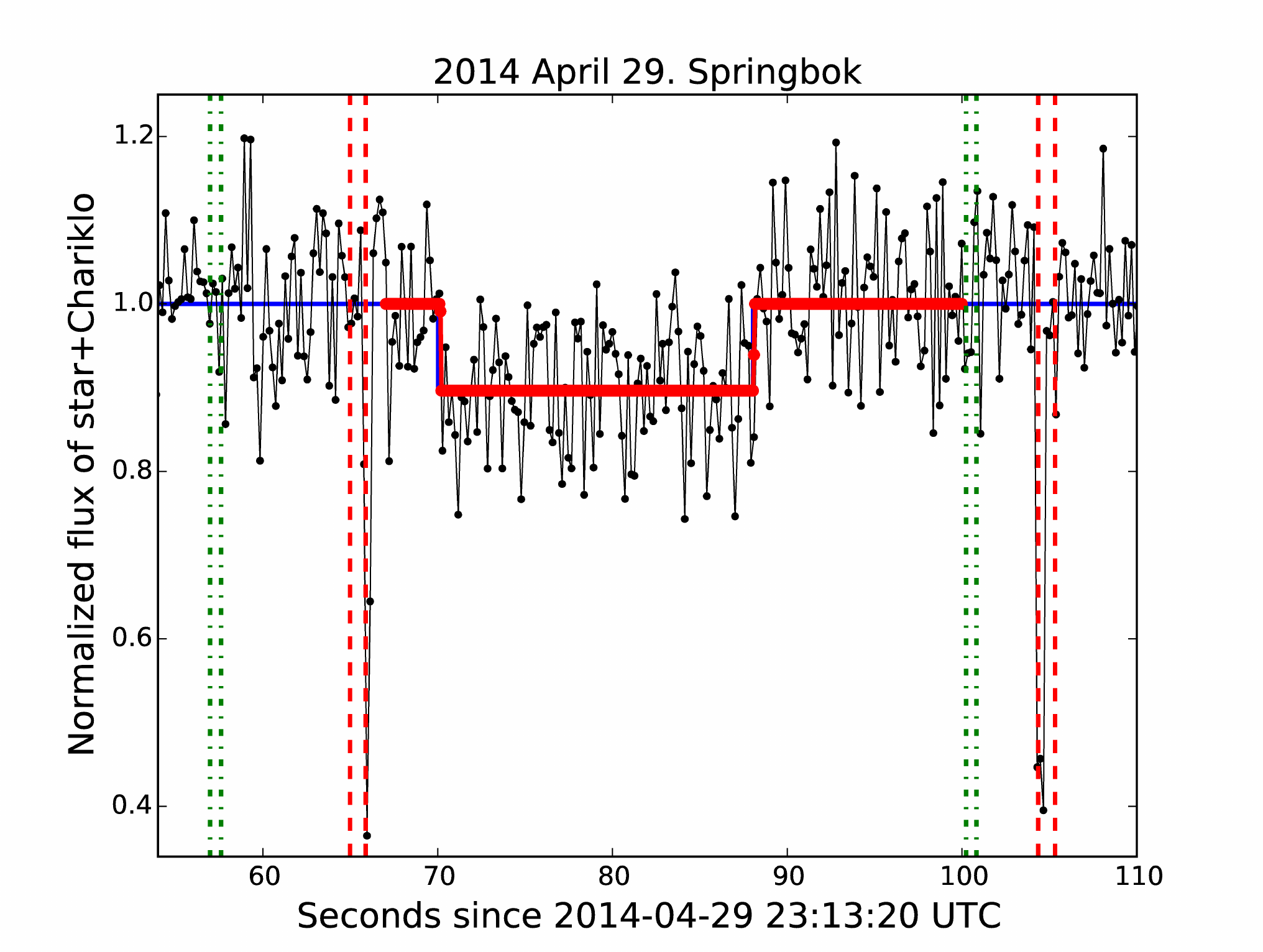}{0.4\textwidth}{(a)}
\fig{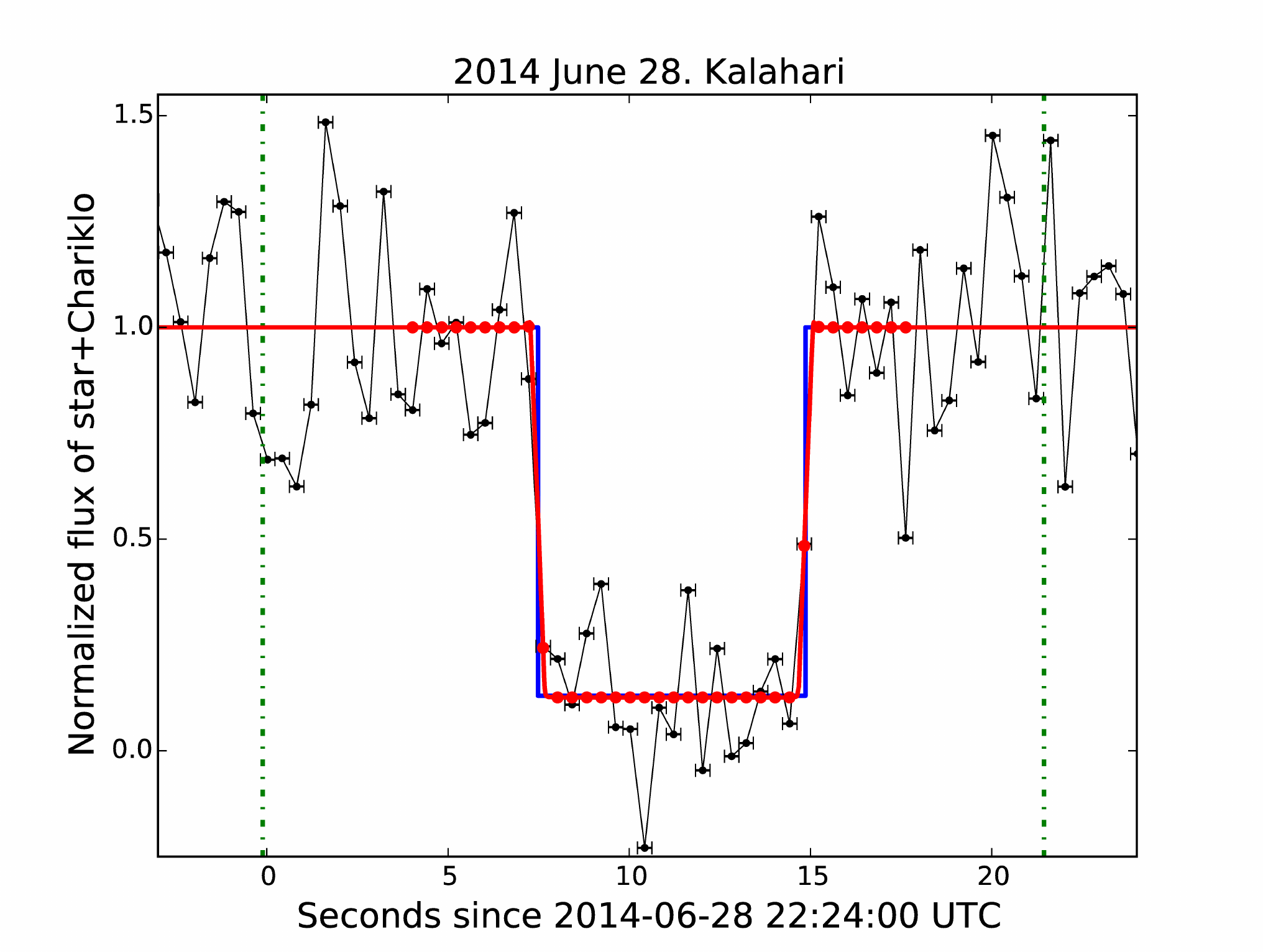}{0.4\textwidth}{(b)}
}
\gridline{
\fig{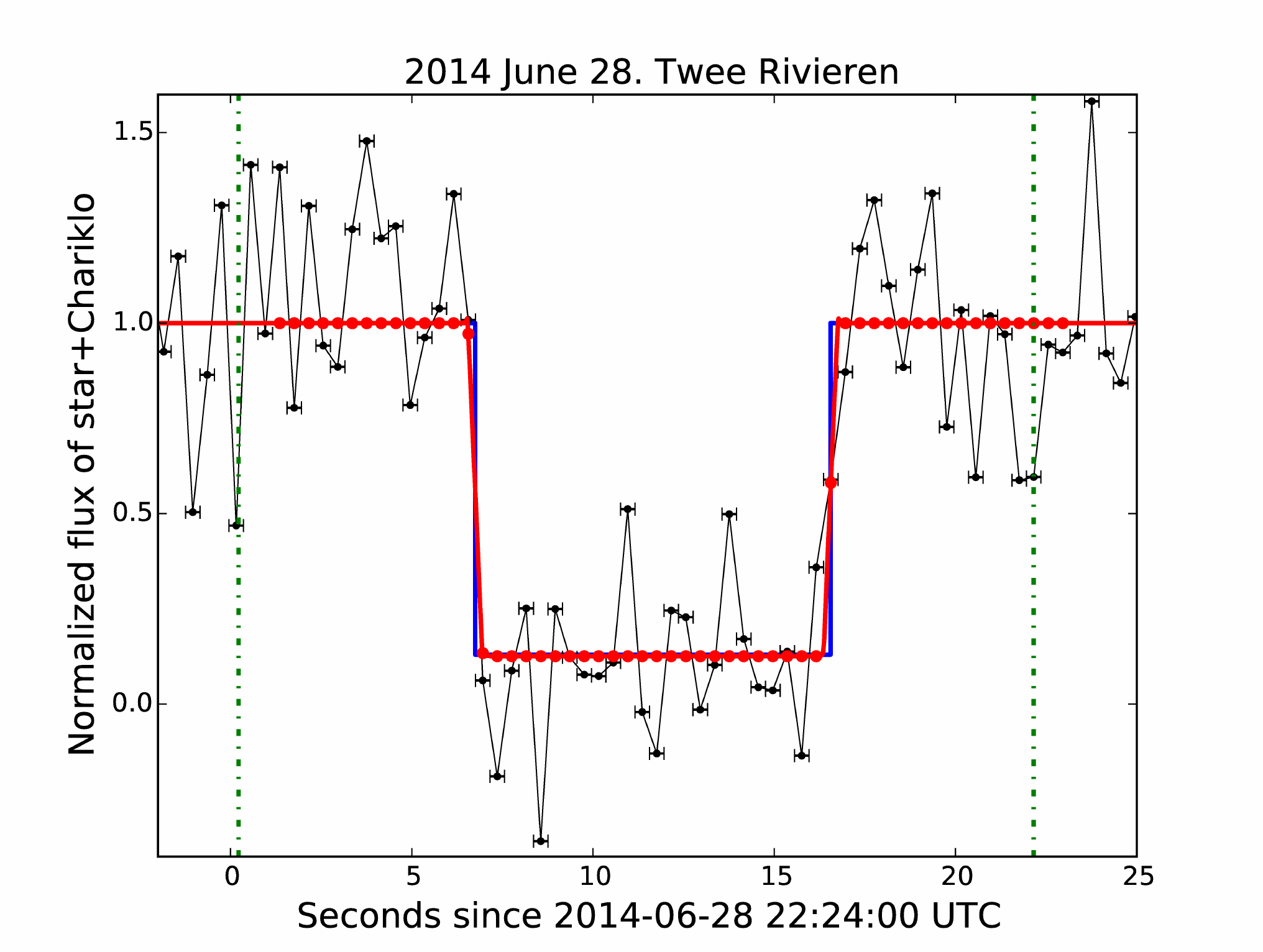}{0.4\textwidth}{(c)}
\fig{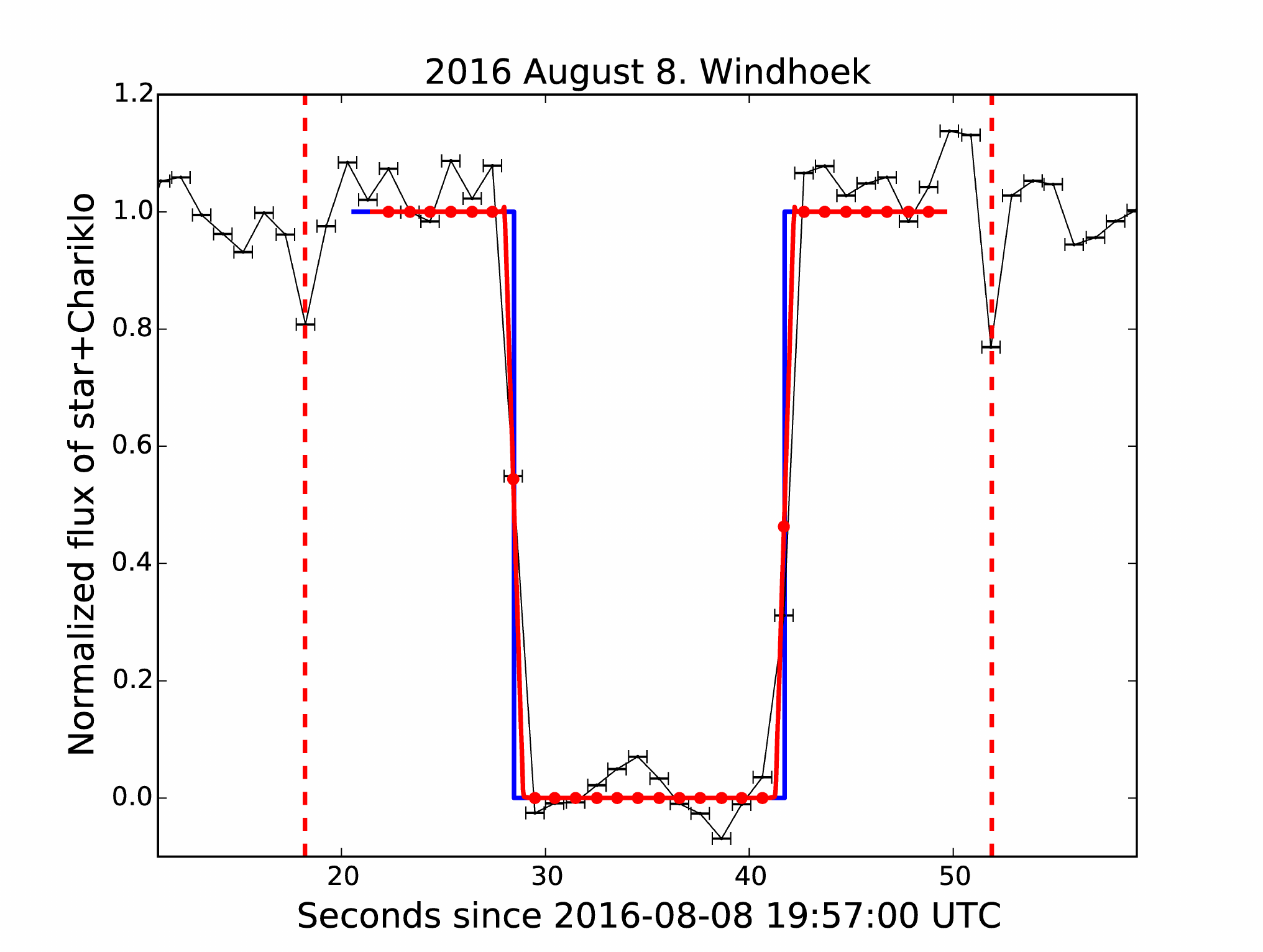}{0.4\textwidth}{(d)}
}
\gridline{
\fig{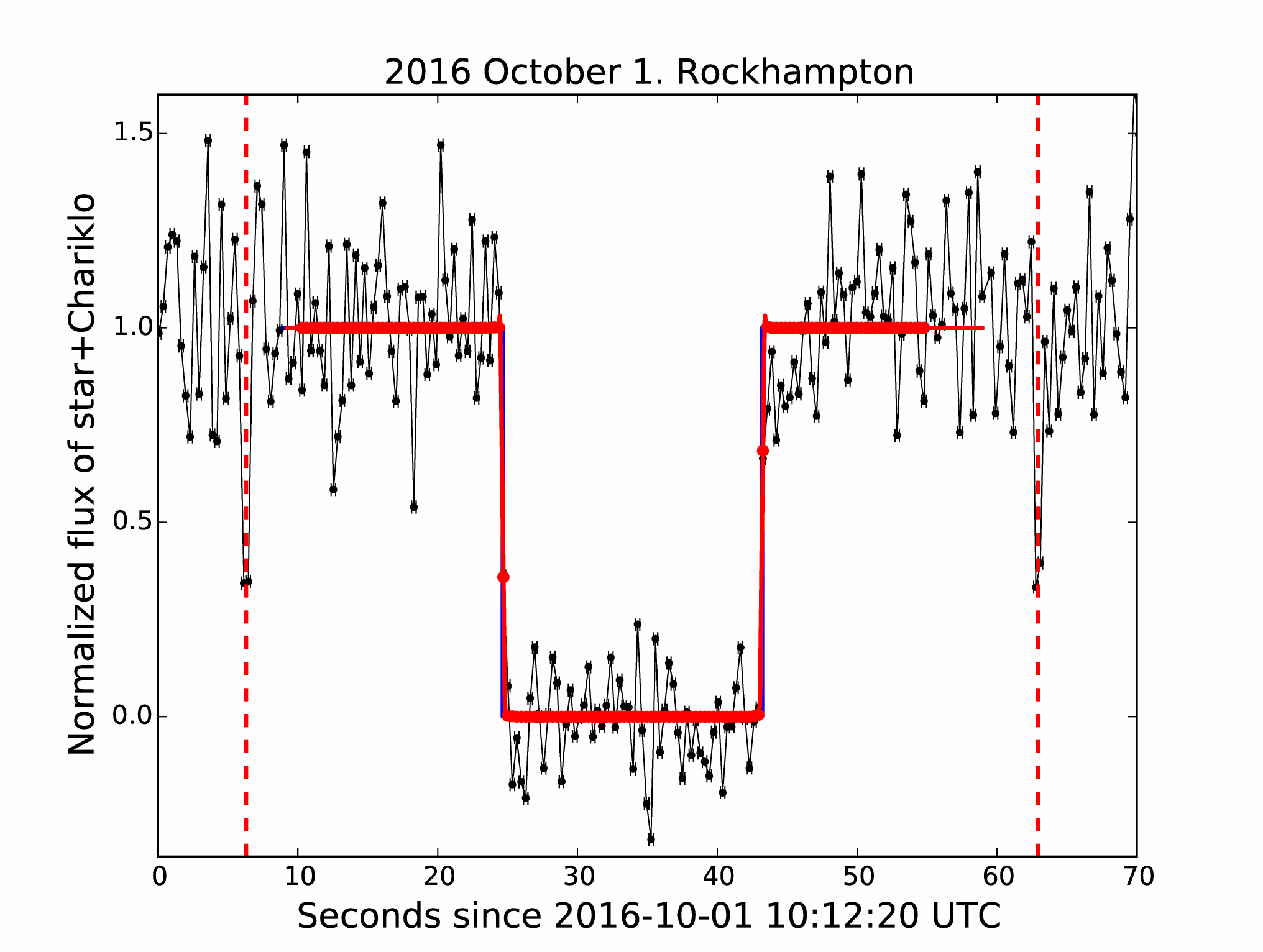}{0.4\textwidth}{(e)}
\fig{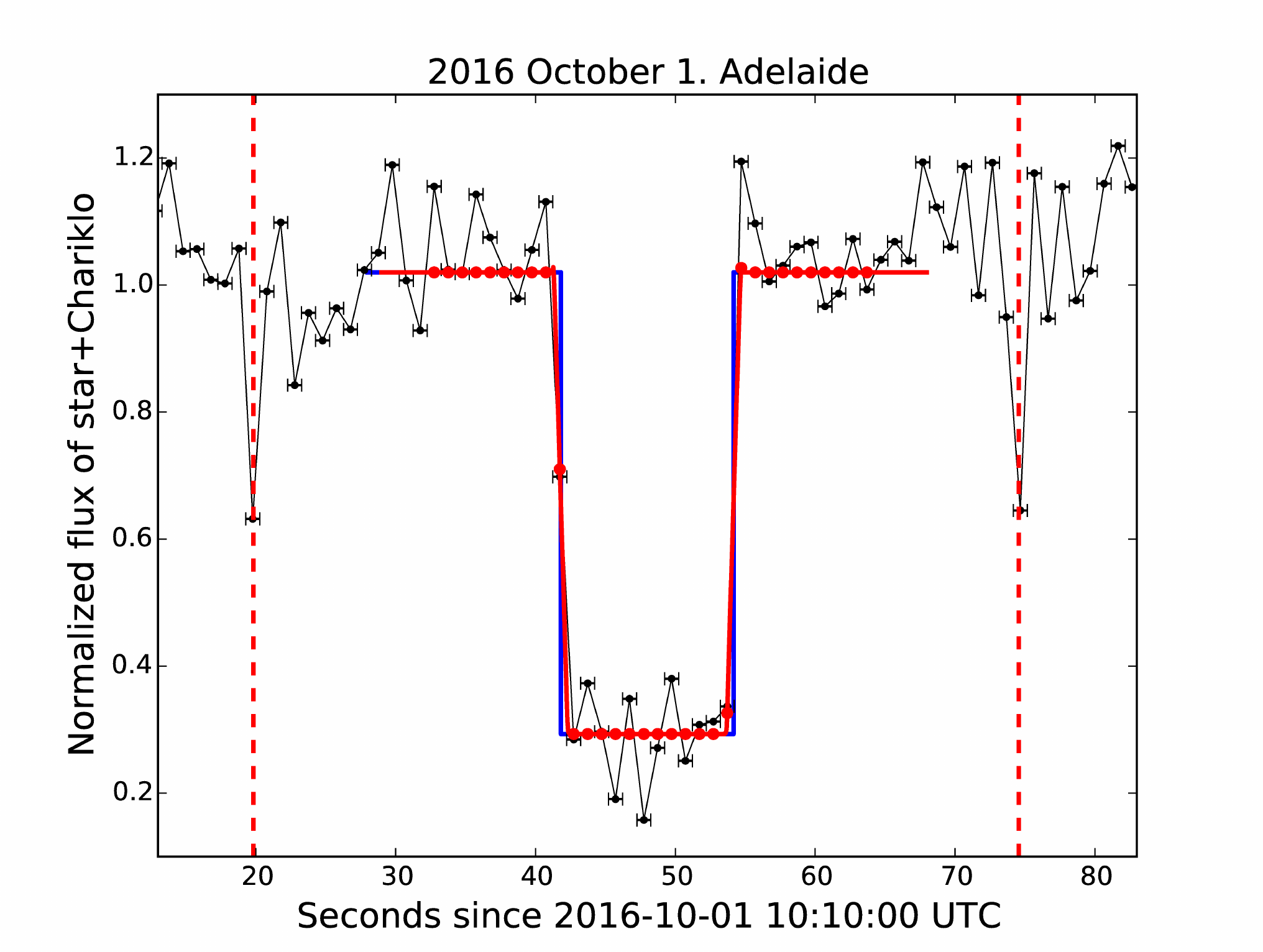}{0.4\textwidth}{(f)}
}
\caption{
Occultation light curves used in this paper.
Black dots are the data points with horizontal bars indicating the time interval of acquisition (the exposure time). 
Blue continuous lines are the geometric best solution indicating the limb of the object. 
Red continuous lines are the best solution after applying the limb diffraction model. 
Red dashed lines show the central times of occultations by C1R and C2R rings.  Notice that only the C1R position is indicated for events with unresolved rings.
Green dot-dashed lines indicate the ring position expected after reconstruction of the geometry of the event in cases if the rings are undetected.
}
\label{fig:light_curves2}
\end{figure*}

For each observation the images were reduced in a standard way applying dark and flat frames.
Then we performed aperture photometry for the occulted star and comparison stars in the same image.
Finally we perform relative photometry between the occulted star and comparison star(s) to correct for variations in the sky transparency.
The optimal aperture size was chosen in each case for the target and comparison star(s) to obtain
 the best signal-to-noise ratio (SNR) in each light curve.
The background flux was estimated near the target and nearby reference stars, and then subtracted,
so that the zero flux corresponds to the sky level. 
The total flux from the unocculted star and Chariklo was normalized to unity after fitting the light curve by a third or forth-degree
polynomial before and after the event.
Figures~\ref{fig:light_curves1} and \ref{fig:light_curves2} show the light curves obtained with this procedure.

\subsection{Occultation timing analysis}
\label{sec:occultation_timing}

For each light curve involving the body detection,
 we determine the times of ingress $t_{\rm ing}$ and egress $t_{\rm egr}$ of the occulted star behind Chariklo by fitting 
a sharp-edge occultation profile.
This profile is convolved by Fresnel diffraction produced by the sharp edge of the body, 
then convolved by the stellar diameters projected at Chariklo's distance,
the integration time and
the bandwidth of the optical system (product of the telescope, detector and filter responses) as described in \cite{Widemann2009}.
The profile takes into account the relative speed of the star with respect to Chariklo in the sky plane $v_{ch}$,
and the orientation angle $\alpha$ between the occultation chord and the normal to the local limb. 
For instance, for a local limb perpendicular to the occultation chord, we have $\alpha$=0.

The angle $\alpha$ and times $t_{\rm ing}$, $t_{\rm egr}$ are obtained by minimizing a classical $\chi^{2}$ function

\begin{equation}
\chi^{2}=\sum_1^N \frac{ \left(\phi_{i,o}-\phi_{i,m} \right)^2  }{\sigma_i^2}, \label{eq:chi2_phot}
\end{equation}

where $\phi_{i,o}$ is the normalized flux observed, 
$\phi_{i,m}$ is the synthetic flux from the diffraction model 
and $\sigma_i$ is the uncertainty in the measured photometry.

Except for the occultation of 2013 June 3 observed with the Danish telescope,
the light curves are dominated by the long integration time instead of diffraction effects,
so that the angle $\alpha$ is unconstrained and only the ingress and egress times are obtained.

Table~\ref{tab:coord_and_eph} gives the adopted stellar diameters projected at Chariklo's distance,
Chariklo's geocentric range,
the adopted coordinates of the occulted stars,
and the predicted coordinates of Chariklo at a reference time. 
For the occultations of 2014 June 28 and 2016, 
the apparent stellar diameters are estimated using the $V$ and $K$ 
apparent magnitudes provided by the NOMAD catalog \citep{Zacharias2004} 
using the V-K relations from \cite{Kervella2004} and considering galactic reddening.
For 2013 June 3 and 2014 April 29 we adopt the apparent stellar diameter derived in \citep{Braga-Ribas2014,Berard2017}.

The occultations by the main inner ring (C1R) and the external fainter ring (C2R) show variable width and radial profiles and are described and analyzed in detail in \cite{Berard2017}.
Here we take the midtimes $t_{\rm mid}$ of the rings detections from that work,
that we use in turn to constrain the body apparent center as explained in \ref{sec:ring_model}.

\begin{figure*}
\gridline{
\fig{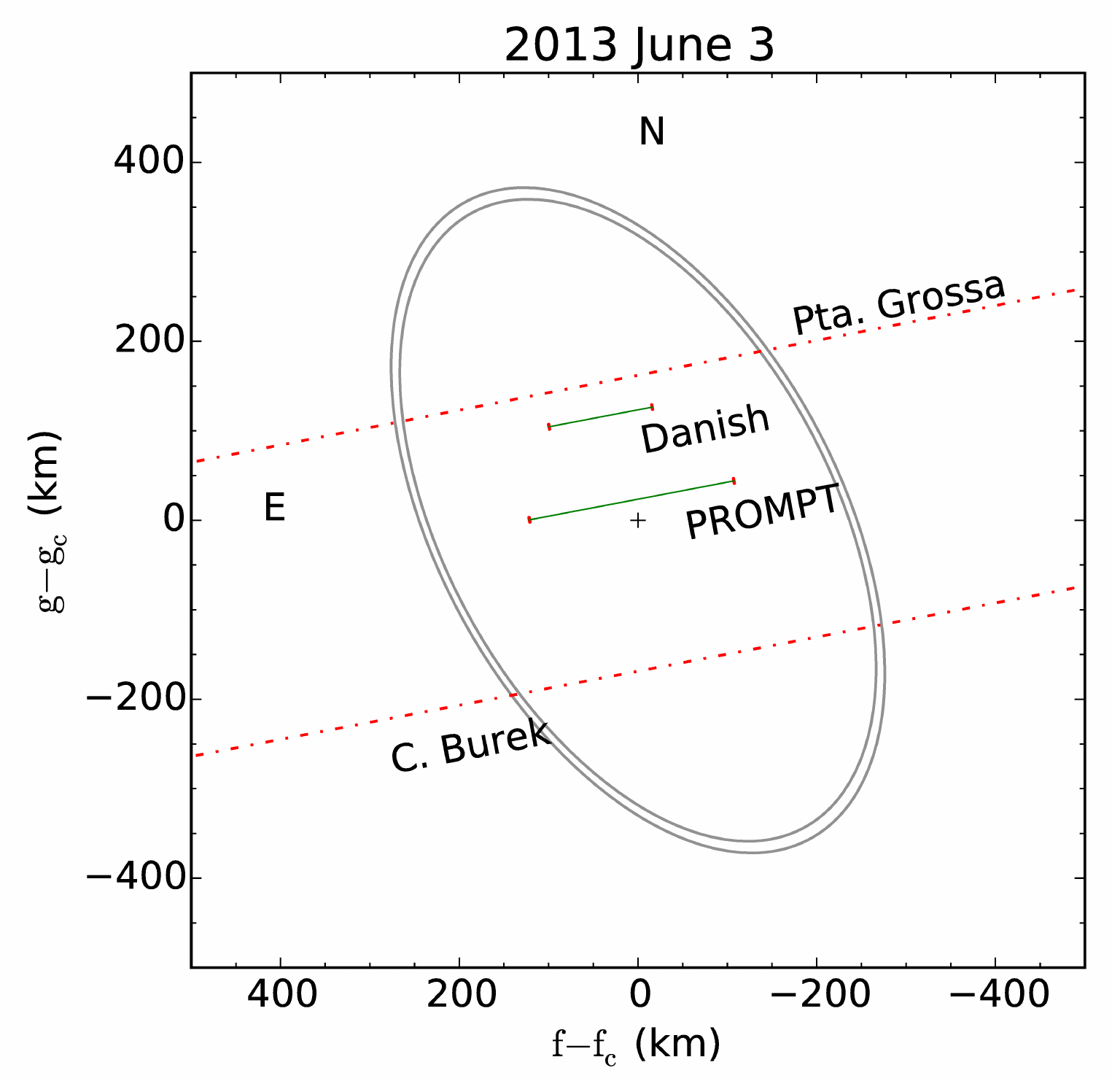}{0.3\textwidth}{(a)}
\fig{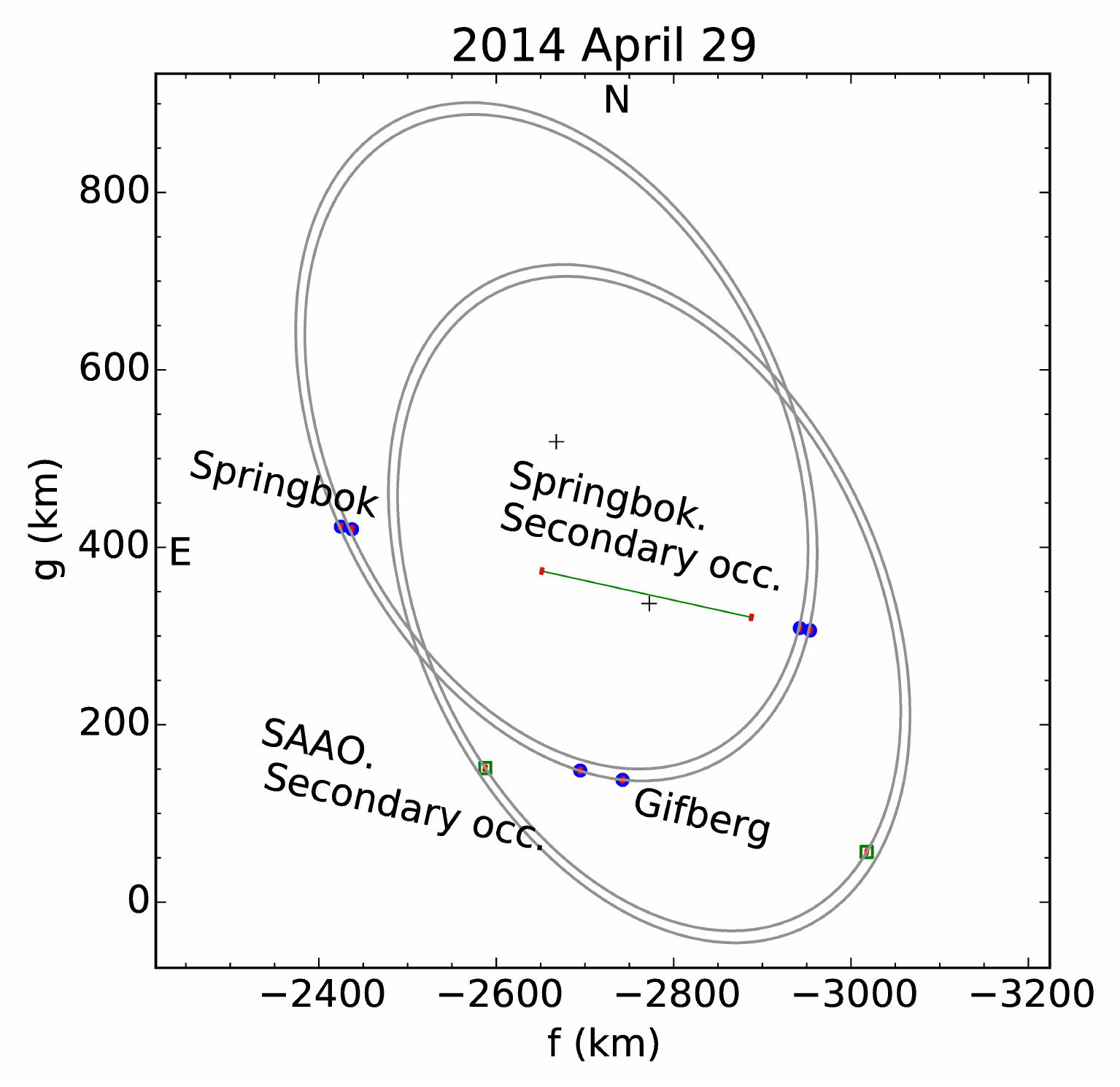}{0.3\textwidth}{(b)}
}
\gridline{
\fig{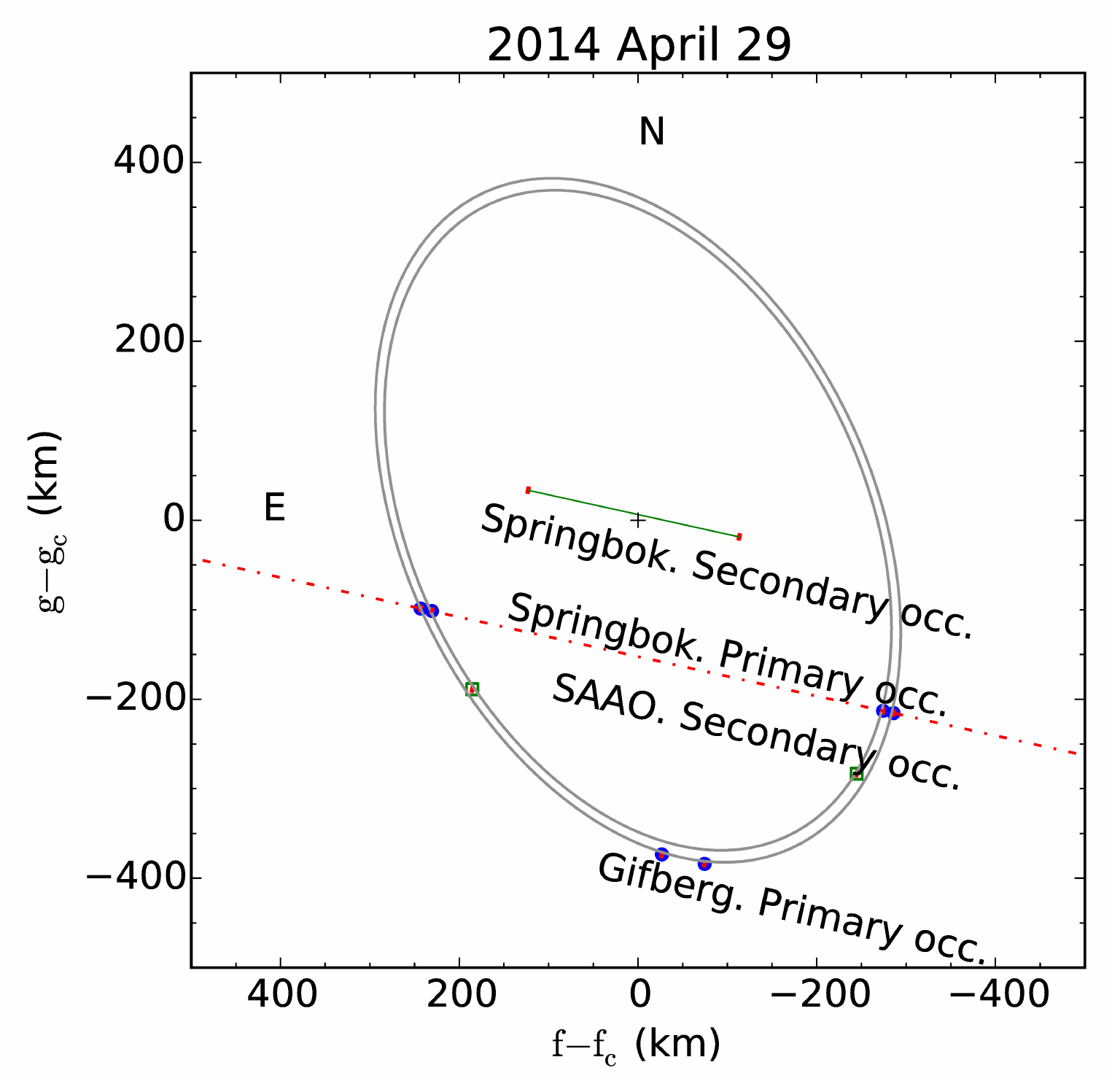}{0.3\textwidth}{(c)}
\fig{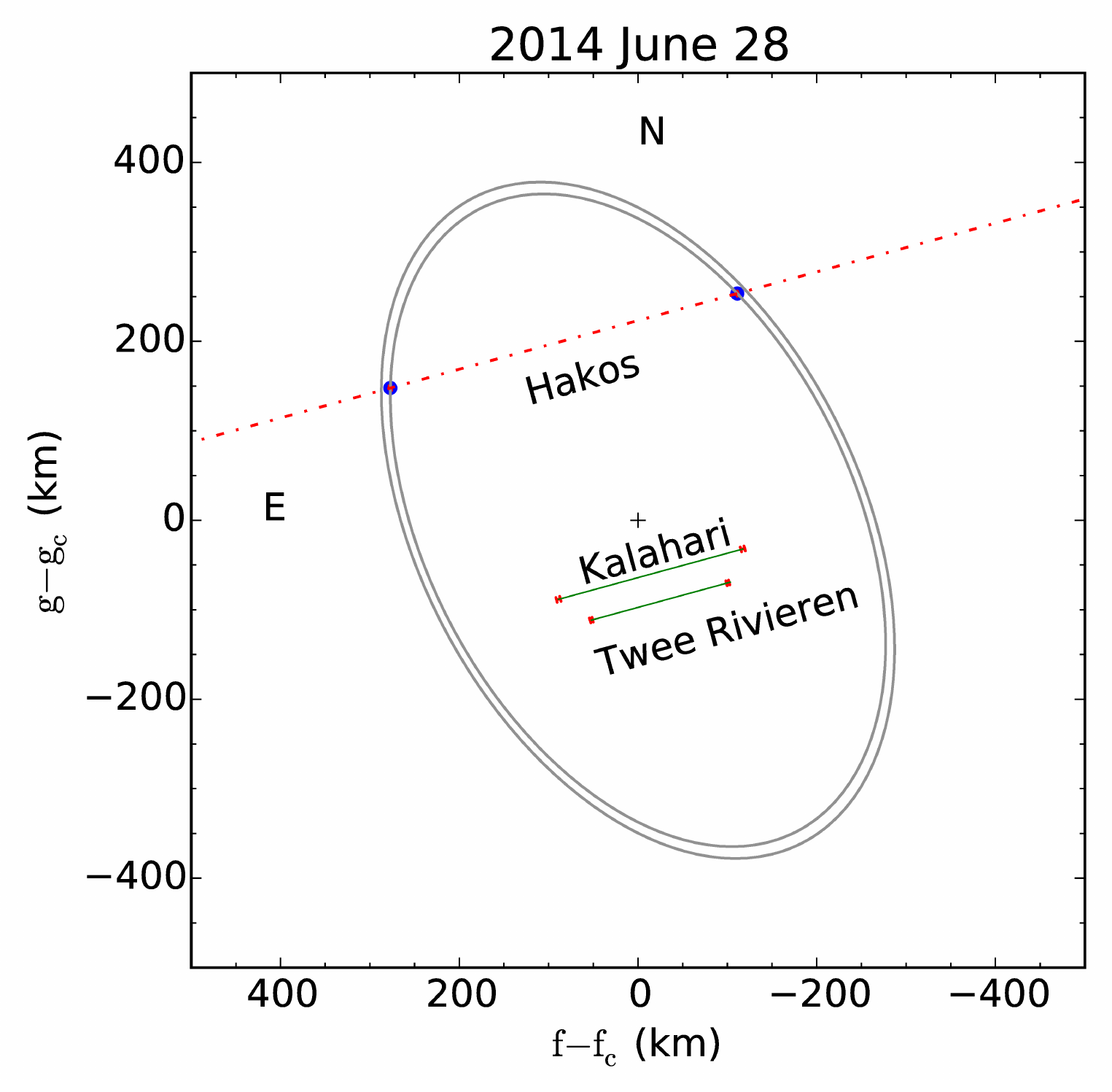}{0.3\textwidth}{(d)}
}
\gridline{
\fig{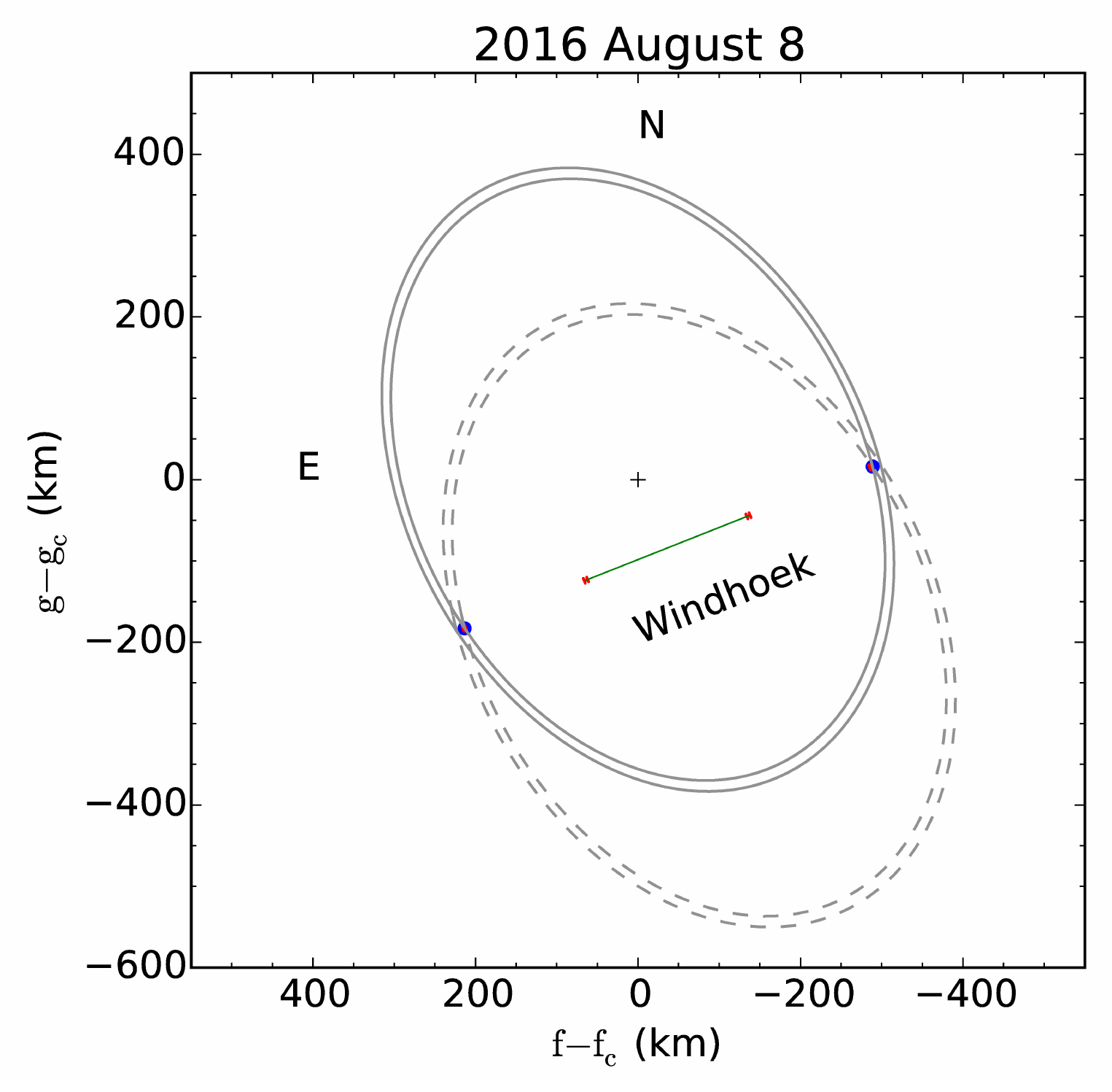}{0.3\textwidth}{(e)}
\fig{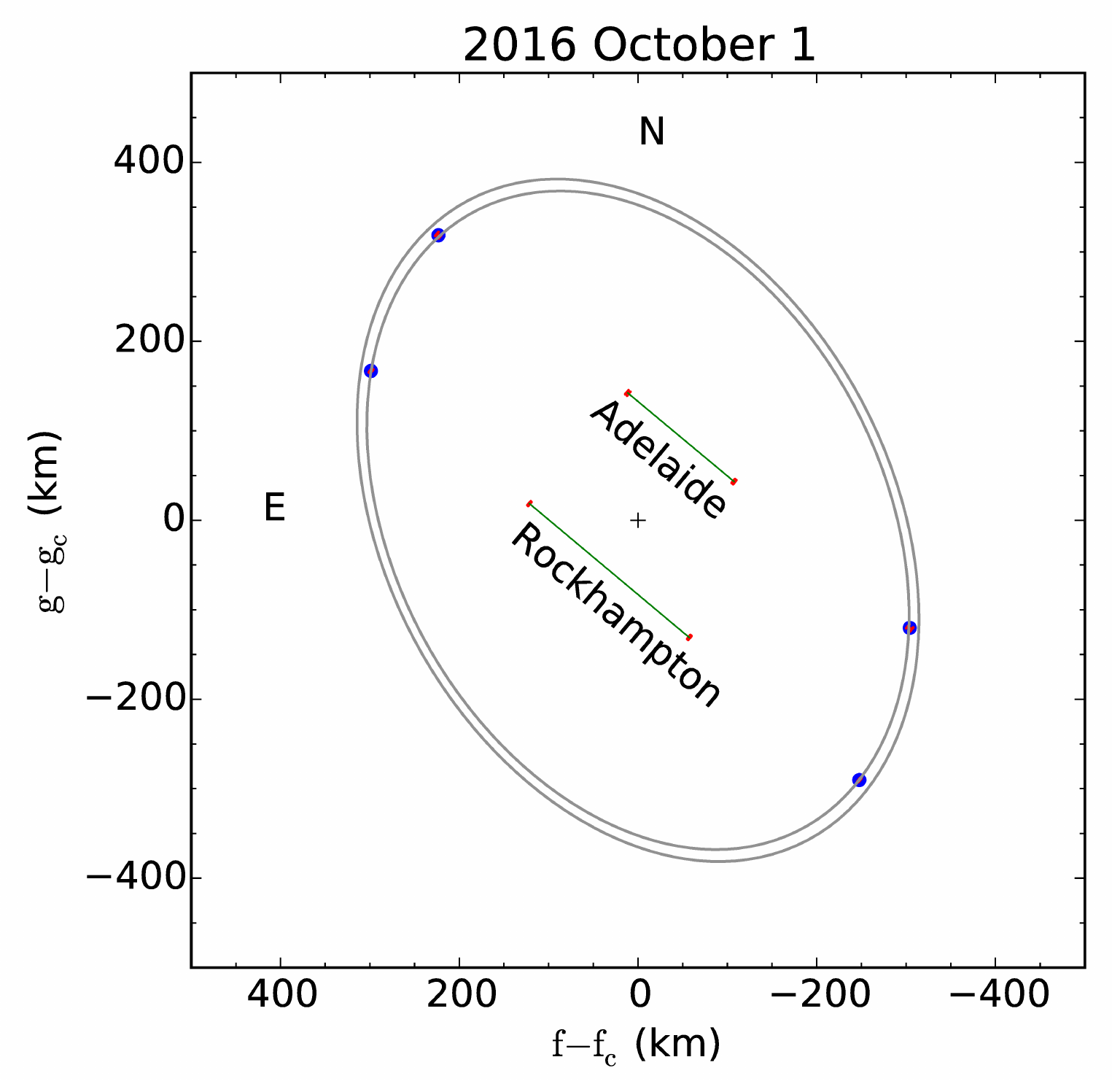}{0.3\textwidth}{(f)}
}
\caption{Geometry for the five occultations by Chariklo's main body between 2013 and 2016.
Positions are given in the sky plane at Chariklo's distance with respect to the center of the system $(f_c,g_c)$
determined in Table~\ref{tab:ring_geometry}.
The ellipses represent the C1R and C2R orbits adopting the diameter and pole position of \cite{Braga-Ribas2014}. 
Dots indicate the rings detections used to fit the center of the system (black crosses).
The continuous green lines are the occultation chords by the main body with uncertainties in red. 
For clarity we only indicate the closest negative detections in red dot-dashed lines used as constraints. 
(a) 2013 June 3, South America.
(b) 2014 April 29, South Africa.  Occultation of a double star. Solid dots and cross are the ring occultation of the primary star and the adopted center of the system.  Open dots correspond to occultations of the secondary star.
(c) 2014 April 29, South Africa. Geometry after applying an offset of $\Delta f$=103~km and $\Delta g$=182.5~km to the secondary events. 
(d) 2014 June 28, South Africa - Namibia.
(e) 2016 August 8, Namibia. Rings solution 1 in continuous line and solution 2 in dashed lines.
(f) 2016 October 1, Australia.
}
\label{fig:occ_geom}
\end{figure*}

The timings corresponding to the occultations by the main body and the rings provide a set of offsets $(f,g)$ 
of the star with respect to the expected body center as seen from each site.
Those offsets are measured in the sky plane at Chariklo's distance
and they are counted positive toward the east and north.
Table~\ref{tab:timing_body} summarizes the timings and offsets for the body detections 
while Table~\ref{tab:timing_ring} lists the ring detections.
Figure~\ref{fig:occ_geom} shows the occultation chords in the sky plane.
Particular conditions during the occultations of 2013 June 3 and 2014 April 29 are discussed below.

\paragraph{South America. 2013 June 3}
\label{sec:20130603}

For this occultation we keep most of the timing analysis reported in \cite{Braga-Ribas2014},
but we re-analyzed the light curve obtained at the Danish telescope.
This light curve has the highest SNR and a high acquisition rate of 10~Hz,
allowing us to resolve diffraction effects
fitting simultaneously the time of ingress/egress 
and the orientation angle $\alpha$.

Figure~\ref{fig:light_curves1} shows the best fits, 
from where we derive $\alpha_{\rm ingress}=60.2\pm0.9^{\circ}$ at ingress 
and $\alpha_{\rm egress}=73.0\pm0.8^{\circ}$ at egress,
while ingress and egress timings are given in Table~\ref{tab:timing_body}.
For illustration, the same figure indicate the modeled light curve for $\alpha$ departing $\pm10^{\circ}$ with respect to the best-fit
showing a clear departure from the data.

\begin{figure}
\centering
\epsscale{.7}
\plotone{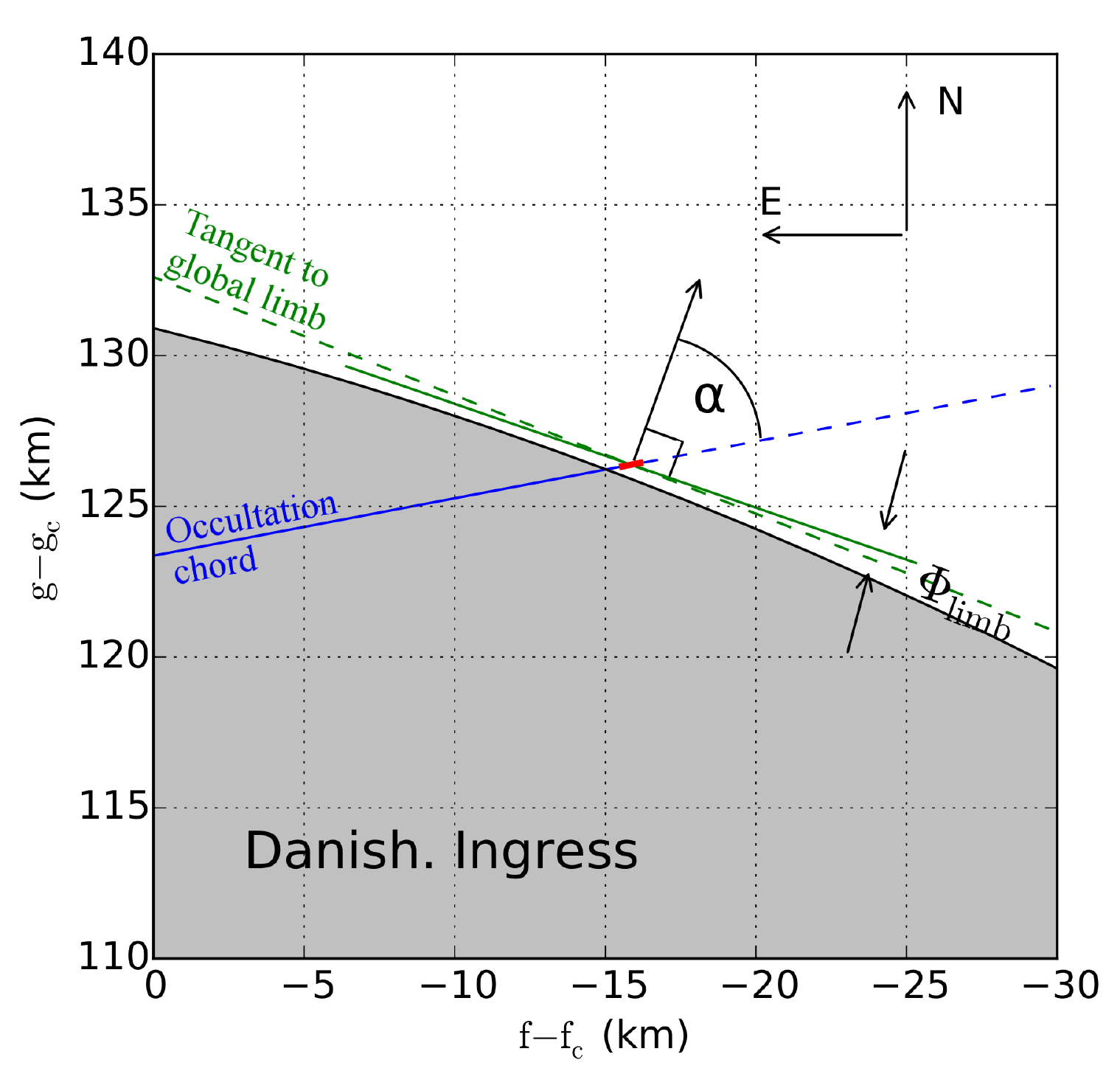}
\plotone{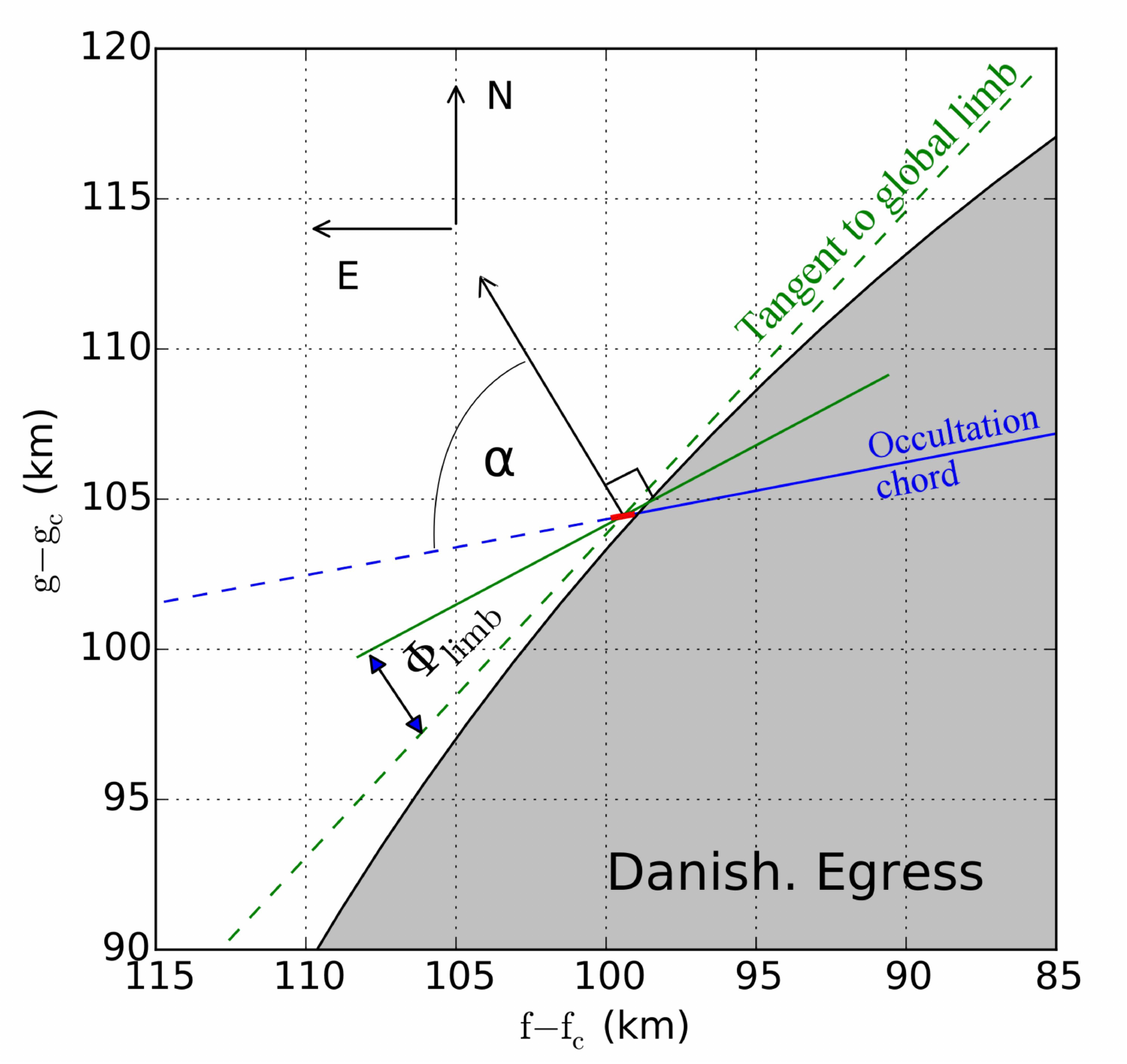}
\caption{ 
Local limb slopes with respect to the generic ellipsoid model as measured at the Danish telescope for the occultation of 2013 June 3.
The tangent to local limb (solid green line) is derived from fitting the angle $\alpha$ in the limb diffraction model to the occultation light curves  (Figure~\ref{fig:light_curves1}). 
The blue-solid lines are the occultation chords with their extremity uncertainties in red.
Depending on Chariklo's main body model, the slopes with respect to the tangent to the global limb of up to $\Phi_{\rm limb}=25\degr$ are observed.
As illustration, this figure shows the local limb orientation for the generic ellipsoidal model.
}
\label{fig:20130603_danish_local_limb_fit}
\end{figure}

From this, we calculate the relative angle $\Phi_{\rm limb}$ between the local and global limb,
the latter understood as the tangent to the projected limb.
Depending on the main body model, we obtain 
$\Phi_{\rm limb}$=2-10$^{\circ}$ at ingress and $\Phi_{\rm limb}$=15-25$^{\circ}$ at egress.
The angle $\Phi_{\rm limb}$ corresponds to what is known as the angle of internal friction, or maximum angle of repose.
As illustration, Figure~\ref{fig:20130603_danish_local_limb_fit} shows 
a local view of the occultation geometry in the sky plane for the generic triaxial ellipsoid model (Section~\ref{sec:ellipsoid})
indicating the occultation chord,
the local limb, 
the global limb
and the angles $\alpha$ and $\Phi_{\rm limb}$.

Finally, the  occultation timings by the main body obtained at the PROMPT telescope (as given by \cite{Braga-Ribas2014}),
as well as the non-detection at Cerro Burek and Ponta Grossa are used as further constraints for the overall shape. 

\paragraph{South Africa. 2014 April 29}
\label{sec:20140429}

This occultation revealed that the occulted star was in fact double.
A stellar atmosphere fit was performed to determine the relative flux and apparent diameter of the stars as detailed in \cite{Berard2017} 
The separation between the main (A) and secondary (B) stars and the diameter of each component projected at the distance of Chariklo are given in Table~\ref{tab:coord_and_eph}.
 
The occultation of the primary star by the rings was detected at Springbok,
while the occultation of the secondary was not detected due to the low SNR.
There were two additional detections of the rings, 
one involving the C2R ring occulting the primary star at Gifberg 
and the other one involving C1R ring occulting the secondary star at SAAO.
The panel b of Figure~\ref{fig:occ_geom} shows the two sets of occultations 
of the primary and secondary stars in the plane of the sky.
The panel c of Figure~\ref{fig:occ_geom} shows the reconstructed geometry of the event after applying the offset between the components of the double star to the occultations of the secondary star. 

\section{RINGS AND BODY MODELS}
\label{sec:phy_model}

\subsection{Ring model}
\label{sec:ring_model}

All the ring occultations observed so far are consistent with two concentric and circular rings 
with fixed pole position and fixed radius,
within $\sim$1$\degr$ and within $\sim$3.3~km of the discovery values, respectively
(see \citealt{Braga-Ribas2014,Berard2017}).
For a particular date, the rings are projected on the sky plane as an ellipse 
with semimajor axis corresponding to the ring radius. 
The ring opening angle $B$ (the elevation of the observer above the ring plane) and the position angle of the semiminor axis $P$ are calculated from the pole position in Table~\ref{tab:ring_geometry}
and Chariklo's position in the sky. 
The single-chord ring occultation of 2016 August 8
provides two possible ring centers while the other events give a unique center.
The ring centers $(f_c,g_c)$ are listed in Table~\ref{tab:ring_geometry},
with error bars that reflect the uncertainties on the ring midtimes given in Table~\ref{tab:timing_ring}.
Occultations indicate that the radial width of the main ring varies between 5 and 7.5~km
adding an additional bias in the determination of the center of $\sim$1~km.
This is of the order of the formal uncertainties for most of the events and 
it is not a dominant effect in the determination of the center of the projected ellipse.
The best fitted ellipses for each event are displayed in Figure~\ref{fig:occ_geom}.

\subsection{Body model}
\label{sec:body_model}

Here we adopt the assumption that the ring system lies in the equatorial plane of the object. 
Indeed, in this situation, a collisional dissipative ring reaches its minimum energy configuration,
while conserving its angular momentum parallel to the body spin axis.
With this assumption (plus the circularity described above) the body and the rings share the same pole position and the same center.

If Chariklo was completely irregular, a simple parametric model (e.g. sphere, ellipsoid) would give a poor estimation of dimensions of the body. 
A well sampled stellar occultation could indicate if Chariklo is irregular but, unfortunately,
 to date we have only a few occultations with one or two positive detections each (Table~\ref{tab:timing_body} and Figure~\ref{fig:occ_geom}).
In that context, we test simple models that can be easily parametrized in order to give
credible intervals for the corresponding parameters.
In practice, we first assume a spherical body that is used to estimate a radius and the scale of topographic features.
Next, we consider a generic triaxial ellipsoid to estimate the length of the semiaxes $a$, $b$ and $c$ and the scale of topographic features. 
Next, considering hydrostatic equilibrium, a homogeneous body will assume either a Maclaurin (spheroid) or a Jacobi (tri-axial ellipsoid) shape for which size, axes ratios, density and topographic features can be evaluated.
The non-spherical models incorporate independent information
as a priori estimates for the model parameters, 
such as the amplitude of the rotational light curve or the long term photometric behavior of the system.
The models mentioned above (sphere, triaxial ellipsoid, Maclaurin and Jacobi) are now discussed in turn.

\subsubsection{Sphere model}
\label{sec:sphere}

This is the simplest case as there is only one free parameter, the sphere radius $R$. 
The projection in the sky plane is a circle with the same radius $R$ for all occultations.
The problem is then reduced to find a circle that best fits all the chords extremities $(f,g)$ provided 
in Table~\ref{tab:timing_body}
with the center located at the $(f_c,g_c)$ values indicated in Table~\ref{tab:ring_geometry}.

\subsubsection{Triaxial ellipsoid}
\label{sec:ellipsoid}

Here we consider a generic triaxial ellipsoid with semiaxes $a>b>c$, rotating around the shortest axis. 
We define the rotation angle $\phi$ as the angle from the central meridian to the prime meridian counted positively along the equator of the object using the right hand rule.
The prime meridian is the one passing through one of the two intersections between the equator and 
the longer axis of the ellipsoid.
Given the current uncertainty in the rotation period, the rotation phase is lost after a few weeks and the rotation angle in each occultation is considered as independent.

The eight free parameters of this model are $a$ (which gives us the size of the object),
the ratios $0<b/a<1$ and $0<c/b<1$,
and the rotation angles $\phi_i$ at each of the five occultations. 

\subsubsection{Maclaurin spheroid}
\label{sec:maclaurin}

A Maclaurin spheroid has an equatorial radius $a$ and polar radius $c$ related by \citep{Chandrasekhar1987}
\begin{equation}
\Omega=\frac{\omega^{2}}{\pi G \rho}=\frac{2\sqrt{1-e^2} }{e^3} \left[ \left(3-2e^2 \right)\arcsin e -3e\sqrt{1-e^2} \right], \label{eq:maclaurin_shape}
\end{equation}
where 
$e^2=1-(c/a)^2$, 
$\omega= 2 \pi/T$  ($T$ being the rotation period of the body), 
$G$ the gravitational constant and 
$\rho$ its (uniform) density.
Note that $\Omega$ is the adimensional rotational parameter that compares
the centrifugal acceleration at the equator of the body to its gravity.
For stable Maclaurin shapes this parameter is between $\Omega=0$ and $\Omega \simeq 0.374$.
The lower limit corresponds to the spherical limit with $\rho\rightarrow \infty$ while the upper limit corresponds to the maximum oblateness and the minimum density.
Here we adopt a rotation period of $T=7.004$~hr \citep{Fornasier2014} giving a minimum density
\begin{equation}
\rho_{\rm min} = 791~ {\rm kg~m^{-3}}. \label{eq:prior_mac_rhol}
\end{equation}

In this case we have two free parameters, the equatorial radius $a$  and the density $\rho$ (that determine $c$, i.e. the shape).

\subsubsection{Jacobi ellipsoid}
\label{sec:jacobi}

A Jacobi ellipsoid is tri-axial with semiaxes $a>b>c$, rotating around the shortest axis. 
The shape and size (given by $a$, $b$ and $c$), the rotation period $T$ and the (uniform) density $\rho$ are related by \citep{Chandrasekhar1987}:
\begin{eqnarray}
\Omega=\frac{\omega^{2}}{\pi G \rho}=2a b c \int_{0}^{\infty }\frac{u}{\left ( a^2+u \right )\left ( b^2+u \right )\Delta }du \label{eq:jacobi_1}\\
a^2 b^2\int_{0}^{\infty } \frac{du}{\left ( a^2+u \right )\left (b^2+u  \right )}=c^2\int_{0}^{\infty}\frac{du}{\left(c^2+u \right )\Delta } \nonumber\\
\Delta =\sqrt{\left (a^2+u  \right )\left (b^2+u  \right )\left (c^2+u  \right )},\nonumber
\end{eqnarray}
which can be solved numerically.  

For a stable Jacobi ellipsoid, the shapes lie between the axi-symmetric spheroid limit with $\Omega=0.374$ and the most elongated solution with $\Omega=0.284$. 
For the adopted rotation period of 7.004 hr the density is in the range $791 < \rho < 1040$~kg~m$^{-3}$. 

We define the prime meridian 
and rotation angle $\phi$ as we did for the generic ellipsoid.

The seven free parameters of this model are $a$ (which give us the size of the object),
the density $\rho$ (which gives us the shape with the Equation~\ref{eq:jacobi_1}),
and the rotation angles $\phi_i$ at each of the five occultations. 

\subsection{Bayesian approach and MCMC implementation}
\label{sec:bayesian_and_mcmc}

Here, we adopt a Bayesian approach to derive probability densities and credible intervals for 
the physical parameters $\theta$ of the models described above given the occultation data $D$.
We are interested in the ``posterior" probability density function (pdf) $p(\theta|D)$ given by:
\begin{equation}
p(\theta|D) \propto \mathcal{L} \times p(\theta), \label{eq:bayes_theorem}
\end{equation}
where $\mathcal{L}$ is the likelihood function,
and $p(\theta)$ is the so-called "prior" distribution.

The likelihood $\mathcal{L}$ determines the probability to obtain the data $D$ given the physical model,
 and a model of the data uncertainties. 
On the other hand, the prior distribution $p(\theta)$ condense the previous knowledge we have about the parameters.
We define the likelihood function $\mathcal{L}$, assuming that the errors stemming from the fits
described in \ref{sec:occultation_timing} have normal distributions.
Moreover, we formally consider statistical uncertainties in our model by introducing an extra 
normally distributed random variable with median zero and standard deviation $\sigma_{m}$, independent from the measurements.
For instance, the parameter $\sigma_{m}$ may account for unmodeled topographic features on an otherwise 
smooth ellipsoidal model for the body.
Thus, $\sigma_{m}$ is estimated from the data and counted as an extra parameter for each physical model.
With these considerations $\mathcal{L}$ is given by \citep[see for example Equation 4.52 of][]{Gregory2005}

\begin{equation}
\mathcal{L}= 
\left ( 2\pi  \right )^{-N/2}
\left \{ \prod_{i=1}^{N}\left ( \sigma_{ri}^{2}+\sigma_{m}^{2}\right )^{-1/2} \right \} 
\exp\left \{-\sum_{i=1}^{N} \frac{\left ( y_i-m\left ( x_i\mid \theta  \right ) \right )^2)}{2\left (\sigma_{ri}^{2}+\sigma_{m}^{2}  \right )} \right \},
\label{eq:likelihood}
\end{equation}
where $N$ is the number of data points $y_{i}=(f_i,g_i)$ derived in \ref{sec:occultation_timing}, corresponding to
the extremities of each occultation chord by the main body,
$x_i$ represent the independent variable (the site and time of each occultation),
$\sigma_{ri}$ are the uncertainties on the chord extremities $\sigma_{ch}$
projected along the radial direction and counted from the center $(f_c,g_c)$ of the body (see Table~\ref{tab:ring_geometry}).
Finally, 
$\theta$ is the vector representing the parameters of the model that describes the object.
In that context, $m\left ( x_i\mid \theta  \right )$ is the position of the chord extremity predicted by each model.
The prior $p(\theta)$ is derived from physical considerations (for instance, the stability criteria for Maclaurin and Jacobi models),
 and the observational evidence as explained below.

\subsubsection{Analysis of photometry}
\label{sec:priors}

Here we consider the rotational light curve amplitude $\Delta m$ and V absolute magnitude $H_V$
which are in turn used to derive a priori estimates for Chariklo's size, shape and/or density, 
depending on the model adopted.
These a priori estimates are incorporated in the Bayesian modeling 
through the prior probability distribution $p(\theta)$.

Notice that in this discussion we neglect the effect of eclipses between the main body and the rings.
For instance, for a spherical body with radius $R=129$~km,
eclipses between the body and the main ring C1R occur for opening angles  $\lesssim20^{\circ}$.
The projected area of the ring eclipsed by Chariklo's main body is $\lesssim 10\%$ of the total ring projected area,
while the projected area of the main body eclipsed by the rings is $\lesssim 1\%$ of the total body projected area.
Both effects are negligible compared to the uncertainty in the ring's width of $\sim 20\%$ \citep{Berard2017}.
Moreover, none of the stellar occultations analyzed in this work involve eclipses.

\paragraph{Rotational light curve amplitude}
\label{sec:rotational_light_curve}

Chariklo exhibits a rotational light curve amplitude that varies in time.
There was no detection of the light curve amplitudes obtained in 1997 and 1999 \citep{Davies1998,Peixinho2001},
when Chariklo was close to its maximum opening angle.
Peak-to-peak amplitudes of 0.13 mag and 0.11 mag were then measured in 2006 and 2013, respectively \citep{Galiazzo2016,Fornasier2014}.
As not given by the authors, we adopt conservative upper limits for the amplitudes in 1997 and 1999, and uncertainties for those of 2006 and 2013,
derived from the uncertainties in the respective photometry.

\begin{figure}
\plotone{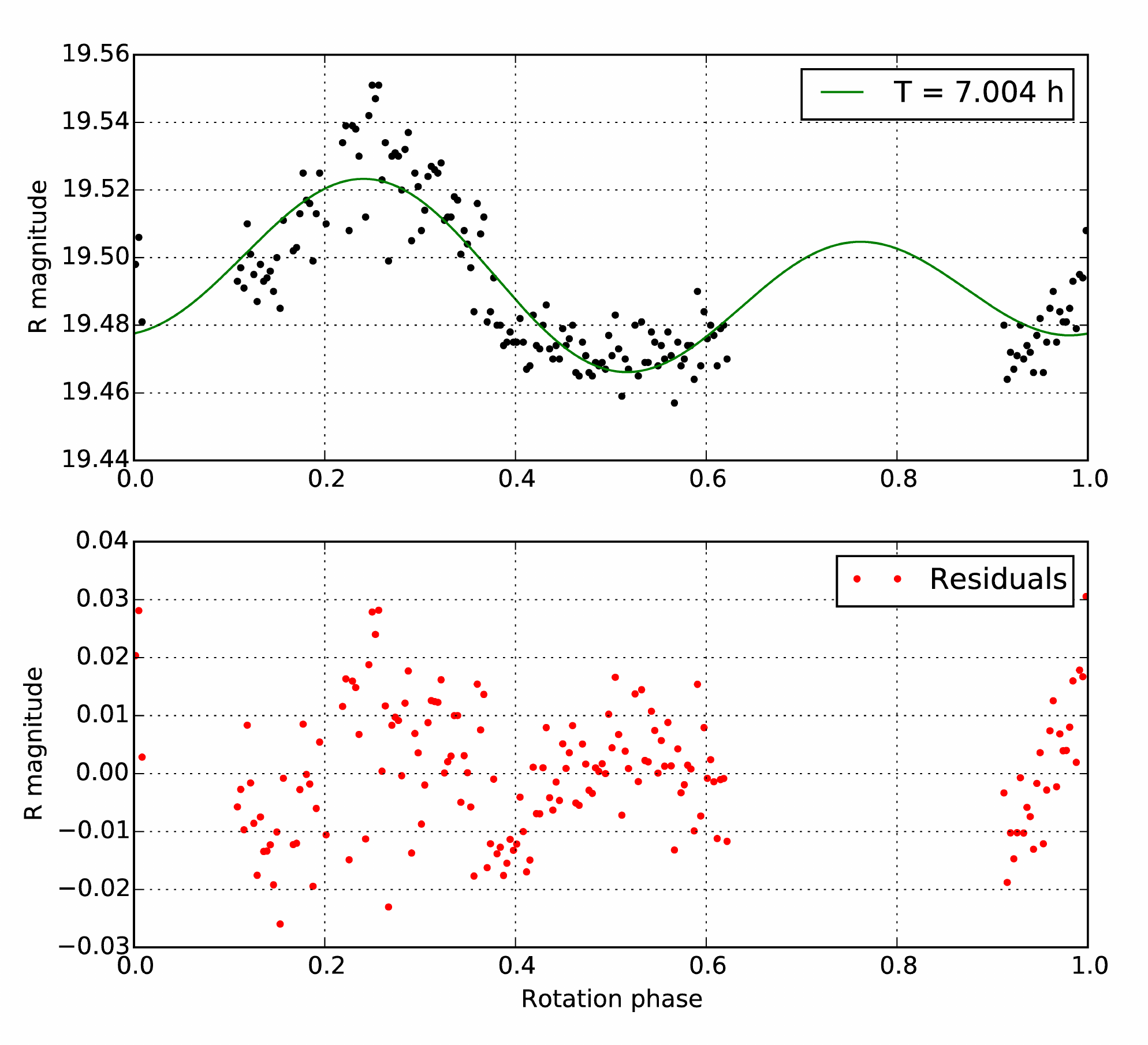}
\caption{
Chariklo's rotational light curve obtained on 2015 July 20 with the SOI camera at SOAR telescope.
The light curve covers about 5 hours of Chariklo's rotation. 
The green-solid line is a fit with a second order polynomial after folding the data with a period $T$=7.004~hr from \cite{Fornasier2014}. 
The peak-to-peak amplitude from this light curve is $\Delta m=0.06\pm0.02$ mag. 
Using the pole position from Table~\ref{tab:ring_geometry}, the opening angle of Chariklo's system at this date is $B=42^\circ$. 
Together with $\Delta m$ and $H_V$ values from the literature, this is used to derive a prior for the size and shape of the models.
}
\label{fig:rotational_light_curve_SOAR_2015}
\end{figure}

Here we add to this data set a partial light curve obtained in July 2015 with the SOAR Optical Imager (SOI).
About 200 images were taken with the R Bessell filter using an exposure time of 80 s.
A bias correction and flat-fielding was performed with the SOAR/SOI IRAF routines. The images were processed using difference image photometry implemented in the IDL code \textit{DanDIA} \citep{Bramich2008}. 
A light curve was obtained using aperture photometry with IRAF routines \citep{Tody1986}.
Figure \ref{fig:rotational_light_curve_SOAR_2015} shows the light curve obtained covering $\sim$5~h from which we determined a peak-to-peak amplitude $\Delta m$=0.06$\pm$0.02.

Values from the literature and from this work are summarized in Table~\ref{tab:rotational_amplitude}.
Using the pole position from Table~\ref{tab:ring_geometry} we calculate the opening angle $B$ for each date
that indicates a correlation between $\Delta m$ and $|B|$.
Additionally, the light curve from \cite{Fornasier2014} is  double-peaked with minimums separated by about half of the rotation phase.
Both facts suggest that the brightness variations are dominated by the variable projected area of an elongated body instead of albedo variegations.

With these considerations, 
we model the peak-to-peak rotational light curve amplitude with the contribution of the main body and rings given by \citep{Fernandez-Valenzuela2016a}

\begin{equation}
\Delta m = -2.5\log\left( \frac{A_{\rm min} p_b+A_r (I/F)}{A_{\rm max} p_b+A_r (I/F)} \right), \label{eq:dm_model}
\end{equation}

where $A_{\rm min}$ and $A_{\rm max}$ are the minimum and maximum projected area of Chariklo's body, respectively,
$p_b$ is the geometric albedo of Chariklo's body,
$A_{r}$ is the projected area of the rings,
and \footnote{Here $I$ is the intensity emitted by the ring surface and $\pi F$ is the incident solar flux density. 
The quantity $I/F$ must not be confused with the geometric albedo of the ring particles $p_p$.
} $(I/F)$ is the ring reflectivity.

\paragraph{Absolute magnitude}
\label{sec:hv_analysis}

Chariklo also exhibits an absolute V magnitude $H_V$ that varies in time \citep{Belskaya2010,Fornasier2014,Duffard2014}.
Considering contributions from the changing aspect of Chariklo's main body and its rings, 
we model $H_V$ with the relation

\begin{equation}
H_{V} = H_{\odot} -2.5 \log\left(\frac{A_{b}~p_{b}+A_{r}~(I/F)}{\pi~au_{km}^{2}} \right ), \label{eq:hv_model}
\end{equation}

where
$H_{\odot}$=-26.74 is the absolute magnitude of the Sun in V,
$A_{b}$ is the projected area of Chariklo's main body,
and
$au_{km}$ is an Astronomical Unit in km.

\paragraph{Priors for the generic ellipsoid}
\label{sec:priors_ellipsoid}

For the generic triaxial ellipsoid we use the $\Delta m$ and $H_V$ to derive estimates for
the semimajor axis $a$ and the ratios $b/a$ and $c/a$.

To derive the ring contribution to the brightness variations,
we consider a ring of radius $\sim$~400~km and
width $w_1\sim$5.5~km (neglecting the contribution from the faint and narrow C2R ring),
and a ring reflectivity $(I/F)$ varying between 0\% (neglecting the ring contribution) up to 9\%,
considering previous estimations of this quantity \citep{Braga-Ribas2014,Duffard2014}.
For the body contributions, 
we adopt a body geometric albedo $p_b=4.2\pm0.5\%$ from \cite{Fornasier2014},

We fit the Equations~\ref{eq:dm_model}
and \ref{eq:hv_model}
to the $\Delta m$ values in Table~\ref{tab:rotational_amplitude} 
and $H_V$ from the literature in a least-squares scheme,
to obtain 

\begin{equation}
a=138\pm16~km, \frac{b}{a}= 0.86 \pm 0.04 ,  \frac{c}{b} = 0.89 \pm 0.30, \label{eq:prior_tri_abc}
\end{equation}

which are used in Section~\ref{sec:results_triaxial_ellipsoid} to define normally distributed priors for those parameters.

\paragraph{Priors for Maclaurin spheroid}
\label{sec:priors_maclaurin}

For the Maclaurin model we use $H_V$ values to derive estimates for the equatorial radius $a=b$ and the density $\rho$.
Adopting the same ring dimensions, body geometric albedo and range of ring reflectivity as before,
we fit the Equation~\ref{eq:hv_model} to the 
$H_V$ from the literature in a least-squares scheme to obtain 

\begin{equation}
a=b= 135\pm25~km, \label{eq:prior_mac_a}
\end{equation}

while we find that the density $\rho$ is unbounded and can take values in all the valid interval between $\rho=791$~kg~m$^{-3}$ and $\rho\rightarrow \infty$.

In practice, we consider a conservative upper limit for the density 
\begin{equation}
\rho_{\rm max}=5000~kg~m^{-3}, \label{eq:prior_mac_rhoh}
\end{equation}
after considering the known density distribution of asteroids and TNOs  \citep{Britt2002,Carry2012a}.

These values are used in Section~\ref{sec:results_maclaurin_spheroid} 
to define normally and uniformly distributed priors for those parameters.

\paragraph{Priors for Jacobi}
\label{sec:priors_jacobi}

For the Jacobi ellipsoid we must impose the binding conditions from Equation~\ref{eq:jacobi_1} to $a$, $b$ and $c$.
As done with the generic ellipsoid, 
we fit the Equations~\ref{eq:dm_model}
and \ref{eq:hv_model}
to the $\Delta m$ and $H_V$ values
to obtain 

\begin{equation}
a=151\pm13~km, \rho = 804 \pm 6~kg~m^{-3}.  \label{eq:prior_jac_rho}
\end{equation}

which is used to define normally distributed priors in Section~\ref{sec:results_jacobi_ellipsoid}.

\paragraph{Rotation angle during occultations}
\label{sec:rot_angle_binding}

It is worth to mention here some words about the rotation angle during the stellar occultations.
The rotation light curve from \cite{Fornasier2014} was obtained in 2013 between June 11 and June 12. 
Assuming that most of the variability is due to shape instead of albedo variegations, 
we define the rotation angle $\phi$=0$^{\circ}$ at one of the brightness minima of that light curve, 
for instance JD=2456455.23.
This is used to determine the rotation angle $\phi(JD,T_{\rm sid})$ at any given date $JD$ for a given sidereal rotation period $T_{\rm sid}$.
For instance, for the stellar occultation of June 3, which is only eight days before this measurement,
we find
 
\begin{equation}
\phi_{\rm 2013Jun03}=356\degr\pm54\degr, \label{eq:phi_2013}
\end{equation}
 adopting $T_{\rm sid}=7.004\pm0.036$~hours.
Unfortunately, given the current accuracy of the rotation period, 
the rotation angle is essentially lost after a few weeks,
preventing us to derive a rotation angle for the occultations in 2014 and 2016 solely from the light curve in \cite{Fornasier2014}.
The rotation angle in the ellipsoidal and Jacobi models
is then considered independent between occultations and explored between 0$^{\circ}$ and 180$^{\circ}$
due to the rotational symmetry.

\subsubsection{MCMC scheme}
\label{sec:mcmc_scheme}

To estimate the posterior probability distribution $p(\theta|D)$,
 we adopt a Markov Chain Monte Carlo (MCMC) scheme to draw samples from it.
The MCMC sampling is done using the library called \textit{emcee} \citep{Foreman-Mackey2013} 
which implements the affine-invariant ensemble sampler by \cite{Goodman2010}.

To generate the posterior samples we follow a standard procedure.
For each model, we run a MCMC with $n_{\rm walk}$ random ``walkers", each of them exploring the parameter space.
To determine the number of random steps $n_{\rm burn}$ necessary to ensure the chain convergence, 
we adopt $n_{\rm burn}>10\times \tau_f$, 
where $\tau_f$ is the integrated autocorrelation ``time" of the chain measured in chain steps \citep{Foreman-Mackey2013}
(for this we use the implementation given in \textit{emcee}). 
Once this is done,
we continue the MCMC for $n_{\rm samp}$ steps from where we obtain the samples
which are representative of the posterior probability of interest $p(\theta|D)$. 
Then, the marginal probability distribution for the parameters $ \theta$ is estimated using the histograms of the samples. 
From the histograms, we derive the best-fit parameter values and credible intervals.
For the credible intervals, we use the highest posterior probability density interval containing 68\% of the samples. 
This is the smallest interval such that any point inside the interval has a higher probability density than any other point outside of the interval.

Additionally, and as an heuristic test for convergence, 
for each model we run several chains starting the $n_{\rm walk}$ walkers at different random positions well spread in the parameter space. 
We repeat this procedure several times to verify that we obtain the same results. 

\section{OCCULTATION RESULTS}
\label{sec:results}

\subsection{Sphere}
\label{sec:results_sphere}

\begin{figure}
\plotone{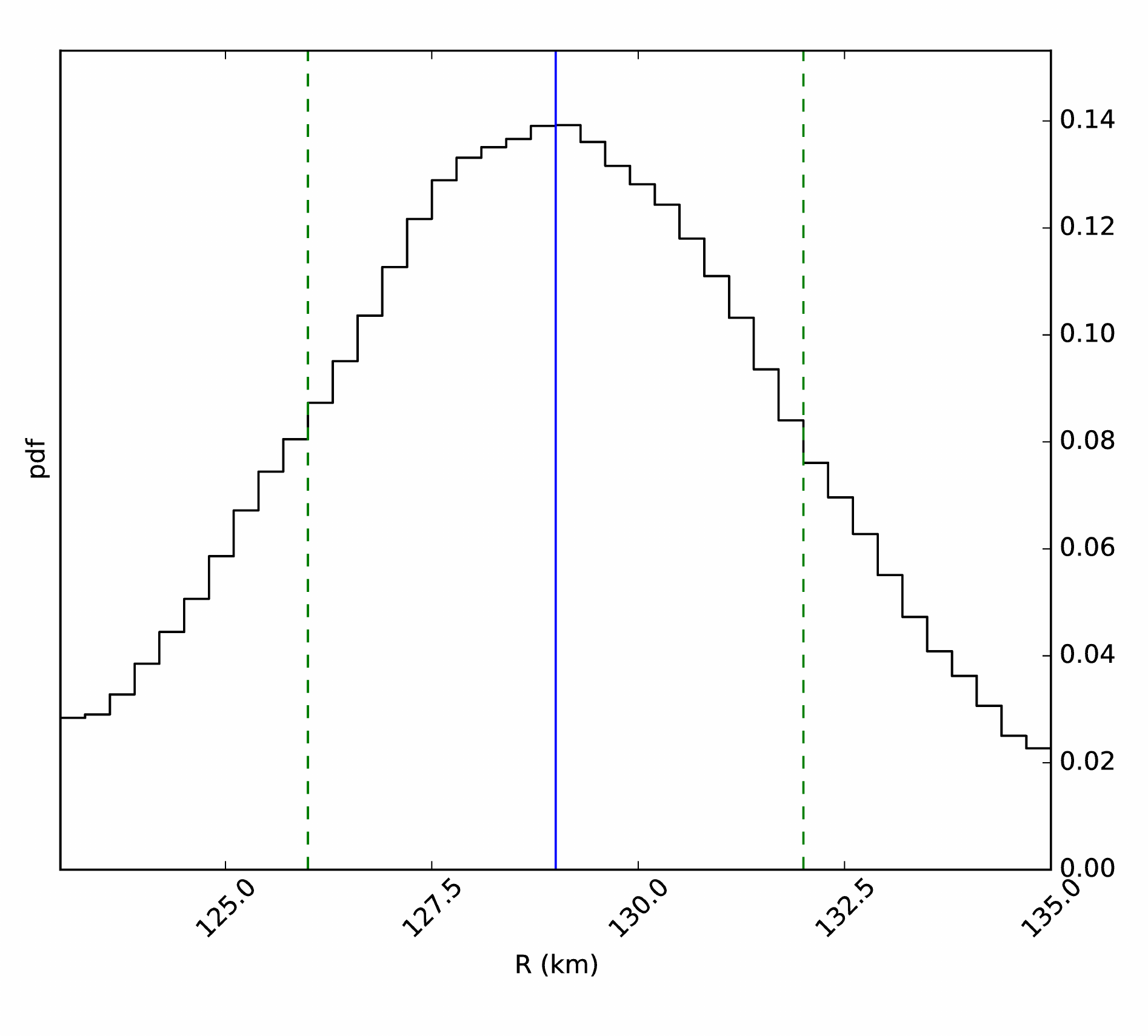}
\caption{ 
Posterior probability function (pdf) for the sphere radius $R$. 
The blue-continuous line is the best value and the green-dashed lines indicate the 68\% credible interval from where we determine the value for $R=129\pm3$~km.
}
\label{fig:posterior_sphere}
\end{figure}

For the spherical model we use as prior a uniform distribution between $R=100$ and $R=150$~km,
and for $\sigma_m$ we adopt a uniform distribution between 0~km and 50~km.
Using $n_{\rm walk}$=500, $n_{\rm burn}=10^4$ steps and $n_{\rm samp}=10^2$ steps,
 we obtain the posterior pdf shown in Figure~\ref{fig:posterior_sphere},
 and eventually a sphere radius of $R = 129 \pm 3$~km (68\% credible interval).

For the best-fit radius, we obtain a ``topographic" parameter $\sigma_m$=11~km.
That is, the radial departures from the best-fit limb can be modeled as normally distributed with standard deviation of 11~km,
which is $\sim$~9\% of the radius $R$.

\begin{figure}
\plotone{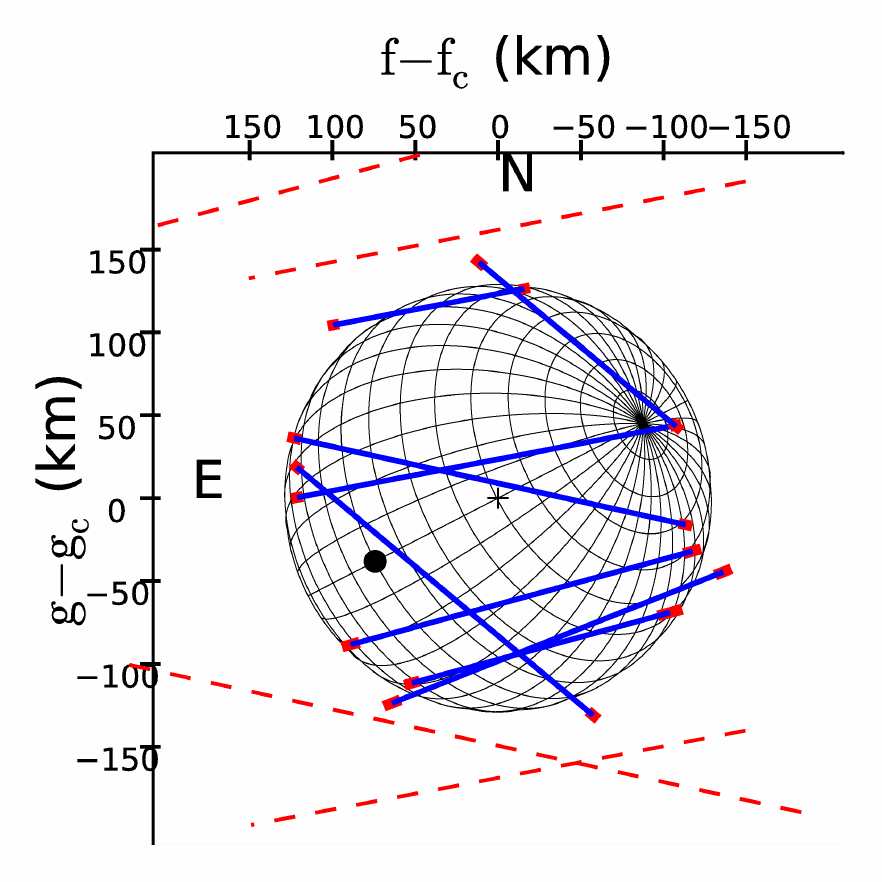}
\caption{
Best-fit for the sphere model with radius $R=129\pm3$~km compared to all the occultation chords projected in the sky plane. 
The sphere model is plotted with the average orientation of $B=40^{\circ}$ and $P=-63^{\circ}$.
Measurement uncertainties are indicated in red.
Red dashed lines outside the body are the multiple negatives detections used to constraint the model. 
Black dot indicate the sphere's equator.
}
\label{fig:results_sphere_allpoints}
\end{figure}

\begin{figure}
\plotone{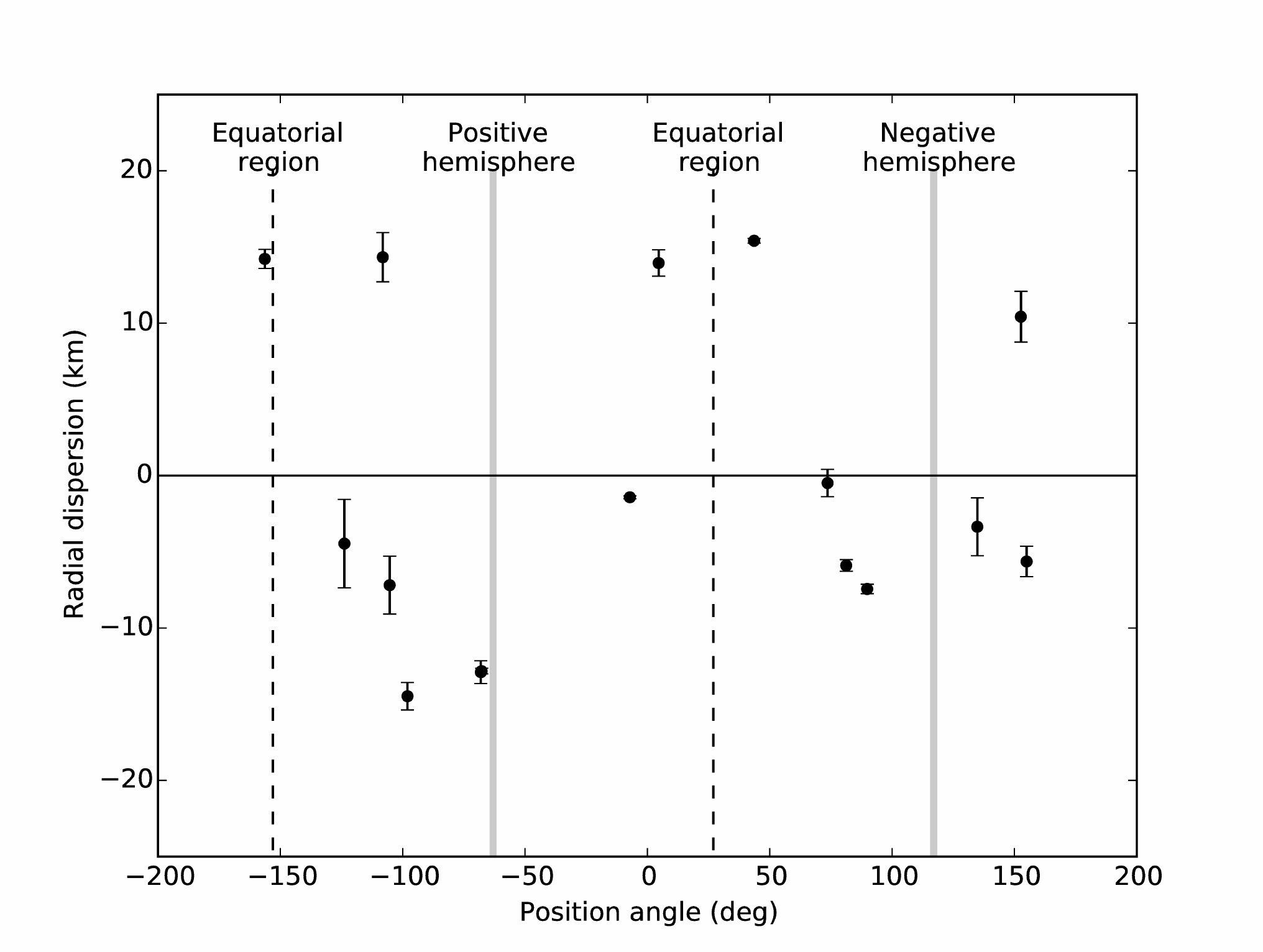}
\caption{ 
Radial residual as a function of position angle (counted positively from celestial north towards east) for the spherical model of Figure~\ref{fig:results_sphere_allpoints}.
Error bars are the uncertainties in the position of the occultation chord extremities projected in the radial direction.
Dashed lines indicate the equatorial region while the solid lines indicate the polar region. 
Positive and negative hemispheres are defined according to the choice of the pole position given in Table~\ref{tab:ring_geometry}.
We observe a clear correlation of the radial residuals with the position angle along the limb, with positive residuals in the equatorial region
and negative residuals in the polar regions.
 }
\label{fig:dispersion_vs_pos_angle}
\end{figure}

In Figure~\ref{fig:results_sphere_allpoints} we compare all the occultation chords with the best fitted limb using the spherical model. 
In Figure~\ref{fig:dispersion_vs_pos_angle} we plot the radial difference of each chord extremity with respect to the best limb as a function of the position angle (counted positively from the north toward the east).
There is a clear tendency for the chord extremities to be inside the sphere limb around the polar regions and outside the limb in the equatorial regions.
As the departures are significantly larger than the uncertainties on the data points,
 this naturally motivates us to test the flattened models below.

\subsection{Tri-axial ellipsoid}
\label{sec:results_triaxial_ellipsoid}

For the triaxial ellipsoid model, we use normally distributed priors for $a$, $b/a$ and $c/b$ with values from
Equation~\ref{eq:prior_tri_abc}.
The normal distributions are truncated such that $a>0$ and the ratios $b/a$ and $c/b$ stay in the open interval $]0,1[$,
keeping the condition $a>b>c$.
For $\sigma_m$ we adopt a uniform distribution between 0~km and 50~km as with the previous model.
Finally, for the rotation angles $\phi_i$ we adopt uniform distributions between 0$^{\circ}$ and 180$^{\circ}$.
We do not explore the range $180-360^{\circ}$ due to the rotational symmetry of the ellipsoid.

\begin{figure}
\plotone{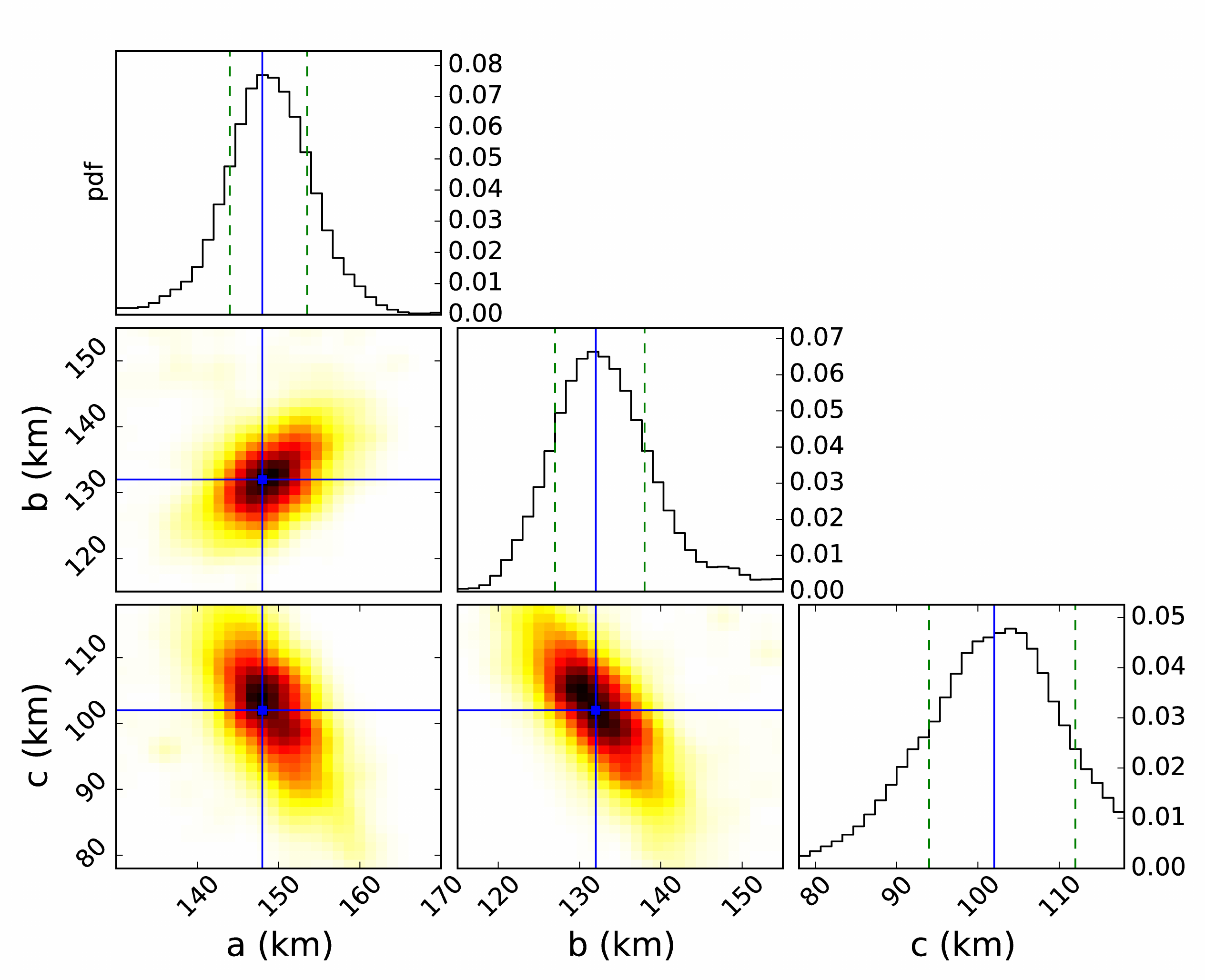}
\caption{
Results for the generic triaxial ellipsoid.
The plots in the diagonals show the marginal posterior pdf $p(\theta|D)$ (Equation~\ref{eq:bayes_theorem}) 
for the semiaxes $a$, $b$ and $c$.
The rest of the plots are the joint posterior pdf for
$a$ vs $b$ (left-center),
$a$ vs $c$ (bottom-left),
and $b$ vs $c$ (bottom-center).
The blue-continuous lines indicate the best-fit values adopted and the green-dashed lines indicate the 68\% credible intervals
given in Table~\ref{tab:results}.
}
\label{fig:posterior_ellipsoid}
\end{figure}

\begin{figure}
\plotone{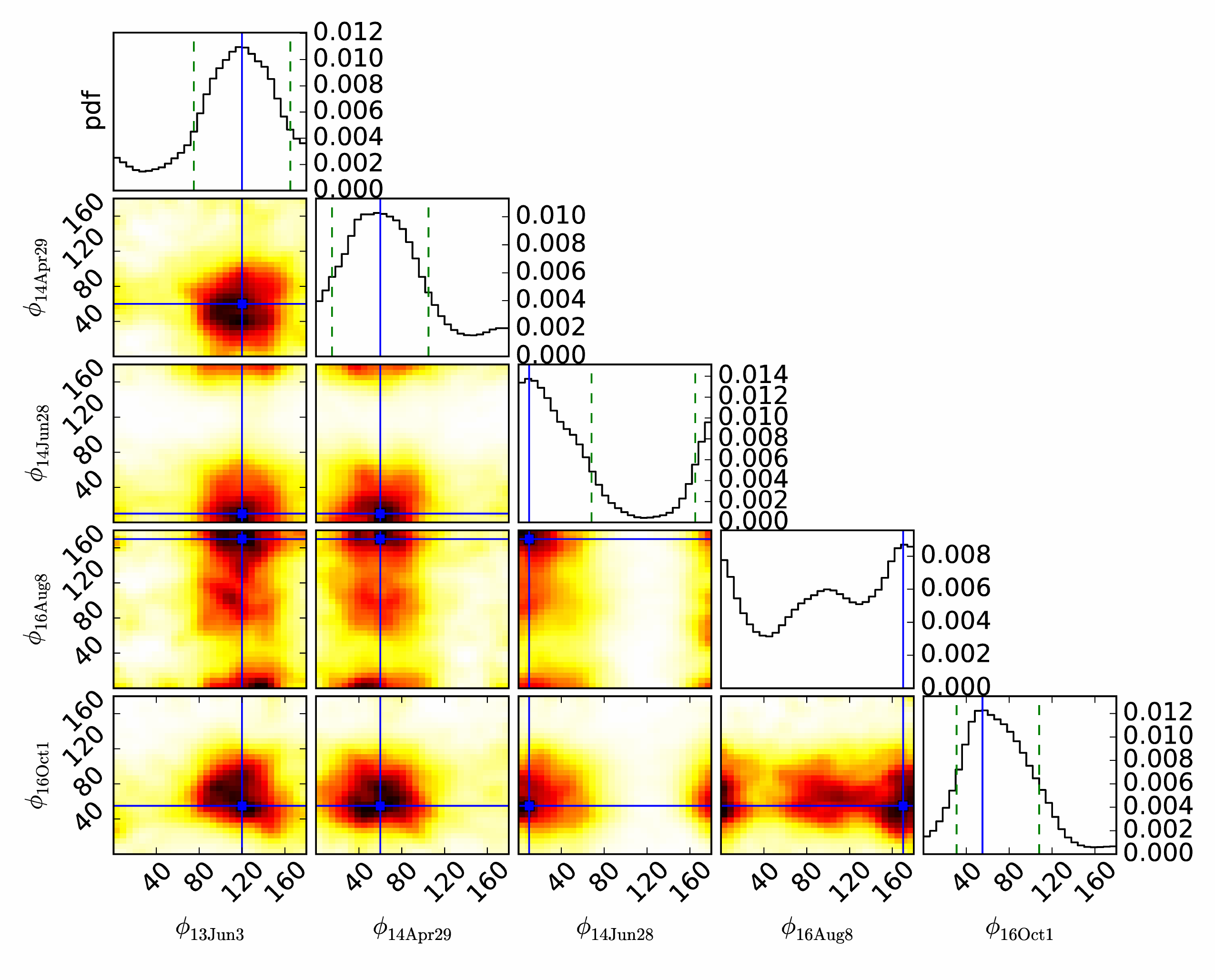}
\caption{
Posterior pdf of the rotation angle $\phi$ for each occultation for the generic ellipsoid model.
The best-fit values used in Figure~\ref{fig:ellipsoid_best_fit_bf} are indicated by blue dots and lines.
Green-dashed lines indicate the 68\% credible intervals, 
except for the occultation of 2016 August 8 where the rotation angle pdf is multimodal.
For August 8, the two peaks in the pdf are due to the single chord.
Each peak in the pdf occur when the body limb is roughly equidistant from the chord extremities.
}
\label{fig:posterior_ellipsoid_angle}
\end{figure}

\begin{figure*}
\gridline{
\fig{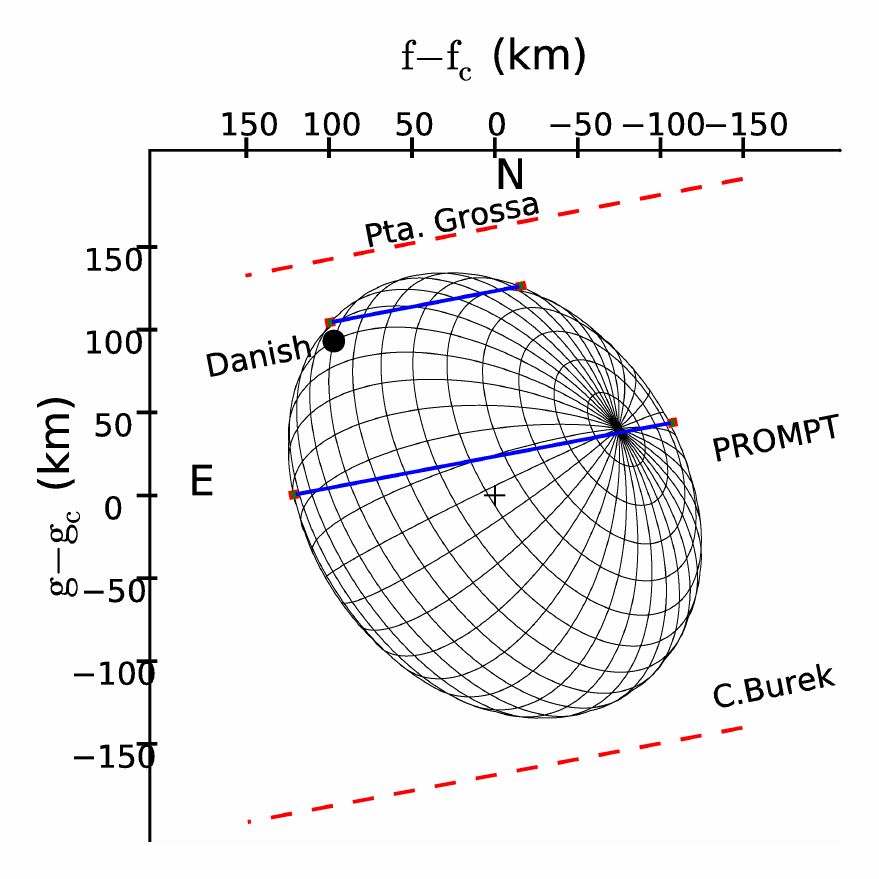}{0.35\textwidth}{(a) 2013 June 3}
\fig{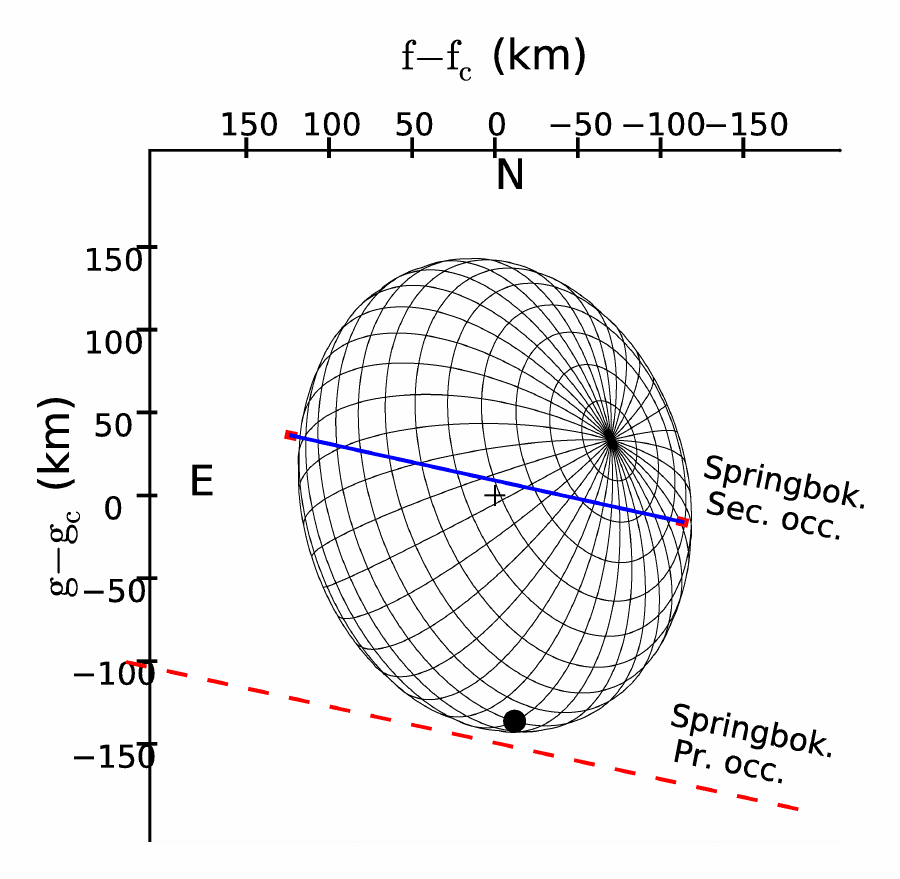}{0.35\textwidth}{(b) 2014 April 29}
}
\gridline{
\fig{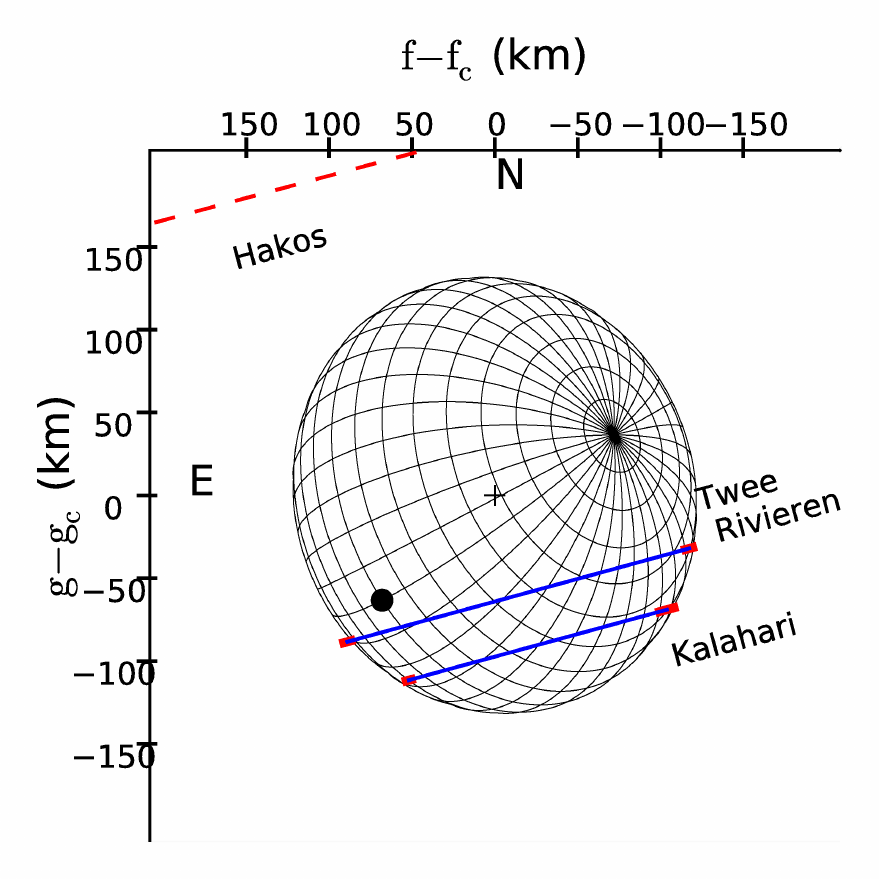}{0.35\textwidth}{(c) 2014 June 28}
\fig{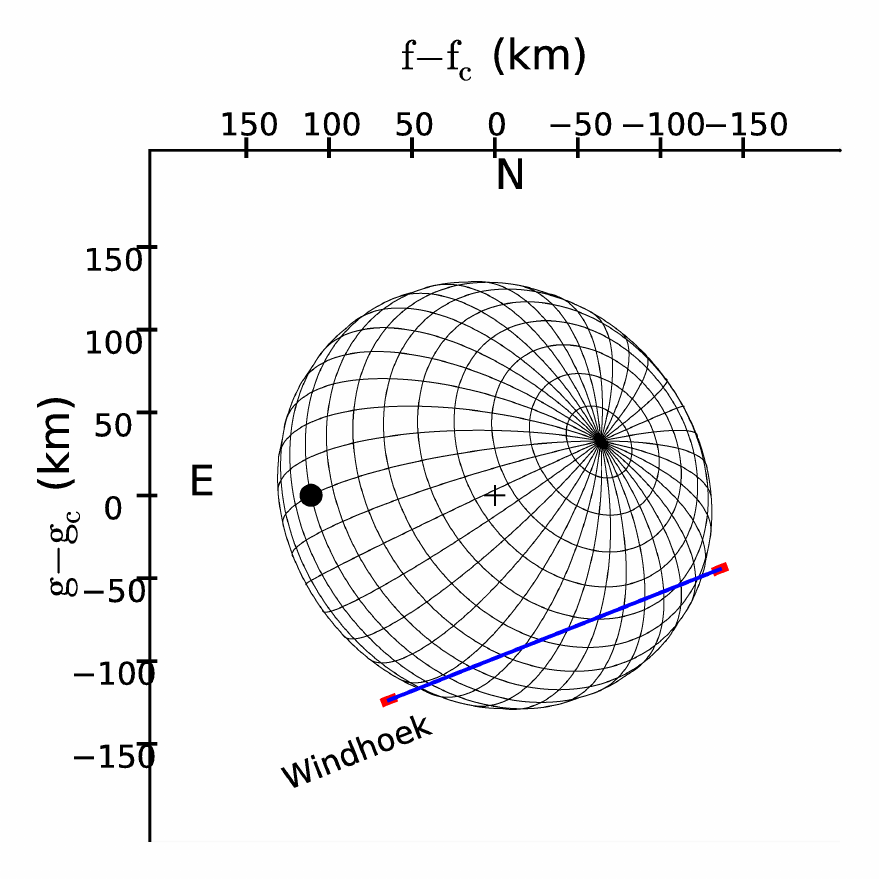}{0.35\textwidth}{(d) 2016 August 8}
}
\gridline{
\fig{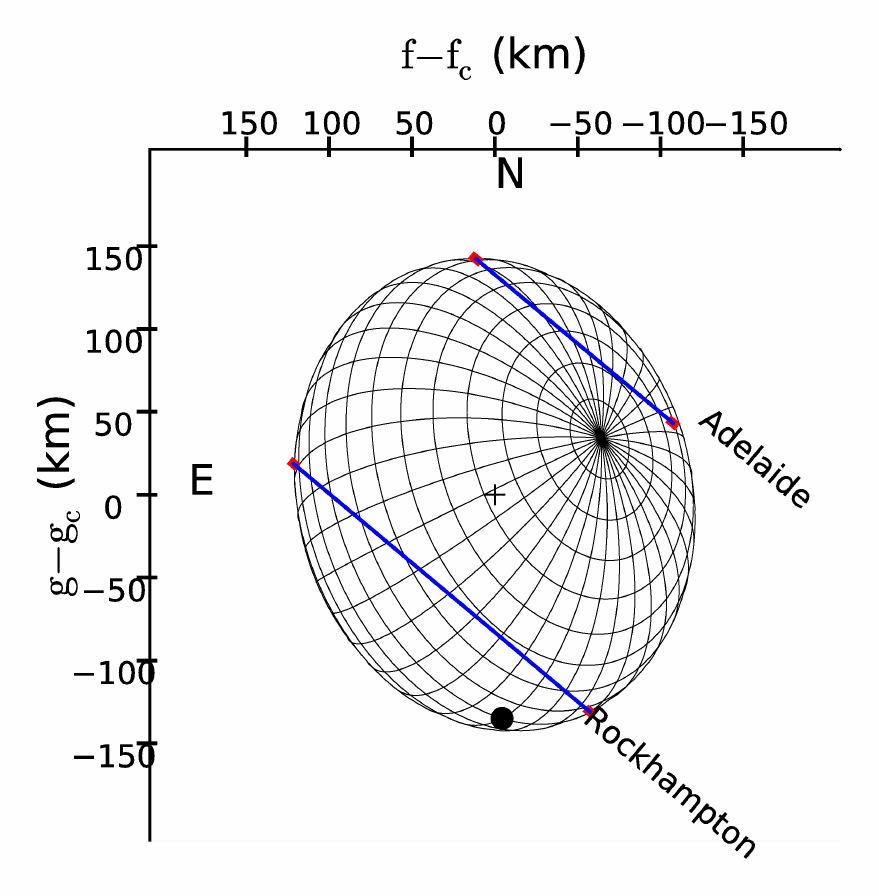}{0.35\textwidth}{(e) 2016 October 1}
}
\caption{
Results of the triaxial model using the best-fit values in Table~\ref{tab:results}.
At each panel, the body has the same pole position $(\alpha_P,\delta_P)$ and apparent center $(f_c,g_c)$ than the rings (not shown) as given in Table~\ref{tab:ring_geometry}. 
Blue lines are the detections of the main body with uncertainties in red. 
Red dashed lines are the negative chords closest to the object used as constraints for the body model. 
The black dot indicates the intersection between the equator and the prime meridian, 
which is used as reference to define the rotation angle $\phi$.
}
\label{fig:ellipsoid_best_fit_bf}
\end{figure*}

Using $n_{\rm walk}$=500, $n_{\rm burn}=10^4$ and $n_{\rm samp}=10^2$,
 we obtain the posterior pdf for $a$, $b$ and $c$ shown in Figure~\ref{fig:posterior_ellipsoid},
 and the posterior for rotation angles shown in Figure~\ref{fig:posterior_ellipsoid_angle},
 from which we determine the parameter values given in Table~\ref{tab:results}.
Figure~\ref{fig:ellipsoid_best_fit_bf} shows the best-fit ellipsoid models compared to the occultation chords. 

From Figure~\ref{fig:posterior_ellipsoid_angle} we obtain a rotation angle $\phi$=120$\degr\pm$45$\degr$ for the occultation of 2013 June 3.
Considering the ellipsoid rotational symmetry, this angle is equivalent to $\phi$=300$\degr\pm$45$\degr$, consistent with the one obtained from the rotational light curve in Section~\ref{sec:priors}.
This validates the assumption that the short term variability is dominated by the projected shape of a rotating elongated body
rather than albedo features.

\paragraph{Sensitivity to priors}
\label{sec:triaxial_prior_sensitivity}

To test the sensitivity to priors, 
we repeat a MCMC run with uniform distribution between 100~km and 200~km for the semimajor axis $a$,
and uniform distribution between 0.1 and 1 for the ratios $b/a$ and $c/b$.
In this case we determine
 $a = 147^{+7}_{-3}$~km, 
 $b=139\pm6$~km, 
$c=98^{+9}_{-8}$~km.

This indicate that the priors have some influence in the results,
particularly in the ratio $b/a=0.95$ which is larger than above.
Nonetheless, the parameters obtained are mainly dominated by the occultation data.

\subsection{Maclaurin spheroid}
\label{sec:results_maclaurin_spheroid}

\begin{figure}
\plotone{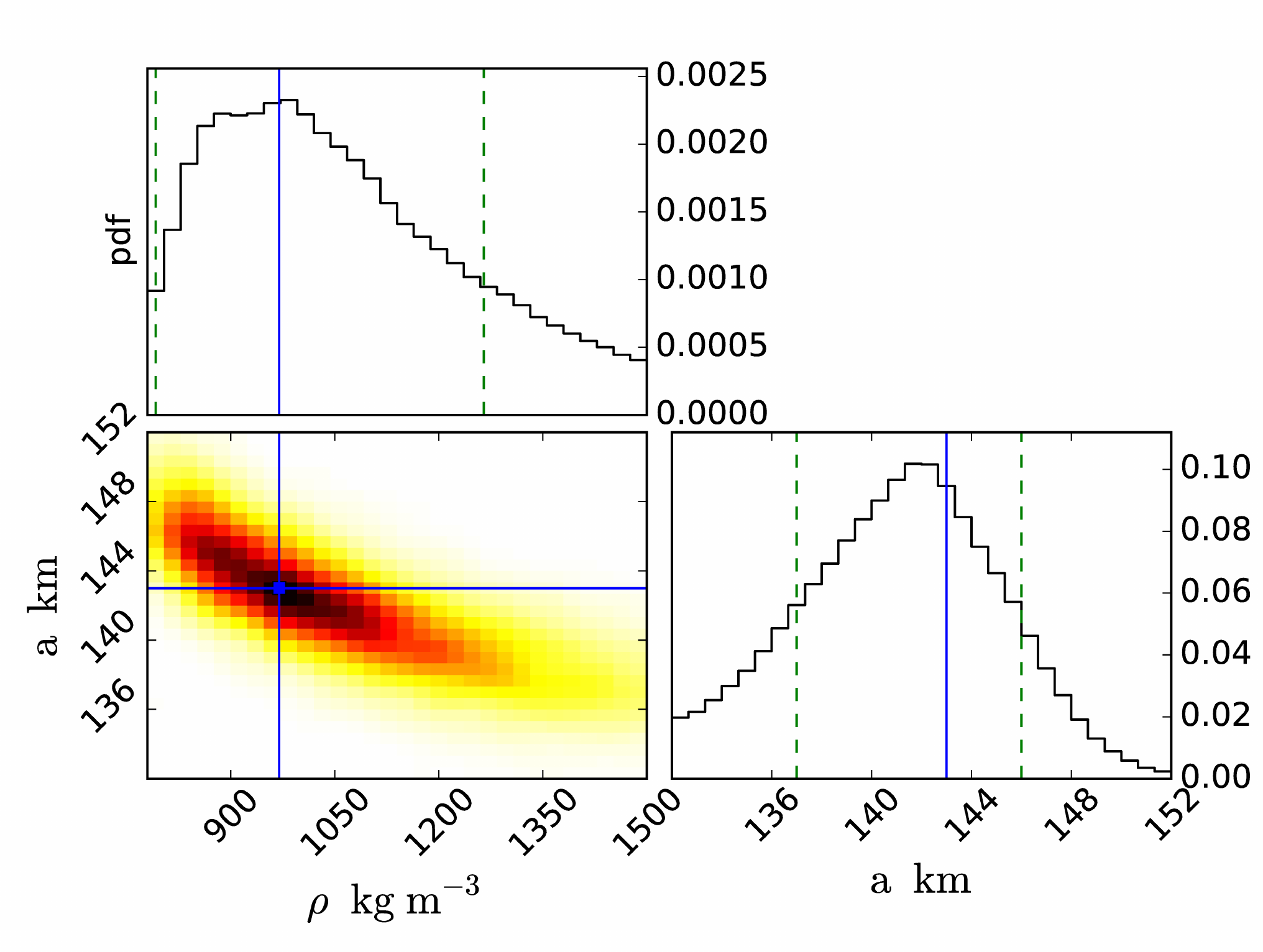}
\caption{
Results for the Maclaurin spheroid model.
Bottom-left: the joint posterior probability $p(\theta|D)$ (Equation~\ref{eq:bayes_theorem}) for the density $\rho$ and equatorial radius $a$.
Top-left: the marginal posterior probability for the density $\rho$. 
Bottom-right: the marginal posterior probability for the equatorial radius  $a$. 
The blue-continuous lines indicate the best parameter values and the dashed vertical lines indicate the 68\% credible intervals
given in Table~\ref{tab:results}.
}
\label{fig:posterior_spheroid}
\end{figure}

For the prior in the density $\rho$, we use a uniform distribution with values from Equation~\ref{eq:prior_mac_rhol} and \ref{eq:prior_mac_rhoh}.
For the equatorial radius $a=b$ we use a normally distributed prior with values from
Equation~\ref{eq:prior_mac_a}.
Using $n_{\rm walk}$=500, $n_{\rm burn}=10^4$ and $n_{\rm samp}=10^2$, 
we obtain the posterior pdf for $\rho$ and $a$ shown in Figure~\ref{fig:posterior_spheroid}.
We take the maximum of the joint distribution of $\rho$ and $a$ as the most probable values,
 while the formal uncertainties are taken from the 68\% credible intervals
from which obtain the values given in Table~\ref{tab:results}.
 
From the joint posterior in the lower-left panel of Figure~\ref{fig:posterior_spheroid},
 we note that $\rho$ and $a$ are correlated.
For lower densities the hydrostatic equilibrium figure is more flattened and, consequently, a larger object is needed to match the occultation data. 

The upper ``wing" for larger densities in the posterior pdf is due to the relation between density and oblateness (Equation~\ref{eq:maclaurin_shape}).
As the density increases, the oblateness changes more slowly and the body approaches asymptotically to a sphere for $\rho \rightarrow \infty $.

\begin{figure*}
\gridline{
\fig{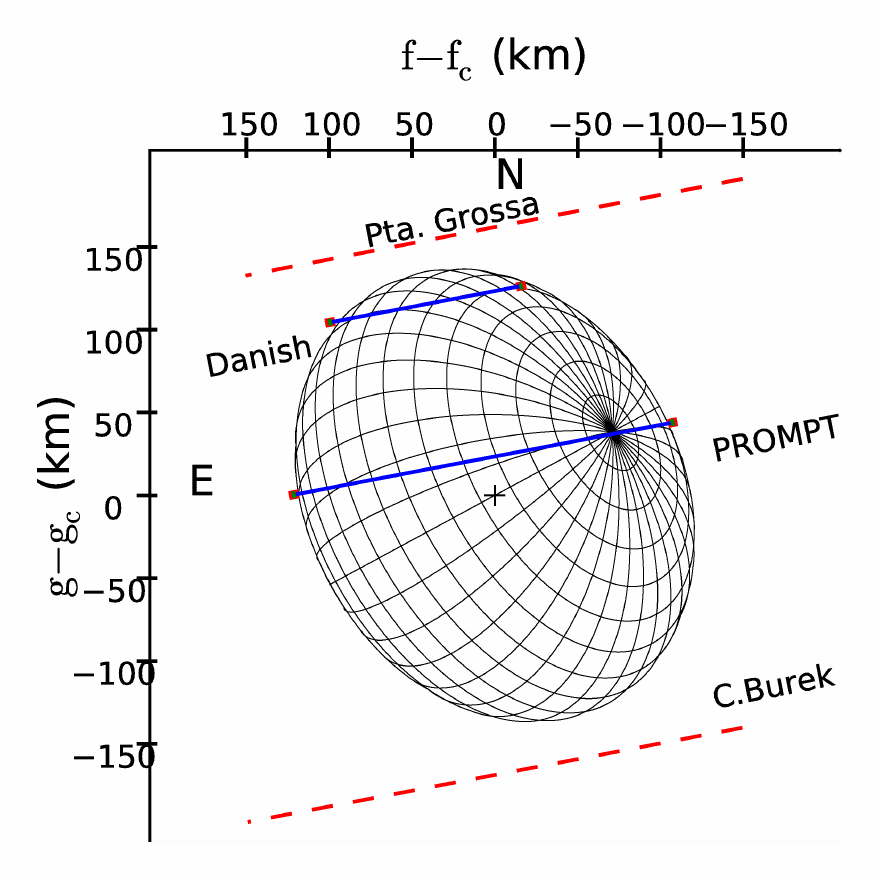}{0.35\textwidth}{(a) 2013 June 3}
\fig{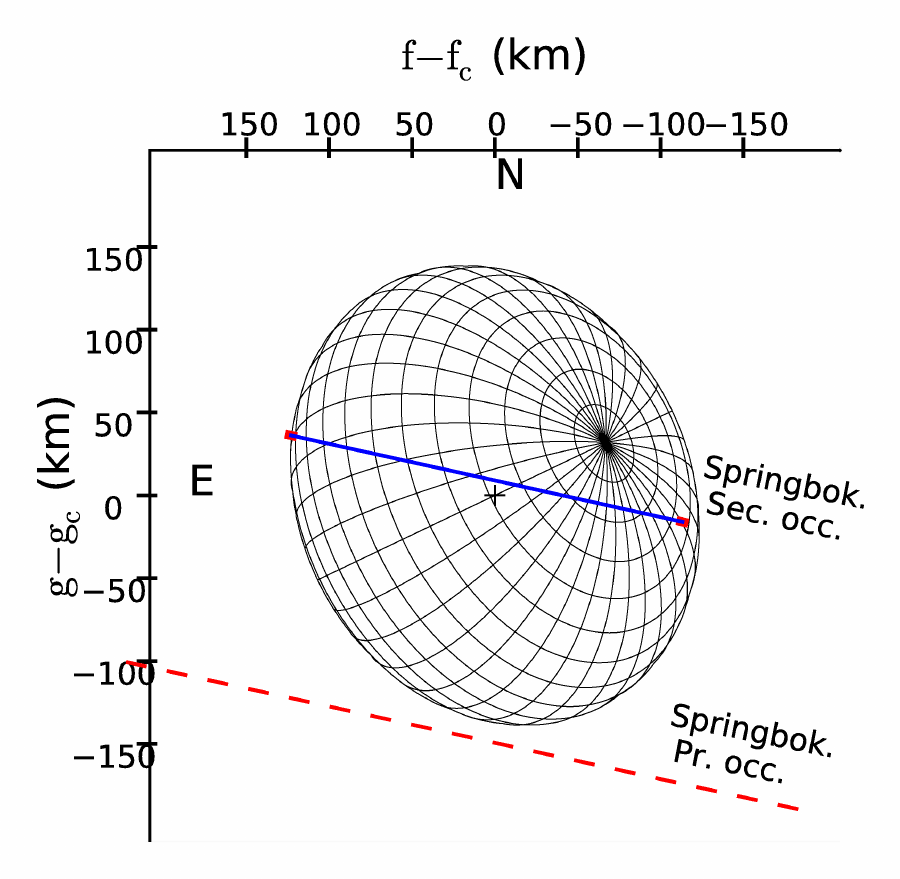}{0.35\textwidth}{(b) 2014 April 29}
}
\gridline{
\fig{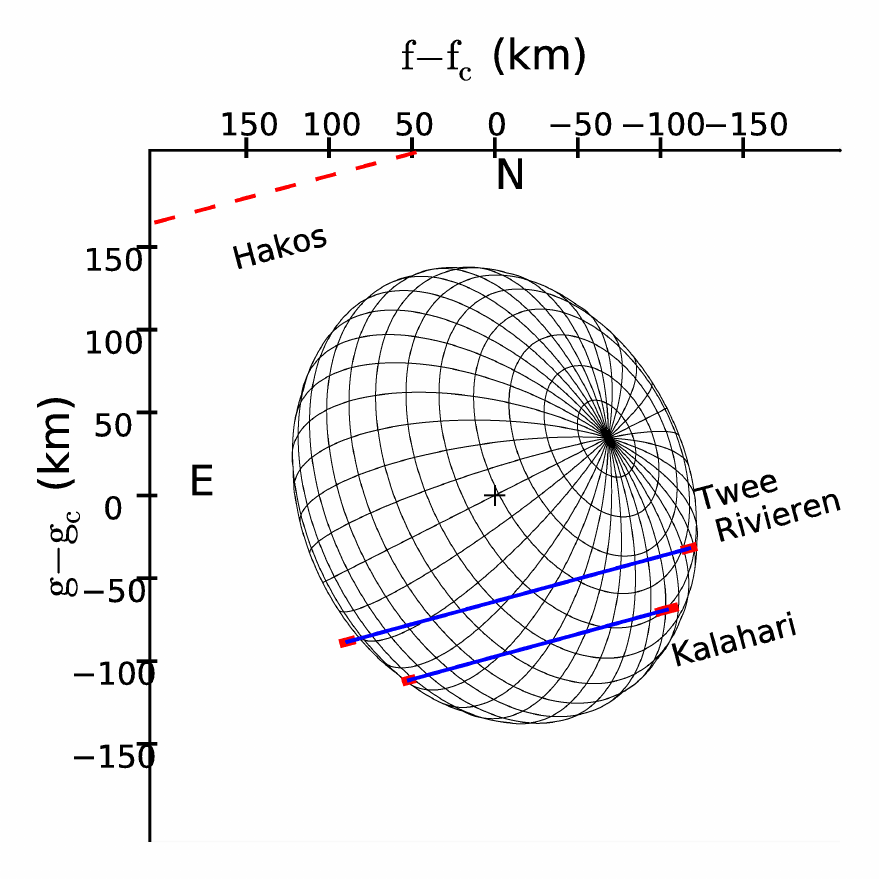}{0.35\textwidth}{(c) 2014 June 28}
\fig{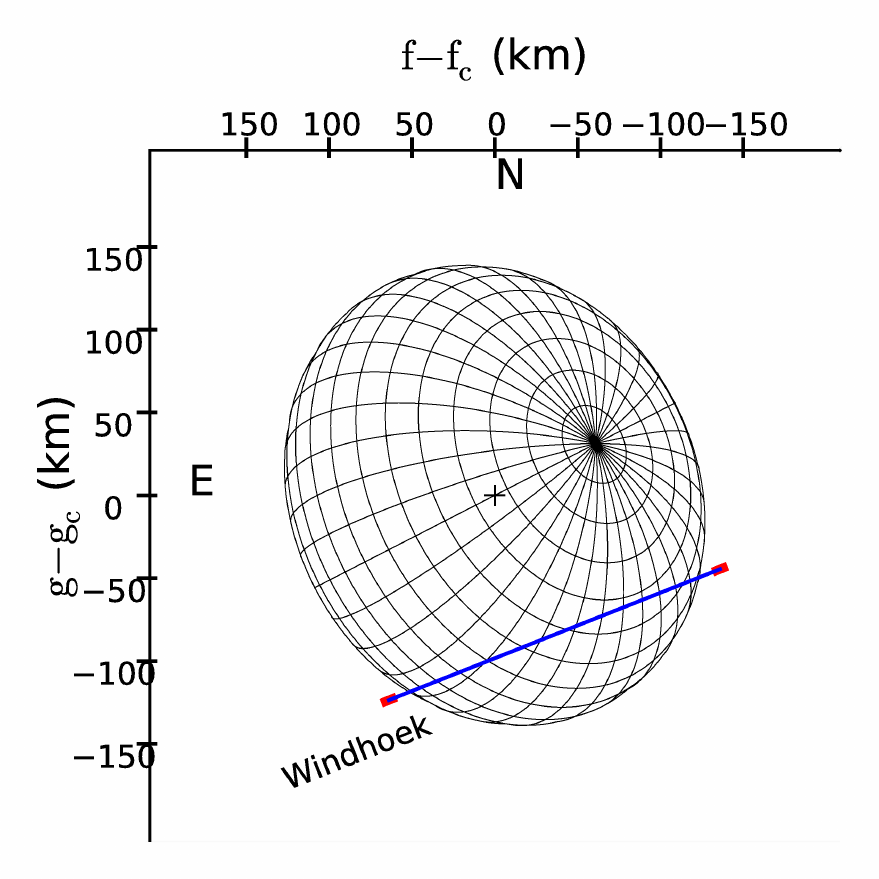}{0.35\textwidth}{(d) 2016 August 8}
}
\gridline{
\fig{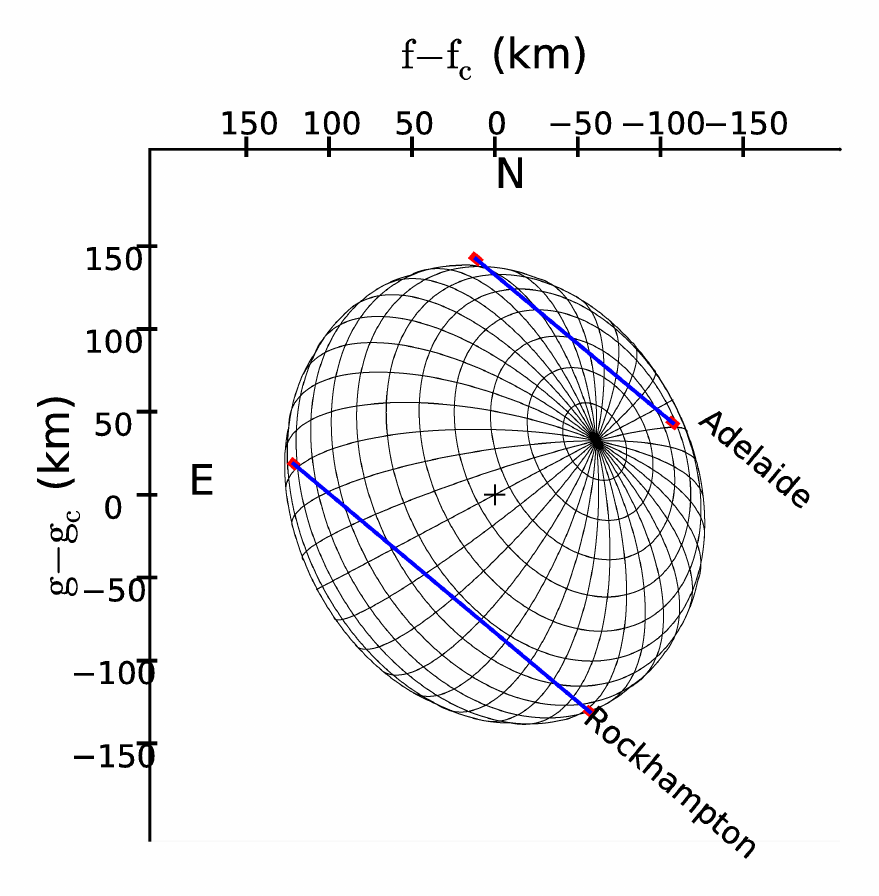}{0.35\textwidth}{(e) 2016 October 1}
}
\caption{
Same as Figure~\ref{fig:ellipsoid_best_fit_bf} for the Maclaurin model using the best-fit values in Table~\ref{tab:results}.
}
\label{fig:mac_bf}
\end{figure*}

Figure~\ref{fig:mac_bf} shows the nominal Maclaurin solution compared to the occultation chords. 
As with the ellipsoidal case, the parameter $\sigma_m=7$~km is smaller than for the spherical model.

\paragraph{Sensitivity to priors}
\label{sec:prior_sens_mac}

As done with the ellipsoidal model,
we repeat a MCMC run using a uniformly distributed prior for the equatorial radius $a$
between 100~km and 150~km.
From the posterior distribution of $\rho$ and $a$ we obtain 
$\rho=950^{+300}_{-150}$~kg m$^{-3}$ and $a$=144$\pm$4~km,
showing that the results are not strongly sensitive to the priors chosen
and are dominated by the occultation data.

\subsection{Jacobi ellipsoid}
\label{sec:results_jacobi_ellipsoid}

For the Jacobi ellipsoid model, we use normally distributed priors for 
the semimajor axis $a$ and density $\rho$ 
with values from Equation~\ref{eq:prior_jac_rho}.
Additionally, the density must satisfy the condition of equilibrium as described in Section~\ref{sec:body_model}, that is 791$<\rho<$1040~kg~m$^{-3}$.

\begin{figure}
\gridline{
\fig{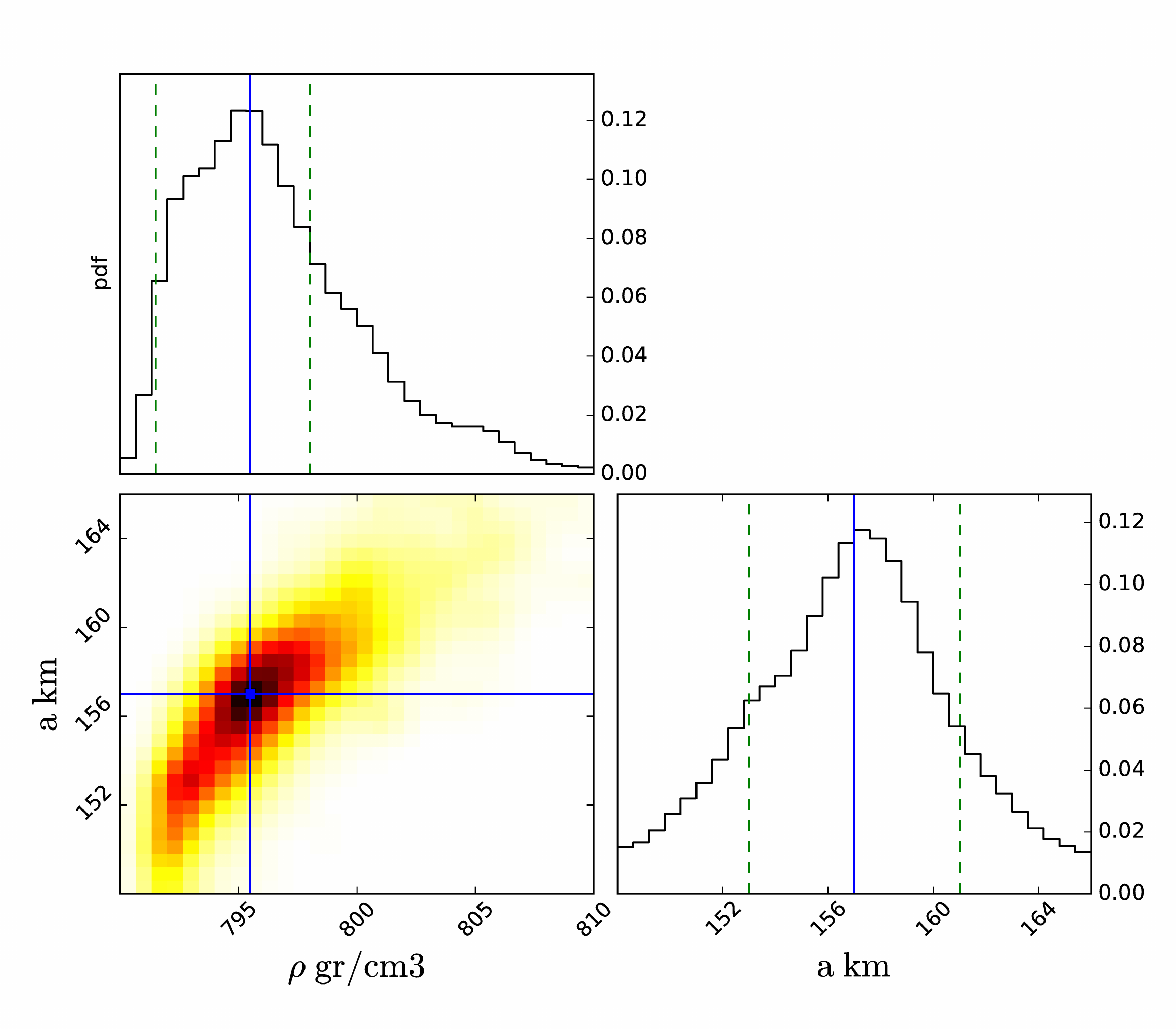}{1\textwidth}{(b) }
}
\caption{
Same as Figure~\ref{fig:posterior_spheroid} for the Jacobi ellipsoid model, 
from where we obtain the parameter values given in Table~\ref{tab:results}.
} 
\label{fig:posterior_jacobi}
\end{figure}

\begin{figure}
\plotone{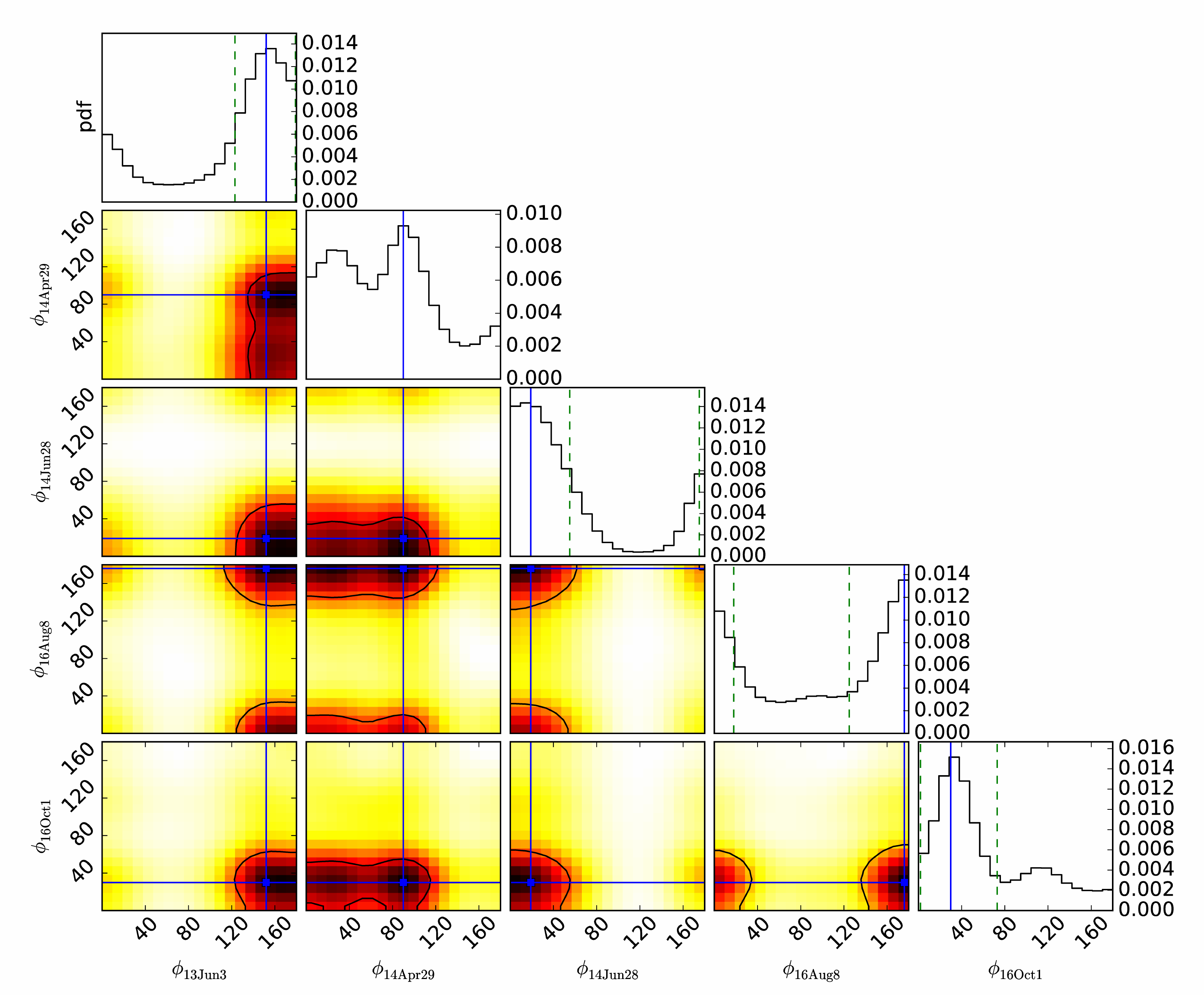}
\caption{
Posterior probability of relative rotation angle $\phi$ for each occultation. 
The best-fit values used in Figure~\ref{fig:jacobi_best_fit_bf} are indicated by blue dots and lines.
}
\label{fig:posterior_jacobi_rho_angle}
\end{figure}

Using $n_{\rm walk}$=500, $n_{\rm burn}=10^4$ and $n_{\rm samp}=10^2$,
 we obtain the posterior pdf for $\rho$ and $a$ shown in Figure~\ref{fig:posterior_jacobi},
 and the posterior for rotation angles shown in Figure~\ref{fig:posterior_jacobi_rho_angle}.
From this and Equation~\ref{eq:jacobi_1} we derive the parameter values given in Table~\ref{tab:results}.

\begin{figure*}
\gridline{
\fig{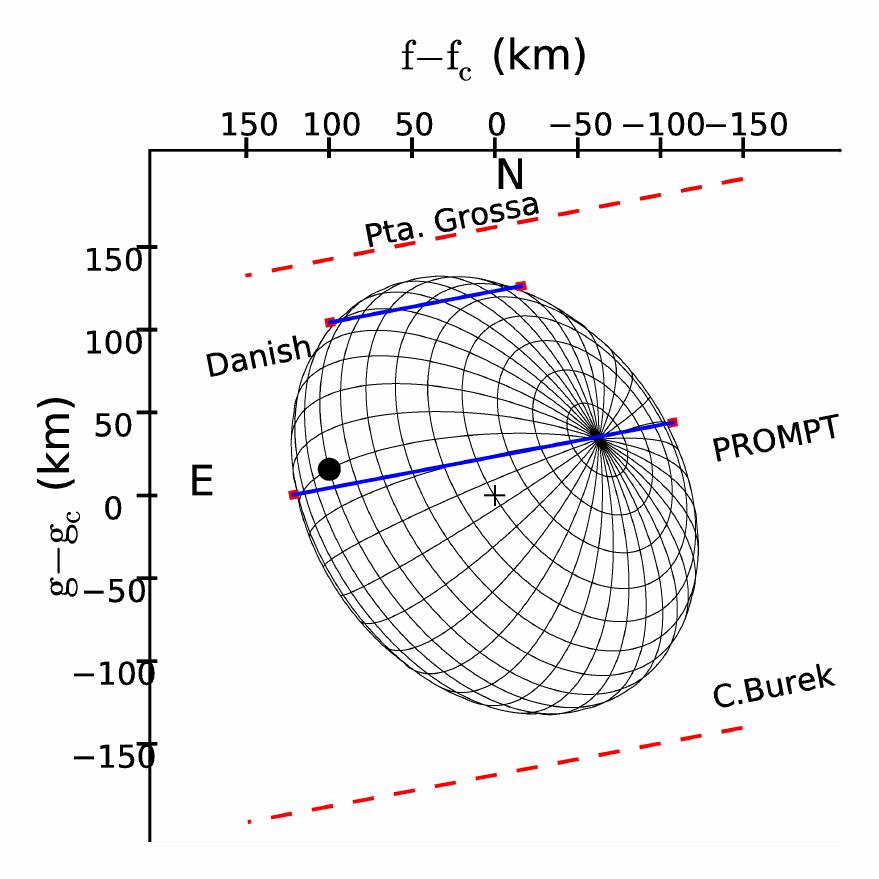}{0.35\textwidth}{(a) 2013 June 3}
\fig{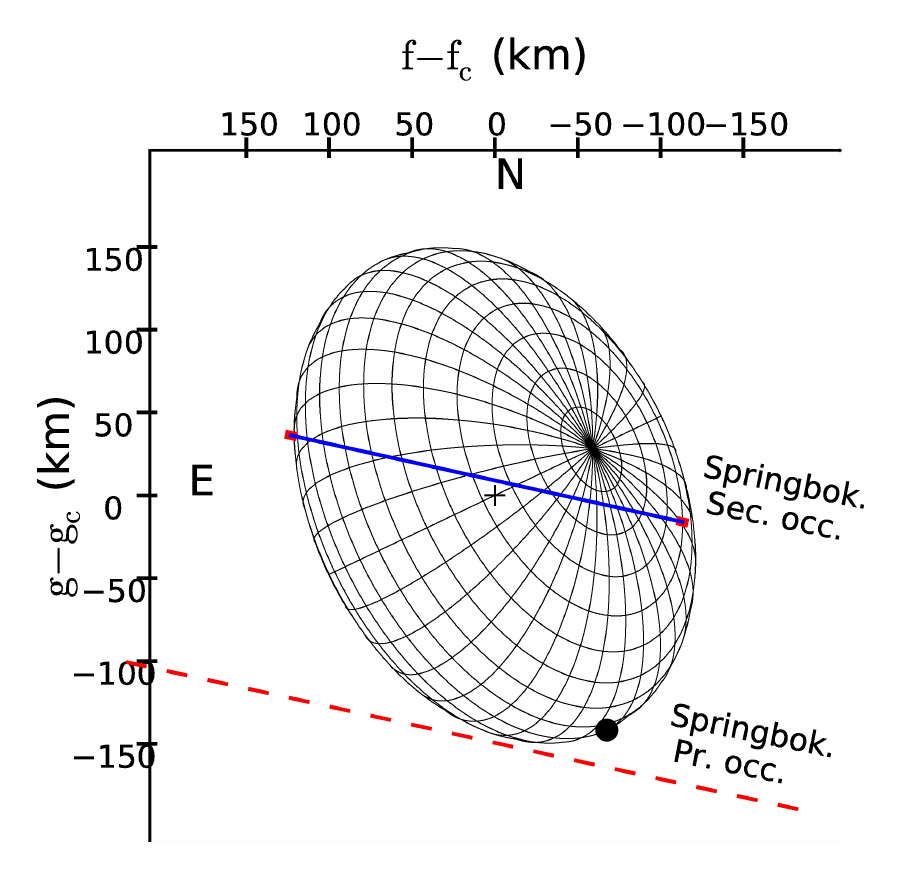}{0.35\textwidth}{(b) 2014 April 29}
}
\gridline{
\fig{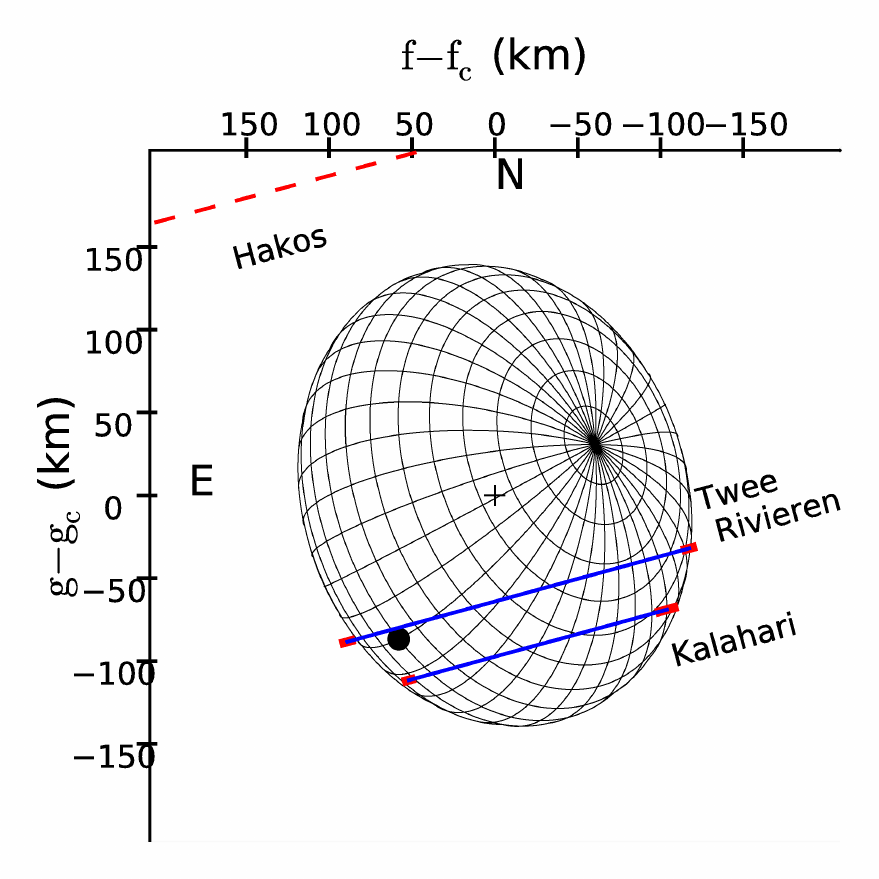}{0.35\textwidth}{(c) 2014 June 28}
\fig{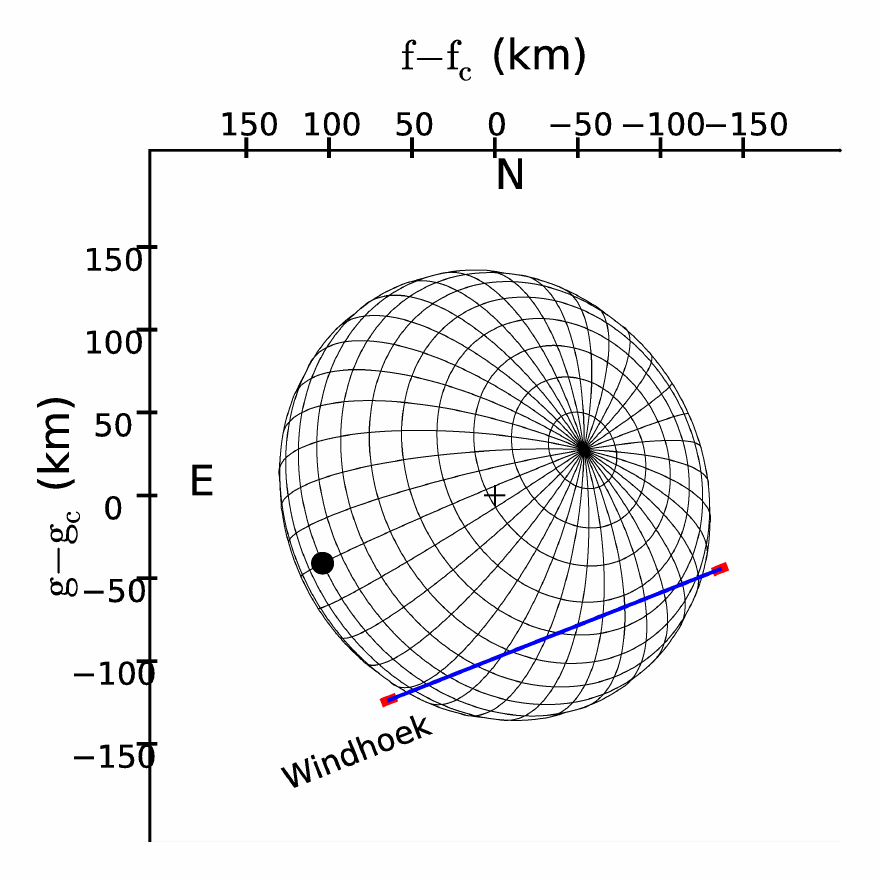}{0.35\textwidth}{(d) 2016 August 8}
}
\gridline{
\fig{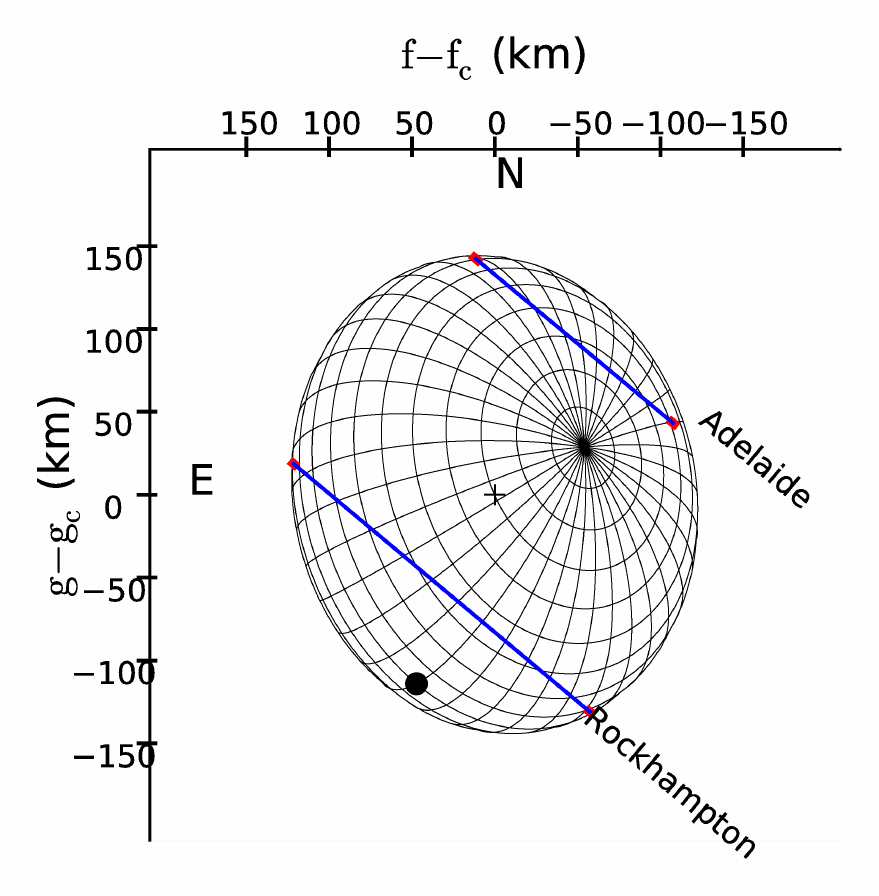}{0.35\textwidth}{(e) 2016 October 1}
}
\caption{
Same as Figure~\ref{fig:ellipsoid_best_fit_bf} for the Jacobi model with the best-fit values from Table~\ref{tab:results}.
}
\label{fig:jacobi_best_fit_bf}
\end{figure*}

Figure~\ref{fig:jacobi_best_fit_bf} shows the best-fit Jacobi models compared to the occultation chords . 
The scattering of the data points with respect to the best-fit limb is given by $\sigma_m$=6~km, similar to the case of the Maclaurin model. 
From Figure~\ref{fig:posterior_jacobi_rho_angle} we obtain a rotation angle $\phi$=152$\degr\pm$20$\degr$ for the occultation of 2013 June 3.
Considering the ellipsoid rotational symmetry, this angle is consistent with the one obtained from the rotational light curve in Section~\ref{sec:priors}.

\paragraph{Sensitivity to priors}
\label{sec:prior_sens_ell}

As done with the ellipsoidal and Maclaurin models,
we repeat a MCMC run using a uniformly distributed prior 
between 100~km and 150~km  for the semimajor axis $a$,
and between 791 and 1040~kg~m$^{-3}$ for the density $\rho$ (the equilibrium condition from Section~\ref{sec:body_model}).
We obtain 
a density $\rho=792^{+4}_{-1}$~kg~m$^{-3}$ 
and semiaxes
$a=152\pm5$~km,
$b=144^{+3}_{-4}$~km 
and $c=86^{+2}_{-1}$~km.
This is similar to the results above, with a slightly smaller object but with same elongation $(a-b)$
showing that, as with the other models, the results are dominated by the occultation data.

\section{DISCUSSION}
\label{sec:discussion}

\subsection{Topographic features and hydrostatic equilibrium}
\label{sec:topographic_features}

The ``topographic" parameter $\sigma_{m}$ (ranging from 6 to 11~km, depending on the model)
 indicates the degree of irregularity of the surface.
Compared to the equivalent radius for each model,
 this irregularities are in the range 5-9\%.
Moreover, the limb slopes measured in 2013 at one station 
(Section~\ref{sec:occultation_timing} and Figure~\ref{fig:20130603_danish_local_limb_fit})
 may reach 25$\degr$, measured with respect to the tangent of the global object limb.
  
Note that Iapetus, with typical density of 1100~kg~m$^{-3}$, can sustain slopes greater than 30$\degr$ \citep{Castillo-Rogez2007} while being more massive than Chariklo.
Similarly, Hyperion's limb profiles show local slopes of up to 20$\degr$, with respect to the fitted elliptical limb \citep{Thomas1989}.
More generally, Hyperion is irregular with topographic features of RMS$\sim$~12\% with respect to the mean radius,
 while Phoebe is close to a spheroid in equilibrium with features of RMS$\sim$~5\% \citep{Castillo-Rogez2012},
 both comparable to $d_{\rm RMS}$ values given in Table~\ref{tab:results}.

In summary, the topographic features and slopes found for Chariklo are typical of small icy satellites with size and density in the same range than Chariklo.

\subsection{Body albedo and rings reflectivity}
\label{sec:albedo}

\begin{figure}
\plotone{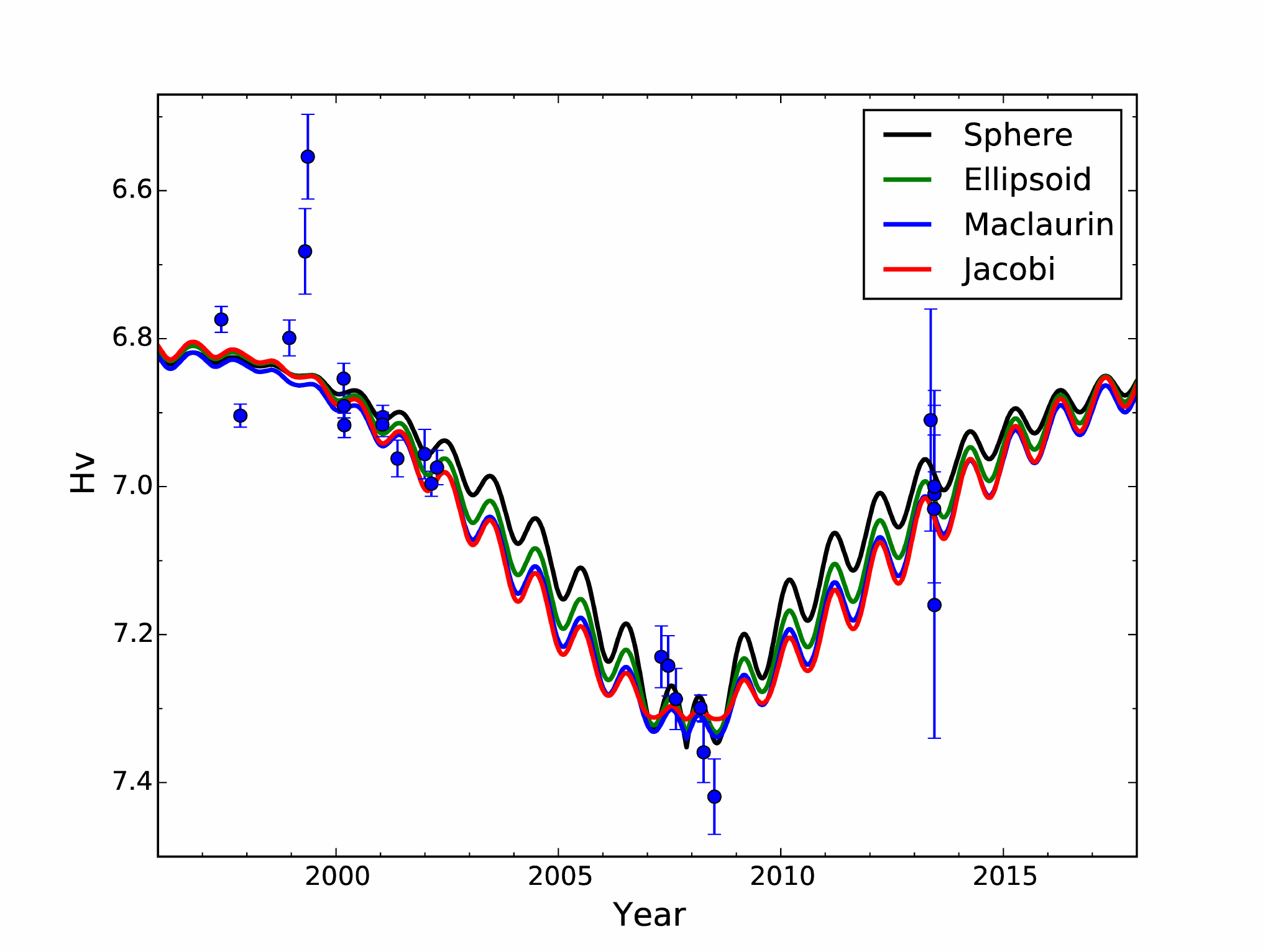}
\caption{
Best-fit to the V absolute magnitude $H_V$ of Chariklo's system.
The main body geometric albedo $p_b$ and ring reflectivity $I/F$ are fitted with a least-squares to the $H_V$ values from the literature  \citep{Belskaya2010,Fornasier2014,Duffard2014}.
In the extreme case of the spherical model,
all the brightness variation is due to the change in the rings aspect angle, with $I/F$=8.9\%.
On the other hand, the change in the projected area of the Jacobi model explains most of the long-term brightness variations,
resulting in very dark rings with $I/F$=0.6\%.
The Maclaurin and generic triaxial ellipsoid models give intermediate results (Table~\ref{tab:results}).
With both contributions, from the main body and rings considered,
all the models fit equally well the $H_V$ values.
}
\label{fig:Hv_shape_albedo}
\end{figure}

We proceed to evaluate the geometric albedo  $p_b$ of Chariklo and
 the ring reflectivity $I/F$ (see Section~\ref{sec:priors}),
considering the long-term brightness variations of Chariklo.
We use the absolute V magnitude $H_V$ from the literature and the same considerations used in Section~\ref{sec:priors}.
Table~\ref{tab:results} summarizes the $p_b$ and $I/F$ using least-squares fits to Equation~\ref{eq:hv_model}. 
Figure~\ref{fig:Hv_shape_albedo} show the fits to the $H_V$ data for the four models,
which are virtually indistinguishable to each other but
give substantially different relative contribution to the brightness variation from the rings and the main body.
The body geometric albedo $p_b$ does not depend strongly on the body model 
with values in the range 3.1\% to 4.2\%.
In contrast, the ring reflectivity $I/F$ depends on the model adopted for the body.
For instance, the spherical model attributes all the photometric variability to the ring 
resulting in $I/F$=8.9\%,
 close to the previously found value for a spherical body \citep{Braga-Ribas2014}.
However, the Jacobi model attribute most of the variability to the changing aspect of Chariklo,
resulting in a significantly darker ring, $I/F=0.6\%$. 
The Maclaurin and the generic ellipsoidal models give intermediate results with $I/F=3.4\%$ and $I/F=4.9\%$ respectively 
but lower than previously estimated values for a non-spherical body \citep{Braga-Ribas2014,Duffard2014}.

The reflectivity $I/F$ can be related to the albedo of the ring particles $p_p$ in two extreme regimes: 
a monolayer ring where the ring thickness is comparable to the particle size and
a polylayer ring where the ring thickness is large compared to the particles.
Currently, there is not enough information to discriminate between these two regimes,
 but it is illustrative to consider them in turn here.

For a monolayer ring,
 the equivalent width is defined as $E_{p}=W(1-f_n)$,
where $W$ is the radial width and $f_n$ is the fractional transmission normal to the ring \citep{Elliot1984}.
This gives the effective area covered by the ring particles, neglecting mutual shadowing.
Taking the typical value of $E_{p}$=2.2~km for the main ring C1R \citep{Berard2017} and the average width W$\sim$5.5~km 
considered here, the geometric albedo of the ring particles is $p_p=(5.5/2.2)\times I/F$.
Depending on the model used here (from sphere to Jacobi), $p_p$ ranges from 22\% to 1.5\%, respectively.

In the polylayer regime, the ring reflectivity $I/F$ can be approximated by a single scattering model \citep{Chandrasekhar1960}:
\begin{equation}
I/F=\frac{p_p}{2}\left[1-\exp\left( \frac{-2 \tau_N}{\mu}\right)\right],
\end{equation} 
where $\mu$=$\sin (B)$ and $\tau_N$ is the ring normal optical depth.
Using an approximate $\tau_N$=0.4 measured for the main ring C1R \citep{Braga-Ribas2014},
we see that 
$p_p$ is 2-3 times $I/F$, similar to the monolayer case.

The ring particles thus can be darker than those of Uranus \citep{Karkoschka2001},
$p_p\sim5\%$,
 but cannot be as bright as Saturn's ring particles \citep{Cuzzi2009}, $p_p\sim50\%$.
For the Jacobi model, the geometric albedo of the ring particles $p_p<2\%$
is lower than those of TNOs \citep{Lacerda2014}, 
with geometric albedo $p\gtrsim4\%$. 
This makes the Jacobi solution less plausible giving preference to the generic ellipsoid model,
which gives a geometric albedo of ring particles of $p_p=10-15\%$.

\subsection{Comparison with radiometric results}
\label{sec:comparison_radiometric}

Chariklo's equivalent 
radius\footnote{Defined as $r_{\rm equiv}=\sqrt{A/\pi}$, where $A$ is the apparent surface area of the body, 
not be confused with the volumetric equivalent radius $R_{\rm equiv}$ of each model,
 which does not depend on orientation.}
$r_{\rm equiv}$ has been estimated from thermal measurements,
with values ranging from 108~km to 151~km \citep{Jewitt1998,Altenhoff2001,Sekiguchi2012,Bauer2013,Fornasier2014}.

For the Maclaurin model, $r_{\rm equiv}$ only depends on the opening angle,
 while for the Jacobi models it also depends on the rotation angle.
Although observed values are compatible at the 2-$\sigma$ level with our models (with $r_{\rm equiv}$ ranging between 110~km and 140~km),
they should be taken with caution because the published radius values have been estimated from simplified models (using NEATM, in some cases even assuming a particular value of the beaming factor) or from more elaborate thermophysical models but without knowledge of pole orientation,
and in particular because all models assumed spherical shapes.

For the comparison to be realistic, a reanalysis of the thermal data is thus necessary to take into account for 
different shape models,
changes in orientation with time,
and to estimate the possible ring contribution to the thermal emission. 
These aspects are explored in \cite{Lellouch2017}.

\section{CONCLUSIONS}
\label{sec:conclusions}

The combination of results from stellar occultations with a quantitative statistical approach is a powerful tool to derive sizes and shapes of small and distant objects.
In the case of the Centaur object Chariklo, this is of great importance for constraining the dynamics of its ring system.

In this work we have explored four models for Chariklo's main body shape: a sphere, a triaxial ellipsoid, a Maclaurin spheroid, and a Jacobi ellipsoid.
Using a Bayesian approach, we combine five stellar occultations observed between 2013 and 2016 with rotation light curves to derive credible intervals for the size, shape, and density of Chariklo.

Using the spherical model, we find that topographic features with height of about 9\% of Chariklo's radius 
can explain our observations. 
This is comparable to the values of small icy bodies of similar size and density as Hyperion and Phoebe.
However, we observe a clear correlation of the radial residuals with the position angle along the limb,
being positive near the equator and negative near the pole.
This strongly suggests that Chariklo is flattened or elongated.

The ellipsoidal and Jacobi model are consistent with
 the stellar occultation data, 
the rotational light curve amplitude and, 
in the case of the occultation of 2013, with the expected rotation phase.
This suggests that Chariklo is an elongated body.

Clearly, an improved value of Chariklo's rotational period will constrain the rotational angle at each occultation date,
 and thus reduce the number of free parameters of the models.

Accounting for the fact that Chariklo may have an oblate or ellipsoidal shape, 
we find that the ring reflectivity is much less constrained than previously considered.
While a spherical shape for the body implies ring particles four times brighter than Uranus ring particles,
the Jacobi model may result in ring particles twice darker.
This large range of uncertainty is a strong incentive for improving our knowledge of Chariklo's size and shape, 
using better predicted events in the Gaia era, thus allowing well-sampled stellar occultations.

The density obtained in the cases of Jacobi and Maclaurin models is in the range 800-1250~kg~m$^{-3}$,
indicative of an icy body.
This must be taken with caution, though, as this assumes a homogeneous body in hydrostatic equilibrium. 
The corresponding mass range is 6-8$\times10^{18}$~kg. 
With that value, it is interesting to note that 
the 3:1 resonance between the mean motion of the particles and the rotation of the body\footnote{ At this resonance a ring particle undergoes one revolution while Chariklo completes three rotations. }
is located at radius 408$\pm$20~km, 
close to the radii of C1R and C2R, respectively 391~km and 405~km.
The potential implications of this resonance will be considered in another work.

\acknowledgments
R. Leiva acknowledges support from CONICYT-PCHA/Doctorado Nacional/2014-21141198. 
The authors acknowledge support from the French grant 
``Beyond Neptune II"  ANR-11-IS56-0002.
Part of the research leading to these results has received funding from the European Research Council under the European Community's H2020 (2014-2020/ ERC Grant Agreement n$^{\circ}$ 669416 ”LUCKY STAR”).
The research leading to these results has received funding from the European
Union's Horizon 2020 Research and Innovation Programme, under Grant
Agreement N$^{\circ}$. 687378, project SBNAF.
E.M. acknowledges support from the Contrato de subvenci\'on 205-2014 Fondecyt - Concytec, Per\'u.
J.I.B.C. acknowledges a CNPq grant n$^{\circ}$ 308150/2016-3.
M.A. thanks the CNPq (Grants 473002/2013-2 and
308721/2011-0) and FAPERJ (Grant E-26/111.488/2013).
G.B-R. acknowledges for the support of the CAPES (203.173/2016) and FAPERJ/PAPDRJ (E26/200.464/2015 - 227833) grants.
R.V-M thanks grants: CNPq-306885/2013, Capes/Cofecub-2506/2015, Faperj:  
PAPDRJ-45/2013 and E-26/203.026/2015
This work is partly based on observations performed at the MPG 2.2 meter telescope, program CN2016A-87.
Based on observations obtained at the SOAR telescope, program SO2015A-015.
The 50 cm telescopes used for the Hakos observations belong to the IAS observatory at Hakos/Namibia. 
This work was partially supported by the National Research Foundation of South Africa and contains data taken at the South African Astronomical Observatory (SAAO).
This work has made use of data from the European Space Agency (ESA)
mission {\it Gaia} (\url{https://www.cosmos.esa.int/gaia}), processed by
the {\it Gaia} Data Processing and Analysis Consortium (DPAC,
\url{https://www.cosmos.esa.int/web/gaia/dpac/consortium}). Funding
for the DPAC has been provided by national institutions, in particular
the institutions participating in the {\it Gaia} Multilateral Agreement.

\facilities{SOAR (SOI), Max Planck:2.2m (WFI), LNA:BC0.6m}
\software{DanDIA \citep{Bramich2008}, IRAF \citep{Tody1986}, emcee \citep{Foreman-Mackey2013}}

\bibliographystyle{aasjournal}
\bibliography{chariklo_shape}

\onecolumngrid
\newpage
\begin{deluxetable}{llll}
\tabletypesize{\scriptsize}
\tablecolumns{4}
\tablewidth{0pc}
\tablecaption{Observations used to constrain Chariklo's main body \label{tab:observations}}
\tablehead{
\colhead{Site} & \colhead{Latitude} & \colhead{Telescope aperture (m)}&\colhead{Observers} \\
\colhead{} & \colhead{Longitude} & \colhead{Camera,Filter} 		& \colhead{} \\
\colhead{} & \colhead{Altitude (m)} & \colhead{Exp. time, cycle (s)} 	& \colhead{} \\
}
\startdata
\multicolumn{4}{c}{\textbf{2013 June 3. South America.  \tablenotemark{a} }}\\
\hline
\multicolumn{4}{c}{\textbf{2014 April 29. South Africa.}}\\
\hline
Springbok 	& $29^{\circ}$ 39' 40.2"S 	& 0.3 & F. Colas \\
South Africa 	& $17^{\circ}$ 52' 58.8"W & Raptor Merlin 247,Clear  	& C. de Witt \\
   			& 951 						& 0.06, 0.06 	& \\   		
\hline 
Gifberg & $31^{\circ}$48'34.6"S & 0.3 & J.-L. Dauvergne \\
South Africa 	& $18^{\circ}$46'59.4"E & Merlin,Clear 	& P. Schoenau  \\
   		& 345 					& 0.05, 0.05  & \\
\hline 
South African Astronomical 		& $32^{\circ}$22'46.0"S & 1.9 & H. Breytenbach \\
Observatory, Sutherland (SAAO) 	& $20^{\circ}$48'38.5"E & SHOC,Clear 	& A. A. Sickafoose  \\
South Africa				& 1760					&  0.0334,0.04	&  \\
\hline   
\multicolumn{4}{c}{\textbf{2014 June 28. South Africa and Namibia.}}\\
\hline   
Kalahari Trails & $26^{\circ}$ 46' 27"S & 0.3 & L. Maquet \\
South Africa   & $20^{\circ}$ 37' 55"E & Merlin, Clear 	&  			\\
   				& 860					& 0.4, 0.4 &           \\ 
\hline 
Twee Rivieren   & $26^{\circ}$28'14"S 	& 0.3 & J.-L. Dauvergne\\
South Africa 		& $20^{\circ}$36'42"E	& Merlin,Clear 	&    \\
                & 885				  	& 0.4, 0.4&			\\
\hline 
IAS-Observatory	& $23^{\circ}$14'10"S	& 0.51	& K.-L. Bath	\\
Hakos			& $16^{\circ}$21'42"E	& Merlin,Clear 	& 			\\
Namibia       & 1695  		  		& 0.2, 0.2&			\\
\hline   
\multicolumn{4}{c}{\textbf{2016 August 8. Namibia.}}\\
\hline   
Windhoek & $22^{\circ}$ 41' 54.9"S & 0.35 & H.-J. Bode \\
Namibia   & $17^{\circ}$ 6' 32.4"E &  ZWO~ASI120MM,Clear 	&  			\\
   				& 1900 m					&1, 1 &           \\ 
\hline   
\multicolumn{4}{c}{\textbf{2016 October 1. Australia.}}\\
\hline   
The Heights Observatory & $34^{\circ}$ 48' 44.7"S  & 0.3 & A. Cool	\\
Adelaide	& $138^{\circ}$ 40' 56.9"E & QHY5L-II, Clear 	&  B. Lade \\
Australia 	& 100 m 			& 1, 1  &           			\\ 
\hline   
Rockhampton & $23^{\circ}$ 16' 9.6"S  & 0.3 & S. Kerr	\\
Australia	& $150^{\circ}$ 30' 0.7"E & Watec 910BD,Clear 	&    \\
 	& 50 m 			& 0.32, 0.32  &           			\\ 
\enddata
\tablecomments{
The stations considered here involve occultation by Chariklo's main body where 
Chariklo's rings were simultaneously detected.
In those cases, the general geometry of the system can be constrained
including the apparent center and orientation of the pole axis.
The Springbok and Hakos stations
provided negative results (that is, no occultation by the body) 
but which occultation chords are sufficiently close to Chariklo's main body giving strong constrains in the extension of it (see Figure~\ref{fig:occ_geom}).
For the same reasons, 
we consider in this analysis the observations from Danish and PROMPT telescopes (body detections)
as well as Ponta Grossa and Cerro Burek (no occultation by the body) of the occultation of 2013 June 3.
\tablenotetext{a}{Circumstances of the these observations are given in \cite{Braga-Ribas2014}.}
}
\end{deluxetable}

\clearpage
\begin{table*}
\centering
\caption{Coordinates of Chariklo and the occulted star.}\label{tab:coord_and_eph}
\begin{tabular}{llllll}
          &  \multicolumn2c{2014 Apr 29} & 2014 Jun 28 & 2016 Aug 8 & 2016 Oct 1 \\ 
\hline
$\alpha$ & \tablenotemark{a} 17$^{\rm h}$39$^{\rm m}$2.1336$^{\rm s}$ & -10.1 mas  \tablenotemark{b} & 17$^{\rm h}$24$^{\rm m}$50.3800$^{\rm s}$ & 18$^{\rm h}$18$^{\rm m}$3.6927$^{\rm s}$ & 18$^{\rm h}$16$^{\rm m}$20.0796$^{\rm s}$ \\ 
$\delta$ & \tablenotemark{a} -38$^{\circ}$52'48.802" & -17.8 mas  \tablenotemark{b} & -38$^{\circ}$41'5.618"  & -33$^{\circ}$52'28.3920" & -33$^{\circ}$ 1' 10.756" \\ 
V & ... & ... & 15.20 & 14.06 & 14.33 \\ 
K & ... & ... & 12.47 & 12.1 & 13.253\\ 
$\theta_{\rm LD}$ (mas)  & ... & ... & 0.011$\pm$0.005 & 0.015$\pm$0.003 & 0.007$\pm$0.002 \\
D$_{\rm star}$ (km)   & 0.2$\pm$0.02 \tablenotemark{c} & 0.09$\pm$ 0.02 \tablenotemark{c} & 0.11$\pm$0.05 & 0.16$\pm$0.03 & 0.08$\pm$0.02 \\ 
t$_{\rm ref}$ UTC &  \multicolumn2c{23:10:00} & 22:24:00 & 19:57:00 & 10:10:00 \\ 
$\alpha_{\rm Ch}$ &  \multicolumn2c{17$^{\rm h}$39$^{\rm m}$2.1566$^{\rm s}$} & 17$^{\rm h}$24$^{\rm m}$50.3991$^{\rm s}$  & 18$^{\rm h}$18$^{\rm m}$3.6908$^{\rm s}$ & 18$^{\rm h}$16$^{\rm m}$20.0982$^{\rm s}$ \\ 
$\delta_{\rm Ch}$ &  \multicolumn2c{-38$^{\circ}$52'48.739"} & -38$^{\circ}$41'5.628"  & -33$^{\circ}$52'28.241" & -33$^{\circ}$1'10.8424" \\ 
d$_{\rm Ch}$  (km)  &  \multicolumn2c{2.109$\times$10$^9$} & 2.075$\times$10$^9$ & 2.193$\times$10$^9$ & 2.319$\times$10$^9$  \\ 
\hline 
\end{tabular} 
\tablecomments{
($\alpha$,$\delta$) are the right ascension and declination of the occulted star,
while d$_{\rm Ch}$ is Chariklo's geocentric range and
($\alpha_{\rm Ch}$,$\delta_{\rm Ch}$)  the predicted Chariklo's right ascension and declination
at the reference time t$_{\rm ref}$.
V, K are the star magnitudes from NOMAD catalog \citep{Zacharias2004}.
$\theta_{\rm LD}$ is stellar angular diameter,
while D$_{\rm star}$ is the stellar diameter projected at the distance of Chariklo,
 from V-K relations in \cite{Kervella2004} after considering galactic reddening.
For the stellar occultation of 2013 June 3 we adopted values from \cite[Supplementary information]{Braga-Ribas2014}
}
\tablenotetext{a}{Primary component of double star.}
\tablenotetext{b}{Offset of the secondary component with respect to primary.}
\tablenotetext{c}{Fitting atmospheric models to primary and secondary stars, for details see \citealt{Berard2017}.}
\end{table*}

\begin{table*}
\begin{center}
\caption{Occultation data used to constrains Chariklo's size and shape.}
\label{tab:timing_body}
\begin{tabular}{lllllll}
\tableline\tableline
Site 	& ing/egr 	& UTC time & f (km) & g (km) & $\sigma_{ch}$(km) \tablenotemark{d}\\
\tableline
\multicolumn{5}{c}{2013 June 3. South America.} 	\\
\tableline
Pta. Grossa \tablenotemark{a} & \multicolumn{2}{l}{No body detection} &		&   		& 		 \\
Danish 	& ing 	& 6:25:27.893$\pm$0.014 s &  -2750.6 & 920.2 & 0.3 \\
Danish 	& egr 	& 6:25:33.188$\pm$0.014 s &  -2635.3 & 898.2 & 0.3 \\
PROMPT\tablenotemark{a} & ing 	& 6:25:24.835$\pm$0.009 s &  -2842.0 & 837.7 & 0.2 \\
PROMPT & egr 	& 	6:25:35.402$\pm$0.015 s & -2613.3 & 794.2 & 0.3 \\
Cerro Burek \tablenotemark{a} & \multicolumn{2}{l}{No body detection} &		&   		& 		 \\
\hline
\multicolumn{5}{c}{2014  April 29. South Africa.} \\
\tableline
Springbok \tablenotemark{b}		& \multicolumn{2}{l}{No body detection} &		&   		& 		 \\
Springbok \tablenotemark{c}	& ing 	& 23:14:30.04$\pm$0.07 s & -2887.6(-2782.6) & 321.0(503.0) & 0.9 \\
Springbok \tablenotemark{c}	& egr 	& 23:14:48.05$\pm$0.07 s & -2651.3(-2546.3) & 373.2(555.2) & 0.9 \\
\tableline
\multicolumn{5}{c}{2014 June 28  South Africa and Namibia} 	\\
\tableline
Hakos		& \multicolumn{2}{l}{No body detection} &		&   		& 		 \\
Kalahari 		& ing 	& 22:24:07.48$\pm$0.20 s & -878.6 	& 1306.6 & 4.4 \\ 
Kalahari 		& egr  	& 22:24:14.86$\pm$0.07 s & -723.2 	& 1264.2 &  1.5 \\ 
Twee Rivieren 	& ing 	& 22:24:06.73$\pm$0.10 s & -892.6 	& 1343.7 & 2.2 \\  
Twee Rivieren 	& egr  	& 22:24:16.54$\pm$0.10 s & -686.3 	& 1287.4 & 2.2 \\   
\tableline
\multicolumn{5}{c}{ 2016 August 8. Namibia. }\\
\tableline
Windhoek 		& ing 	& 19:57:28.460$\pm$0.13 s & 631.1 & -520.5 & 2.1 \\ 
Windhoek 		& egr  	& 19:57:41.870$\pm$0.14 s & 831.1 & -599.6 & 2.2 \\ 
\tableline
\multicolumn{5}{c}{ 2016 October 1.  Australia.} \\
\tableline
Rockhampton 	& ing 	& 10:12:44.66$\pm$0.04 s & -497.8 &    0.4 & 0.5 \\ 
Rockhampton 	& egr  	& 10:13:03.20$\pm$0.06 s & -676.7 & -149.2 & 0.8 \\ 
Adelaide 		& ing 	& 10:10:41.82$\pm$0.10 s & -607.9 & 124.1 & 1.3 \\ 
Adelaide 		& egr  	& 10:10:54.16$\pm$0.08 s & -726.8 & 25.1 & 1.0 \\ 
\end{tabular}
\tablecomments{
Here we list only the positive occultations by Chariklo's body and
the negative occultations close enough to the body used to constrain Chariklo's size and shape.
The second column indicate the detection of the ingress and egress in Chariklo's occultation shadow at each site. 
The values $(f,g)$ are offset of the star with respect to the expected body center $(f,g)=(0,0)$ as seen from each site at ingress and egress.
Those offsets are measured in the sky plane at Chariklo's distance and are counted respectively positive toward the east and north.
\tablenotetext{a}{From \citep[extended data table 5]{Braga-Ribas2014}.}
\tablenotetext{b}{Occultation of primary star (see text).}
\tablenotetext{c}{Occultation of secondary star (see text).}
\tablenotetext{d}{The uncertainty $\sigma_{ch}$ measured in the direction of the occultation chord as derived from the timing uncertainties.}
}
\end{center}
\end{table*}

\begin{table}
\begin{center}
\caption{Midtime of the occultations by Chariklo's rings.}
\label{tab:timing_ring}
\begin{tabular}{llllllll}
\tableline\tableline
Site 	& ing/egr 	& $t_{\rm mid}$ & f  		& g 		& $v_{ch}$ & $\sigma_{ch}$ \\
		& 				& UTC & (km)	& (km)	& (km~s$^{-1}$) & (km) \\
\tableline
\multicolumn{6}{c}{2014  April 29. South Africa. } \\
\tableline
\multicolumn{6}{c}{C1R}\\
Springbok \tablenotemark{a}	& ing 	& 23:14:25.884 $\pm$ 0.007 s & -2942.16 & 308.95 & 13.4 & 0.1 \\ 
Springbok \tablenotemark{a}	& egr 	& 23:15:04.362 $\pm$ 0.006 s & -2437.27 & 420.48 & 13.4 & 0.1 \\  
SAAO \tablenotemark{b}		& ing 	& 23:13:56.191 $\pm$ 0.007 s & -3017.66 &  56.58 & 13.4 & 0.1 \\
SAAO \tablenotemark{b}		& egr 	& 23:14:28.964 $\pm$ 0.008 s & -2587.52 & 151.12 & 13.4 & 0.1 \\
\multicolumn{6}{c}{C2R}\\
Springbok \tablenotemark{a}	& ing 	& 23:14:24.990 $\pm$ 0.020 s & -2953.89 & 306.36 & 13.4 & 0.3 \\ 
Springbok \tablenotemark{a}	& egr 	& 23:15:05.324 $\pm$ 0.019 s & -2424.65 & 423.27 & 13.4 & 0.3 \\  
Gifberg \tablenotemark{a}	& ing 	& 23:14:30.109$^{+0.015}_{-0.008}$  s & -2742.39 &  137.87 & 13.4 & $^{+0.2}_{-0.1}$ \\
Gifberg \tablenotemark{a}	& egr 	& 23:14:33.750 $\pm$ 0.008  s & -2694.62 & 148.40 & 13.4 & 0.1 \\
\tableline
\multicolumn{6}{c}{2014 June 28. Namibia. }\\
\tableline
\multicolumn{6}{c}{C1R+C2R unresolved}\\
Hakos 		& ing 	& 22:24:25.796 $\pm$ 0.041 s & -886.44 & 1629.24 & 21.8 & 0.9 \\ 
Hakos 		& egr 	& 22:24:44.218 $\pm$ 0.035 s & -498.24 & 1523.61 & 21.8 & 0.8 \\
\tableline
\multicolumn{6}{c}{ 2016 August 08. Namibia. }\\
\tableline
\multicolumn{6}{c}{C1R+C2R unresolved}\\
Windhoek 		& ing 	& 19:57:18.209 $\pm$ 0.249 s & 478.18 & -459.92 & 16.0 & 4.0 \\ 
Windhoek 		& egr 	& 19:57:51.892 $\pm$ 0.109 s & 980.60 & -658.82 & 16.0 & 1.8 \\
\tableline
\multicolumn{6}{c}{ 2016 October 1. Australia. }\\
\tableline
\multicolumn{6}{c}{C1R+C2R unresolved}\\
Rockhampton & ing & 10:12:26.284 $\pm$ 0.072 s & -320.56 & 148.77 & 12.6 & 0.91 \\ 
Rockhampton	& egr & 10:13:22.928 $\pm$ 0.049 s & -876.01 & -308.47 & 12.6 & 0.62 \\ 
\hline
Adelaide & ing & 10:10:19.826 $\pm$ 0.186 s & -396.05 & 300.41 & 12.5 & 2.3 \\
Adelaide & egr & 10:11:14.558 $\pm$ 0.218 s & -923.38 & -138.48 & 12.5 & 2.7 \\
\end{tabular}

\tablecomments{
C1R is the internal and wider ring while C2R is the external ring. 
$t_{\rm mid}$ are the midtimes of the occultations by the rings, from \citealt{Berard2017}.
Ingress/egress indicates the first and second detection of the respective ring.
The values $(f,g)$ are offset of the star with respect to the expected body center $(f,g)=(0,0)$ as seen from each site derived from the midtimes $t_{\rm mid}$.
Those offsets are measured in the sky plane at Chariklo's distance and are counted respectively positive toward the east and north.
$v_{ch}$ is the speed of the star relative to Chariklo, projected in the sky plane.
$\sigma_{ch}$ is the uncertainty of the chord extremity measured along the chord direction as given by the timing uncertainty.
\tablenotetext{a}{Occultation of primary star (see text).}
\tablenotetext{b}{Occultation of secondary star (see text).}
}
\end{center}
\end{table}

\begin{table}
\begin{center}
\caption{Adopted ring geometry.}
\label{tab:ring_geometry}
\begin{tabular}{lcccc}
\tableline\tableline 
Pole position \tablenotemark{a} 	& \multicolumn{4}{c}{$\alpha_{P}=151.30^{\circ} \pm 0.5^{\circ}$ and $\delta_{P}=41.48^{\circ} \pm 0.2^{\circ}$} \\
C1R radius\tablenotemark{a}  (km) & \multicolumn{4}{c}{ 390.6 $\pm$ 3.3 }\\
C2R radius \tablenotemark{a} (km) & \multicolumn{4}{c}{ 404.8 $\pm$ 3.3 }\\
\tableline
Date & Opening angle - $B$ & Position angle - $P$ & $f_c$ & $g_c$ \\
	     &  ($^{\circ} $)         &  ($^{\circ} $)  		& (km) 	  & (km) \\
\tableline
2013 Jun 3 \tablenotemark{a} 	& 33.8 $\pm$ 0.4  & -61.6 $\pm$ 0.1 & -2734.7$\pm$ 0.5 & 793.8 $\pm$ 1.4\\
2014 Apr 29  & 40.4 $\pm$ 0.4   	& -64.5 $\pm$ 0.4 & -2669.4$\pm$ 0.2 & 519.0$\pm$ 0.1\\
2014 Jun 28 & 37.8 $\pm$ 0.4   	& -63.1 $\pm$ 0.4 & -775.3$\pm$ 0.5 & 1375.8$\pm$ 0.5\\
2016 Aug 8  \tablenotemark{b}	& 45.2 $\pm$ 0.4	& -62.9 $\pm$ 0.4 & 767.1(691.7) $\pm$ 4.0 & -475.95(-642.8) $\pm$ 7.0 \\
2016 Oct 1	& 44.5 $\pm$ 0.4	& -62.0 $\pm$ 0.4 & -619.3 $\pm$ 1.0 & -17.8 $\pm$ 3.4 	\\
\tableline
\end{tabular}
\tablecomments{
Opening angle $B$ is the elevation of the observer above the ring plane.
$P$ is the position angle of the semiminor axis of the ring projected in the sky plane, counted positively from celestial north towards east. 
With the assumptions used here, $B$ and $P$ corresponds to
the planetocentric declination of the Earth
and the position angle of the pole axis respectively (see Section~\ref{sec:phy_model}).
$f_c$ and $g_c$ are the coordinates of the center of the ring in the sky plane
measured with respect to the expected body center $(f,g)=(0,0)$.
\tablenotetext{a}{From \citealt[ED Table 4]{Braga-Ribas2014} }
\tablenotetext{b}{For the occultation of 2016 August 8 there are two possible solutions for the center of the system.}
}
\end{center}
\end{table}

\begin{table}
\begin{center}
\caption{Rotational light curve amplitudes.}\label{tab:rotational_amplitude}
\begin{tabular}{llll}
\hline
Date & $\Delta m$ & B & Reference \\ 
        & (mag) & $(\degr)$ &  \\ 
\hline
1997 May & $<$0.02  & -56 & \citep{Davies1998} \\ 
1999 Mar & $<$0.05 & -53 & \citep{Peixinho2001} \\ 
2006 Jun & 0.13$\pm$0.03 & -13 & \citep{Galiazzo2016} \\ 
2013 Jun & 0.11$\pm$0.02  & 34 & \citep{Fornasier2014} \\ 
2015 Jul & 0.06$\pm$0.02 & 42   & This work \\ 
\hline
\end{tabular}
\tablecomments{
Peak-to-peak amplitude $\Delta_m$ of the rotational light curve measured for Chariklo at different opening angles $B$. Upper limits for the amplitude in 1997 and 1999, and uncertainties in 2006 and 2013 are estimated from uncertainties in the photometry given by the authors. The amplitude in 2015 is the one obtained from data taken with the SOI camera at SOAR telescope. 
}
\end{center}
\end{table}

\begin{table}
\begin{center}
\caption{Physical parameters of Chariklo from stellar occultations.}
\label{tab:results}
\begin{tabular}{lccccc}
\tableline
Parameter 			& Sphere 		& Maclaurin 	& Ellipsoid			& Jacobi \\
\tableline
$\rho$  (kg~m$^{-3}$) & ...	 		    & $970^{+300}_{-180}$ 	& ...	                             &  $796^{+2}_{-4}$  \\
a (km) 			          & 129 $\pm$ 3  & $143^{+3}_{-6}$ 	        & $ 148^{+6}_{-4}$ & $157 \pm4 $   \\
b (km)				       & 129 $\pm$ 3  & $143^{+3}_{-6}$ 	        & $132^{+6}_{-5}$   & $139 \pm 4 $   \\
c (km)                    & 129 $\pm$ 3  & $96^{+14}_{-4}$ 			&  $102^{+10}_{-8}$ &  $86 \pm 1 $    \\
$R_{\rm equiv}$ (km) & 129 $\pm$ 3  & $126\pm2$                  & $126\pm2$               & $123^{+3}_{-1}$	 \\
$\sigma_{m}$ (km) 		& 11 			& 7 						            & 6                             &  6  \\
\hline
$\rm d_{ RMS}$ (km)  & 10			& 7 						            &   5          & 5 \\
$\rm d_{ max}$ (km) 	& +15			& +11					            &   +12      & +9  \\
\tableline
Mass (kg) 				   &  ...	                 & $ 8\pm 1\times 10^{18}$ & ... &$6.1 \pm 0.1 \times 10^{18}$ \\
$p_b$ (\%)				   & $3.1\pm0.1$  & $3.8\pm0.1$        &   3.7$\pm$0.1    & 4.2 $\pm$ 0.1 \\
$I/F$ (\%) 			       & $8.9\pm0.3$  & $3.4\pm0.3$         &  4.9$\pm$0.3     & 0.6 $\pm$ 0.4  \\
\tableline
\end{tabular}
\tablecomments{
Best parameter values and formal uncertainties from 68\% credible intervals obtained with prior as defined in Section~\ref{sec:priors}.
See Section~\ref{sec:results} for the sensitivity of results to the priors chosen.
$d_{\rm RMS}$ is the RMS dispersion in the radial direction with respect to the nominal body limb. 
$\rm d_{ max}$ is the maximum distance in the radial direction with respect to the nominal body limb.
$p_b$ is the geometric albedo of the body while $I/F$ is the ring reflectivity considering only the main ring with a width $W$=5.5 km (not to be confused with the geometric albedo of the ring particles $p_p$) as determined in Section~\ref{sec:albedo}.
$R_{\rm equiv}$=$(a \times b \times c)^{1/3}$ is the volumetric equivalent radius.
}
\end{center}
\end{table}


\clearpage

\end{document}